\documentclass[english]{article}
\usepackage[utf8]{inputenc}
\usepackage{duomasterforside}
\usepackage{hyperref}

\usepackage{graphicx}
\usepackage{amsmath}
\usepackage{enumerate}
\usepackage{amsthm}
\usepackage{amssymb}
\usepackage{paralist}
\usepackage{dsfont}
\usepackage{booktabs}
\usepackage{tabularx}
\usepackage{mathtools}
\usepackage{verbatim}
\usepackage{ dsfont }
\usepackage{ xcolor }
\usepackage{environ}
\usepackage{lipsum}
\usepackage{wrapfig}
\usepackage{graphicx}
\usepackage{algorithm}
\usepackage{algpseudocode}
\usepackage{caption}
\usepackage{float}
\usepackage[acronym]{glossaries}
\usepackage{afterpage}

\graphicspath{ {./images/} }

\counterwithin{figure}{section}
\numberwithin{equation}{section}

\DeclareMathOperator*{\argmax}{arg\,max}

\newcommand\numberthis{\addtocounter{equation}{1}\tag{\theequation}}

\newcommand\blankpage{%
    \null
    \thispagestyle{empty}%
    \addtocounter{page}{-1}%
    \newpage}

\makeglossaries
\newacronym{ml}{ML}{Machine Learning}
\newacronym{dl}{DL}{Deep Learning}
\newacronym{rl}{RL}{Reinforcement Learning}
\newacronym{sl}{SL}{Supervised Learning}
\newacronym{ann}{ANN}{Artificial Neural Network}
\newacronym{dnn}{DNN}{Deep Neural Network}
\newacronym{ffn}{FFN}{Feedforward Network}
\newacronym{rnn}{RNN}{Recurrent Neural Network}
\newacronym{lstm}{LSTM}{Long Short-Term Memory}
\newacronym{cnn}{CNN}{Convolutional Neural Network}
\newacronym{mse}{MSE}{Mean Squared Error}
\newacronym{mdp}{MDP}{Markov Decision Process}
\newacronym{pomdp}{POMDP}{Partially Observable Markov Decision Process}
\newacronym{mpt}{MPT}{Modern Portfolio Theory}
\newacronym{emh}{EMH}{Efficient Market Hypothesis}
\newacronym{amh}{AMH}{Adaptive Market Hypothesis}
\newacronym{iid}{IID}{Independent and Identically Distributed}
\newacronym{cv}{CV}{Cross-Validation}
\newacronym{sgd}{SGD}{Stochastic Gradient Descent}
\newacronym{relu}{ReLU}{Rectified Linear Unit}
\newacronym{pg}{PG}{Policy Gradient}
\newacronym{ac}{AC}{Actor-Critic}
\newacronym{mdd}{MDD}{Maximum Drawdown}
\newacronym{dqn}{DQN}{Deep Q-Network}
\newacronym{drqn}{DRQN}{Deep Recurrent Q-Network}
\begin{document}

\title{Deep Policy Gradient Methods in Commodity Markets}
\author{Jonas Rotschi Hanetho}

\duoforside[dept={Institute for Informatics}, program={Informatics: Programming and System Architecture}, long]

\pagenumbering{roman} 

\section*{Acknowledgements}

This thesis would not have been possible without my supervisors, Dirk Hesse and Martin Giese. My sincere thanks are extended to Dirk for his excellent guidance and mentoring throughout this project and to Martin for his helpful suggestions and advice. 
Finally, I would like to thank the Equinor data science team for insightful discussions and for providing me with the tools needed to complete this project.

\afterpage{\blankpage}
\newpage
\section*{Abstract}

The energy transition has increased the reliance on intermittent energy sources, destabilizing energy markets and causing unprecedented volatility, culminating in the global energy crisis of 2021. In addition to harming producers and consumers, volatile energy markets may jeopardize vital decarbonization efforts. Traders play an important role in stabilizing markets by providing liquidity and reducing volatility. Forecasting future returns is an integral part of any financial trading operation, and several mathematical and statistical models have been proposed for this purpose. However, developing such models is non-trivial due to financial markets' low signal-to-noise ratios and nonstationary dynamics.

This thesis investigates the effectiveness of deep reinforcement learning methods in commodities trading. 
It presents related work and relevant research in algorithmic trading, deep learning, and reinforcement learning. 
The thesis formalizes the commodities trading problem as a continuing discrete-time stochastic dynamical system. 
This system employs a novel time-discretization scheme that is reactive and adaptive to market volatility, providing better statistical properties for the sub-sampled financial time series. 
Two policy gradient algorithms, an actor-based and an actor-critic-based, are proposed for optimizing a transaction-cost- and risk-sensitive trading agent. The agent maps historical price observations to market positions through parametric function approximators utilizing deep neural network architectures, specifically CNNs and LSTMs.

On average, the deep reinforcement learning models produce an $83$ percent higher Sharpe ratio than the buy-and-hold baseline when backtested on front-month natural gas futures from 2017 to 2022. 
The backtests demonstrate that the risk tolerance of the deep reinforcement learning agents can be adjusted using a risk-sensitivity term. 
The actor-based policy gradient algorithm performs significantly better than the actor-critic-based algorithm, and the CNN-based models perform slightly better than those based on the LSTM. 
The backtest results indicate the viability of deep reinforcement learning-based algorithmic trading in volatile commodity markets.

\afterpage{\blankpage}
\newpage
\listoffigures

\afterpage{\blankpage}
\newpage
\listoftables

\afterpage{\blankpage}
\newpage
\tableofcontents{}

\afterpage{\blankpage}
\newpage
\cleardoublepage\pagenumbering{arabic}
\section{Introduction}\label{sec:intro}
%Overview of themes addressed. 

\subsection{Motivation}

The transition to sustainable energy sources is one of the most critical challenges facing the world today. 
By 2050, the European Union aims to become carbon neutral \cite{europeancommission}. However, rising volatility in energy markets, culminating in the 2021 global energy crisis, complicates this objective. 
Supply and demand forces determine price dynamics, where an ever-increasing share of supply stems from intermittent renewable energy sources such as wind and solar power. 
Increasing reliance on intermittent energy sources leads to unpredictable energy supply, contributing to volatile energy markets \cite{sinsel2020challenges}. 
Already volatile markets are further destabilized by evolutionary traits such as fear and greed, causing human commodity traders to overreact \cite{lo2004adaptive}.
Volatile markets are problematic for producers and consumers, and failure to mitigate these concerns may jeopardize decarbonization targets.

Algorithmic trading agents can stabilize commodity markets by systematically providing liquidity and aiding price discovery \cite{isichenko2021quantitative, narang2013inside}.
Developing these methods is non-trivial as financial markets are non-stationary with complicated dynamics \cite{taleb1997dynamic}. 
\textit{Machine learning} (ML) has emerged as the preferred method in algorithmic trading due to its ability to learn to solve complicated tasks by leveraging data \cite{isichenko2021quantitative}. 
The majority of research on ML-based algorithmic trading has focused on forecast-based \textit{supervised learning} (SL) methods, which tend to ignore non-trivial factors such as transaction costs, risk, and the additional logic associated with mapping forecasts to market positions \cite{fischer2018reinforcement}.
\textit{Reinforcement learning} (RL) presents a suitable alternative to account for these factors. 
In reinforcement learning, autonomous agents learn to perform tasks in a time-series environment through trial and error without human supervision. 
Around the turn of the millennium, Moody and his collaborators \cite{moody1997optimization, moody1998performance, moody2001learning} made several significant contributions to this field, empirically demonstrating the advantages of reinforcement learning over supervised learning for algorithmic trading.

In the last decade, the \textit{deep learning} (DL) revolution has made exceptional progress in areas such as image classification \cite{he2015delving} and natural language processing \cite{vaswani2017attention}, characterized by complex structures and high signal-to-noise ratios. 
The strong representation ability of deep learning methods has even translated to forecasting low signal-to-noise financial data \cite{xiong2015deep, hiransha2018nse, mcnally2018predicting}.
In complex, high-dimensional environments, deep reinforcement learning (deep RL), i.e., integrating deep learning techniques into reinforcement learning, has yielded impressive results.
Noteworthy contributions include achieving superhuman play in Atari games \cite{mnih2013playing} and chess \cite{silver2017mastering}, and training a robot arm to solve the Rubik's cube \cite{akkaya2019solving}. 
A significant breakthrough was achieved in 2016 when the deep reinforcement learning-based computer program AlphaGo\cite{silver2016mastering} beat top Go player Lee Sedol. In addition to learning by reinforcement learning through self-play, AlphaGo uses supervised learning techniques to learn from a database of historical games. In 2017, an improved version called AlphaGo Zero\cite{silver2017masteringGo}, which begins with random play and relies solely on reinforcement learning, comprehensively defeated AlphaGo. 
Deep RL has thus far been primarily studied in the context of game-playing and robotics, and its potential application to financial trading remains largely unexplored. 
Combining the two seems promising, given the respective successes of reinforcement learning and deep learning in algorithmic trading and forecasting.

\subsection{Problem description}\label{intro:problem-description}

This thesis investigates the effectiveness of deep reinforcement learning methods in commodities trading. It examines previous research in algorithmic trading, state-of-the-art reinforcement learning, and deep learning algorithms. 
The most promising methods are implemented, along with novel improvements, to create a transaction-cost- and risk-sensitive parameterized agent directly outputting market positions. 
The agent is optimized using reinforcement learning algorithms, while deep learning methods extract predictive patterns from raw market observations. 
These methods are evaluated out-of-sample by backtesting on energy futures.

Machine learning relies on generalizability. 
A common criticism against algorithmic trading approaches is their alleged inability to generalize to ``extreme'' market conditions \cite{narang2013inside}.
This thesis investigates the performance of algorithmic trading agents out-of-sample under unprecedented market conditions caused by the energy crisis during 2021-2022. 
It will address the following research questions
\begin{enumerate}
\item Can the risk of algorithmic trading agents operating in volatile markets be controlled? 
\item What reinforcement learning algorithms are suitable for optimizing an algorithmic training agent in an online, continuous time setting?
\item What deep learning architectures are suitable for modeling noisy, non-stationary financial data? 
\end{enumerate}

\subsection{Thesis Organization}
The thesis consists of three parts: the background (part \ref{part:background}), the methodology (part \ref{part:methodology}), and the experiments (part \ref{part:experiments}). 
The list below provides a brief outline of the chapters in this thesis:
\begin{itemize}
\item \textbf{Chapter \ref{c:algotrading}:} Overview of relevant concepts in algorithmic trading. 
\item \textbf{Chapter \ref{c:DL}:} Overview of relevant machine learning and deep learning concepts.
\item \textbf{Chapter \ref{c:RL}:} Overview of relevant concepts in reinforcement learning. 
\item \textbf{Chapter \ref{part:problemsetting}:} Formalization of the problem setting. 
\item \textbf{Chapter \ref{c:m-RL-algo}:} Description of reinforcement learning algorithms. 
\item \textbf{Chapter \ref{c:network-topology}:} Description of the neural network function approximators. 
\item \textbf{Chapter \ref{sec:expandres}:} Detailed results from experiments.
\item \textbf{Chapter \ref{future-work}:} Suggested future work. 
\item \textbf{Chapter \ref{sec:conclusion}:} Summary of contributions, results, and main conclusions. 
\end{itemize}

%\afterpage{\newpage}
\afterpage{\blankpage}
\newpage
\part{Background}\label{part:background}

%\afterpage{\newpage}
\newpage
\section{Algorithmic trading}\label{c:algotrading}

A phenomenon commonly described as an arms race has resulted from fierce competition in financial markets. In this phenomenon, market participants compete to remain on the right side of information asymmetry, which further reduces the signal-to-noise ratio and the frequency at which information arrives and is absorbed by the market \cite{isichenko2021quantitative}. 
An increase in volatility and the emergence of a highly sophisticated breed of traders called high-frequency traders have further complicated already complex market dynamics.
In these developed, modern financial markets, the dynamics are so complex and change at such a high frequency that humans will have difficulty competing. 
Indeed, there is reason to believe that machines already outperform humans in the world of financial trading. 
The algorithmic hedge fund Renaissance Technologies, founded by famed mathematician Jim Simons, is considered the most successful hedge fund ever. From 1988 to 2018, Renaissance Technologies' Medallion fund generated 66 percent annualized returns before fees relying exclusively on algorithmic strategies \cite{zuckerman2019man}. 
In 2020, it was estimated that algorithmic trading accounts for around 60-73 percent of U.S. and European equity trading, up from just 15 percent in 2003 \cite{intelligence2020algorithmic}. 
Thus, it is clear that algorithms already play a significant role in financial markets. 
Due to the rapid progress of computing power\footnote{Moore's law states that the number of transistors in an integrated circuit doubles roughly every two years.} relative to human evolution, this importance will likely only grow.

This chapter provides an overview of this thesis's subject matter, algorithmic trading on commodity markets, examines related work, and justifies the algorithmic trading methods described in part \ref{part:methodology}.
Section \ref{c:at-commarkets} presents a brief overview of commodity markets and energy futures contracts. 
Sections \ref{c:at-fintrading}, \ref{c:at-mpt}, and \ref{c:at-emh} introduce some basic concepts related to trading financial markets that are necessary to define and justify a trading agent's goal of maximizing risk-adjusted returns. 
This goal has two sub-goals: forecasting returns and mapping forecasts to market positions, which are discussed separately in sections \ref{c:at-forecasting} and \ref{c:at-mapping2pos}.
Additionally, these sections provide an overview of how the concepts introduced in the following chapters \ref{c:DL} and \ref{c:RL} can be applied to algorithmic trading and provide an overview of related research. 
The sections \ref{c:at-featureeng} and \ref{c:at-subsampl} describe how to represent a continuous financial market as discrete inputs to an algorithmic trading system. 
To conclude, section \ref{c:at-backtest} introduces backtesting, a form of cross-validation used to evaluate algorithmic trading agents.

\subsection{Commodity markets}\label{c:at-commarkets}

Energy products trade alongside other raw materials and primary products on commodity markets. The commodity market is an exchange that matches buyers and sellers of the products offered at the market. 
Traditionally trading was done in an open-outcry manner, though now an electronic limit order book is used to maintain a continuous market. Limit orders specify the direction, quantity, and acceptable price of a security. Limit orders are compared to existing orders in the limit order book when they arrive on the market. A trade occurs at the price set by the first order in the event of an overlap.
The exchange makes money by charging a fee for every trade, usually a small percentage of the total amount traded.

The basis of energy trade is energy futures, a derivative contract with energy products as the underlying asset \cite{chan2019financial}. 
Futures contracts are standardized forward contracts listed on stock exchanges. 
They are interchangeable, which improves liquidity. 
Futures contracts obligate a buyer and seller to transact a given quantity of the underlying asset at a future date and price. 
The quantity, quality, delivery location, and delivery date are all specified in the contract. 
Futures contracts are usually identified by their expiration month. The ``front-month'' is the nearest expiration date and usually represents the most liquid market. 
Natural gas futures expire three business days before the first calendar day of the delivery month. 
To avoid physical delivery of the underlying commodity, the contract holder must sell their holding to the market before expiry. 
Therefore, the futures and underlying commodity prices converge as the delivery date approaches. 
A futures contract achieves the same outcome as buying a commodity on the spot market on margin and storing it for a future date. 
The relative price of these alternatives is connected as it presents an arbitrage opportunity.
The difference in price between a futures contract and the spot price of the underlying commodity will therefore depend on the financing cost, storage cost, and convenience yield of holding the physical commodity over the futures contract. 
Physical traders use futures as a hedge while transporting commodities from producer to consumer. 
If a trader wishes to extend the expiry of his futures contract, he can ``roll" the contract by closing the contract about to expire and entering into a contract with the same terms but a later expiry date \cite{chan2019financial}. 
The ``roll yield" is the difference in price for these two contracts and might be positive or negative. 
The exchange clearinghouse uses a margin system with daily settlements between parties to mitigate counterparty risk \cite{chan2019financial}.

\subsection{Financial trading}\label{c:at-fintrading}

Financial trading is the act of buying and selling financial assets. 
Owning a financial asset is called being \textit{long} that asset, which will realize a profit if the asset price increases and suffer a loss if the asset price decreases. 
\textit{Short}-selling refers to borrowing, selling, and then, at a later time, repurchasing a financial asset and returning it to the lender with the hopes of profiting from a price drop during the loan term. 
Short-selling allows traders to profit from falling prices.

\subsection{Modern portfolio theory}\label{c:at-mpt}

Harry Markowitz laid the groundwork for what is known as \acrfull{mpt} \cite{markowitz1968portfolio}. MPT assumes that investors are risk-averse and advocates maximizing risk-adjusted returns. The \textit{Sharpe ratio} \cite{sharpe1998sharpe} is the most widely-used measurement of risk-adjusted return developed by economist William F. Sharpe. The Sharpe ratio compares excess return with the standard deviation of investment returns and is defined as 
\begin{equation}\label{eq:sharpe}
Sharpe \; ratio =\dfrac{\mathbb{E}[r_t - \bar{r}]}{\sqrt{var[r_t - \bar{r}]}} \simeq \dfrac{\mathbb{E}[r_t]}{\sigma_{r_t}}
\end{equation} 
where $\mathbb{E}[r_t]$ is the expected return over $T$ samples, $\bar{r}$ is the risk-free rate, and $\sigma_{r_t} > 0$ is the standard deviation of the portfolio's excess return. 
Due to negligible low interest rates, the risk-free rate is commonly set to $\bar{r}=0$. 
The philosophy of MPT is that the investor should be compensated through higher returns for taking on higher risk. 
The St. Petersburg paradox\footnote{For an explanation of the paradox, see the article \cite{sep-paradox-stpetersburg}.} illustrates why maximizing expected reward in a risk-neutral manner might not be what an individual wants. 
Although market participants have wildly different objectives, this thesis will adopt the MPT philosophy of assuming investors want to maximize risk-adjusted returns. 
Hence, the goal of the trading agent described in this thesis will be to maximize the risk-adjusted returns represented by the Sharpe ratio. Maximizing future risk-adjusted returns can be broken down into two sub-goals; forecasting future returns and mapping the forecast to market positions. However, doing so in highly efficient and competitive financial markets is non-trivial.

\subsection{Efficient market hypothesis}\label{c:at-emh}

Actively trading a market suggests that the trader is dissatisfied with market returns and believes there is potential for extracting excess returns, or alpha. 
Most academic approaches to finance are based on the \acrfull{emh} \cite{fama1970efficient}, which states that all available information is fully reflected in the prices of financial assets at any time. 
According to the EMH, a financial market is a stochastic martingale process. 
As a consequence, searching for alpha is a futile effort as the expected future return of a non-dividend paying asset is the present value, regardless of past information, i.e., 
\begin{equation}
\mathbb{E}[R_{t+1} | I_t] = R_t
\end{equation}

Practitioners and certain parts of academia heavily dispute the EMH. 
Behavioral economists reject the idea of rational markets and believe that human evolutionary traits such as fear and greed distort market participants' decisions, creating irrational markets. 
The \acrfull{amh} \cite{lo2004adaptive} reconciles the efficient market hypothesis with behavioral economics by applying evolution principles (competition, adaptation, and natural selection) to financial interactions. 
According to the AHM, what behavioral economists label irrational behavior is consistent with an evolutionary model of individuals adapting to a changing environment. 
Individuals within the market are continually learning the market dynamics, and as they do, they adapt their trading strategies, which in turn changes the dynamics of the market. 
This loop creates complicated price dynamics. 
Traders who adapt quickly to changing dynamics can exploit potential inefficiencies. 
Based on the AHM philosophy, this thesis hypothesizes that there are inefficiencies in financial markets that can be exploited, with the recognition that such opportunities are limited and challenging to discover.

\subsection{Forecasting}\label{c:at-forecasting}

Unless a person is gambling, betting on the price movements of volatile financial assets only makes sense if the trader has a reasonable idea of where the price is moving. 
Since traders face non-trivial transaction costs, the expected value of a randomly selected trade is negative. 
Hence, as described by the gambler's ruin, a person gambling on financial markets will eventually go bankrupt due to the law of large numbers. 
Forecasting price movements, i.e., making predictions based on past and present data, is a central component of any financial trading operation and an active field in academia and industry. 
Traditional approaches include fundamental analysis, technical analysis, or a combination of the two \cite{secanalysis}. 
These can be further simplified into qualitative and quantitative approaches (or a combination). 
A qualitative approach, i.e., fundamental analysis, entails evaluating the subjective aspects of a security \cite{secanalysis}, which falls outside the scope of this thesis. 
Quantitative (i.e., technical) traders use past data to make predictions \cite{secanalysis}. 
The focus of this thesis is limited to fully quantitative approaches.

Developing quantitative forecasts for the price series of financial assets is non-trivial as financial markets are non-stationary with a low signal-to-noise ratio \cite{taleb1997dynamic}. 
Furthermore, modern financial markets are highly competitive and effective. 
As a result, easily detectable signals are almost certainly arbitraged out. 
Researchers and practitioners use several mathematical and statistical models to identify predictive signals leading to excess returns. 
Classical models include the \textit{autoregressive integrated moving average} (ARIMA) and the \textit{generalized autoregressive conditional heteroskedasticity} (GARCH). 
The ARIMA is a linear model and a generalization of the \textit{autoregressive moving average} (ARMA) that can be applied to time series with nonstationary mean (but not variance) \cite{shumway2000time}. 
The assumption of constant variance (i.e., volatility) is not valid for financial markets where volatility is stochastic \cite{taleb1997dynamic}.
The GARCH is a non-linear model developed to handle stochastic variance by modeling the error variance as an ARMA model \cite{shumway2000time}. 
Although the ARIMA and GARCH have practical applications, their performance in modeling financial time series is generally unsatisfactory \cite{xiong2015deep, mcnally2018predicting}.

Over the past 20 years, the availability and affordability of computing power, storage, and data have lowered the barrier of entry to more advanced algorithmic methods.
As a result, researchers and practitioners have turned their attention to more complex machine learning methods because of their ability to identify signals and capture relationships in large datasets. 
Initially, there was a flawed belief that the low signal-to-noise ratio leaves viable only simple forecasts such as those based on low-dimensional ordinary least squares \cite{isichenko2021quantitative}. 
With the recent deep learning revolution, deep neural networks have demonstrated strong representation abilities when modeling time series data \cite{sutskever2014sequence}. 
The Makridakis competition evaluates time series forecasting methods. In its fifth installment held in 2020, all 50 top-performing models were based on deep learning architectures \cite{makridakis2022m5}. 
A considerable amount of recent empirical research suggests that deep learning models significantly outperform traditional models like the ARIMA and GARCH when forecasting financial time series \cite{xiong2015deep, mcnally2018predicting, siami2018forecasting, sezer2020financial}. 
These results are somewhat puzzling. The risk of overfitting is generally higher for noisy data like financial data. Moreover, the loss function for DNNs is non-convex, which makes finding a global minimum impossible. 
Despite the elevated overfitting risk and the massive overparameterization of DNNs, they still demonstrate stellar generalization. 
Thus, based on recent research, the thesis will apply deep learning techniques to model financial time series.

A review of deep learning methods in financial time series forecasting \cite{sezer2020financial} found that LSTMs were the preferred choice in sequence modeling, possibly due to their ability to remember both long- and short-term dependencies. 
Convolutional neural networks are another common choice. CNNs are best known for their ability to process 2D grids such as images; however, they have shown a solid ability to model 1D grid time series data.
Using historical prices, Hiransha et al. \cite{hiransha2018nse} tested FFNs, vanilla RNNs, LSTMs, and CNNs on forecasting next-day stock market returns on the National Stock Exchange (NSE) of India and the New York Stock Exchange (NYSE). 
In the experiment, CNNs outperformed other models, including the LSTM. 
These deep learning models can extract generalizable patterns from the price series alone \cite{sezer2020financial}.

\subsection{Mapping forecasts to market positions}\label{c:at-mapping2pos}

Most research on ML in financial markets focuses on forecast-based supervised learning approaches \cite{fischer2018reinforcement}. 
These methods tend to ignore how to convert forecasts into market positions or use some heuristics like the Kelly criterion to determine optimal position sizing \cite{isichenko2021quantitative}. 
The forecasts are usually optimized by minimizing a loss function like the \acrfull{mse}.
An accurate forecast (in the form of a lower MSE) may lead to a more profitable trader, but this is not always true. 
Not only does the discovered signal need adequate predictive power, but it must consistently produce reliable directional calls. 
Moreover, the mapping from forecast to market position needs to consider transaction costs and risk, which is challenging in a supervised learning framework \cite{moody1997optimization}. 
Neglecting transaction costs can lead to aggressive trading and overestimation of returns. 
Neglecting risk can lead to trading strategies that are not viable in the real world.
Maximizing risk-adjusted returns is only feasible when accounting for transaction costs and risk. These shortcomings are addressed using reinforcement learning \cite{moody1997optimization, moody1998performance}. 
Using RL, deep neural networks can be trained to output market positions directly. 
Moreover, the DNN can be jointly optimized for risk- and transaction-cost-sensitive returns, thus directly optimizing for the true goal: maximizing risk-adjusted returns.

Moody and Wu \cite{moody1997optimization} and Moody et al. \cite{moody1998performance} empirically demonstrated the advantages of reinforcement learning relative to supervised learning. 
In particular, they demonstrated the difficulty of accounting for transaction costs using a supervised learning framework. 
A significant contribution is their model-free policy-based RL algorithm for trading financial instruments \textit{recurrent reinforcement learning} (RRL). 
The name refers to the recursive mechanism that stores the past action as an internal state of the environment, allowing the agent to consider transaction costs. 
The agent outputs market positions and is limited to a discrete action space $a_t \in \{ -1, 0, 1 \}$, corresponding to maximally short, no position, and maximally long. 
At time $t$, the previous action $a_{t-1}$ is fed into the policy network $f_\theta$ along with the external state of the environment $s_t$ in order to make the trade decision, i.e.,
$$ a_t = f_\theta(s_t, a_{t-1}) $$
where $f_\theta$ is a linear function, and the external state is constructed using the past $8$ returns. 
The return $r_t$ is realized at the end of the period $(t-1, t]$ and includes the returns resulting from the position $a_{t-1}$ held through this period minus transaction costs incurred at time $t$ due to a difference in the new position $a_t$ from the old $a_{t-1}$. 
Thus, the agent learns the relationship between actions and the external state of the environment and the internal state.

Moody and Saffel \cite{moody2001learning} compared their actor-based RRL algorithm to the value-based Q-learning algorithm when applied to financial trading. 
The algorithms are tested on two real financial time series; the U.S. dollar/British pound foreign exchange pair and the S\&P 500 stock index. 
While both perform better than a buy-and-hold baseline, the RRL algorithm outperforms Q-learning on all tests. 
The authors argue that actor-based algorithms are better suited to immediate reward environments and may be better able to deal with noisy data and quickly adapt to non-stationary environments. 
They point out that critic-based RL suffers from the curse of dimensionality and that when extended to function approximation, it sometimes fails to converge even in simple MDPs.

Deng et al. \cite{deng2016deep} combine Moody's direct reinforcement learning framework with a recurrent neural network to introduce feature learning through deep learning. Another addition is the use of continuous action space. 
To constrain actions to the interval $[-1,1]$, the RNN output is mapped to a $\tanh{}$ function. 
Jiang et al. \cite{jiang2017deep} presents a deterministic policy gradient algorithm that trades a portfolio of multiple financial instruments. 
The policy network is modeled using CNNs and LSTMs, taking each period's closing, lowest, and highest prices as input. The DNNs are trained on randomly sampled mini-batches of experience. 
These methods account for transaction costs but not risk. 
Zhang et al. \cite{zhang2020cost} present a deep RL framework for a risk-averse agent trading a portfolio of instruments using both CNNs and LSTMs. 
Jin and El-Saawy \cite{jin2016portfolio} suggest that adding a risk-term to the reward function that penalizes the agent for volatility produces a higher Sharpe ratio than optimizing for the Sharpe ratio directly. Zhang et al. \cite{zhang2020cost} apply a similar risk-term penalty to the reward function.

\subsection{Feature engineering}\label{c:at-featureeng}

Any forecast needs some predictor data, or features, to make predictions. 
While ML forecasting is a science, feature engineering is an art and arguably the most crucial part of the ML process. 
Feature engineering and selection for financial forecasting are only limited by imagination. 
Features range from traditional technical indicators (e.g., Moving Average Convergence Divergence, Relative Strength Index) \cite{zhang2020deep} to more modern deep learning-based techniques like analyzing social-media sentiment of companies using Natural Language Processing \cite{zhang2010trading} or using CNNs on satellite images along with weather data to predict cotton yields \cite{tedesco2020convolutional}. 
Research in feature engineering and selection is exciting and potentially fruitful but beyond this thesis's scope. The most reliable predictor of future prices of a financial instrument tends to be its past price, at least in the short term \cite{isichenko2021quantitative}.
Therefore, in this thesis, higher-order features are not manually extracted. Instead, only the price series are analyzed.

\subsection{Sub-sampling schemes}\label{c:at-subsampl}

Separating high- and low-frequency trading can be helpful, as they present unique challenges. 
High-frequency trading (HFT) focuses on reducing software and hardware latency, which may include building a \$300 million fiber-optic cable to reduce transmission time by four milliseconds between exchanges to gain a competitive advantage \cite{lewis2014flash} \footnote{Turns out they forgot that light travels about $30\%$ slower in glass than in air, and they lost their competitive advantage to simple line-of-sight microwave networks \cite{lewis2014flash}.}. 
This type of trading has little resemblance to the low-frequency trading examined in this thesis, described in minutes or hours rather than milliseconds.

Technical traders believe that the prices of financial instruments reflect all relevant information \cite{taleb1997dynamic}. 
From this perspective, the market's complete order history represents the financial market's state. 
This state representation would scale poorly, with computational and memory requirements growing linearly with time. 
Consequently, sub-sampling schemes for periodic feature extraction are almost universally employed. 
While sampling information at fixed intervals is straightforward, there may be more effective methods.
As exchange activity varies throughout the day, sampling at fixed intervals may lead to oversampling during low-activity periods and undersampling during high-activity periods. 
In addition, time-sampled series often exhibit poor statistical properties, such as non-normal returns, autocorrelation, and heteroskedasticity \cite{de2018advances}.

The normality of returns assumption underpins several mathematical finance models, e.g., Modern Portfolio Theory \cite{markowitz1968portfolio}, and the Sharpe-ratio \cite{sharpe1998sharpe}. 
There is, however, too much peaking and fatter tails in the actual observed distribution for it to be relative to samples from Gaussian populations \cite{mandelbrot1997variation} \footnote{Assuming that the S\&P 500 index returns were normally distributed, the probability of daily returns being below five percent between 1962 and 2004 ($10 \; 698$ observations) would be approximately $0.0005$. However, it happened $8$ times \cite{hasbrouck2007empirical}.}. 
Mandelbrot showed in 1963 \cite{mandelbrot1997variation} that a Lévy alpha-stable distribution with infinite variance can approximate returns over fixed periods. 
In 1967, Mandelbrot and Taylor \cite{mandelbrot1967distribution} argued that returns over a fixed number of transactions may be close to \acrfull{iid} Gaussian.
Several empirical studies have since confirmed this \cite{clark1973subordinated, ane2000order}. 
Clark \cite{clark1973subordinated} discovered that sampling by volume instead of transactions exhibits better statistical properties, i.e., closer to IID Gaussian distribution. 
Sampling by volume instead of ticks has intuitive appeal. 
While tick bars count one transaction of $n$ contracts as one bar, $n$ transactions of one contract count as $n$ bars.
Sampling according to transaction volume might lead to significant sampling frequency variations for volatile securities. 
When the price is high, the volume will be lower, and therefore the number of observations will be lower, and vice versa, even though the same value might be transacted. 
Therefore, sampling by the monetary value transacted, also called dollar bars, may exhibit even better statistical properties \cite{de2018advances}.
Furthermore, for equities, sampling by monetary value exchanged makes an algorithm more robust against corporate actions like stock splits, reverse splits, stock offerings, and buybacks. 
To maintain a suitable sampling frequency, the sampling threshold may need to be adjusted if the total market size changes significantly.

Although periodic feature extraction reduces the number of observations that must be processed, it scales linearly in computation and memory requirements per observation. 
A history cut-off is often employed to represent the state by only the $n$ most recent observations to tackle this problem. 
Representing the state of a partially observable MDP by the $n$ most recent observations is a common technique used in many reinforcement learning applications. 
Mnih et al. \cite{mnih2013playing} used $4$ stacked observations as input to the DQN agent that achieved superhuman performance on Atari games to capture the trajectory of moving objects on the screen. 
The state of financial markets is also usually approximated by stacking past observations \cite{jiang2017deep, zhang2020deep, zhang2020cost}.

\subsection{Backtesting}\label{c:at-backtest}

Assessing a machine learning model involves estimating its generalization error on new data. The most widely used method for estimating generalization error is \acrfull{cv}, which assumes that observations are IID and drawn from a shared underlying data-generating distribution. 
However, the price of a financial instrument is a nonstationary time series with an apparent temporal correlation. 
Conventional cross-validation ignores this temporal component and is thus unsuitable for assessing a time series forecasting model. 
Instead, backtesting, a form of cross-validation for time series, is used. Backtesting is a historical simulation of how the model would have performed should it have been run over a past period. The purpose of backtesting is the same as for cross-validation; to determine the generalization error of an ML algorithm. 

To better understand backtesting, it is helpful to consider an algorithmic trading agent's objective and how it operates to achieve it. The algorithmic trading process involves the agent receiving information and, based on that information, executing trades at discrete time steps. 
These trades are intended to achieve a specific objective set by the stakeholder, which, in the philosophy of modern portfolio theory, is maximizing risk-adjusted returns. 
Thus, assessing an algorithmic trading agent under the philosophy of modern portfolio theory entails estimating the risk-adjusted returns resulting from the agent's actions. However, when testing an algorithmic trading agent, it cannot access data ahead of the forecasting period, as that would constitute information leakage to the agent. For this reason, conventional cross-validation fails in algorithmic trading.

The most precise way to assess the performance of an algorithmic trading agent is to deploy it to the market, let it trade with the intended amount of capital, and observe its performance. 
However, this task would require considerable time since low-frequency trading algorithms are typically assessed over long periods, often several years\footnote{There are a couple of reasons for this; first, there are, on average, 252 trading days per year. Low-frequency trading algorithms typically make fewer than ten trades per day. In order to obtain sufficient test samples, the agent must trade for several years. Second, testing the algorithmic trading agent under various market conditions is crucial. A successful model in particular market conditions may be biased towards those conditions and fail to generalize to other market conditions. }.
Additionally, any small error would likely result in devastating losses, making algorithmic trading projects economically unfeasible.

\begin{figure}[ht]
\centering
\includegraphics[width=\textwidth]{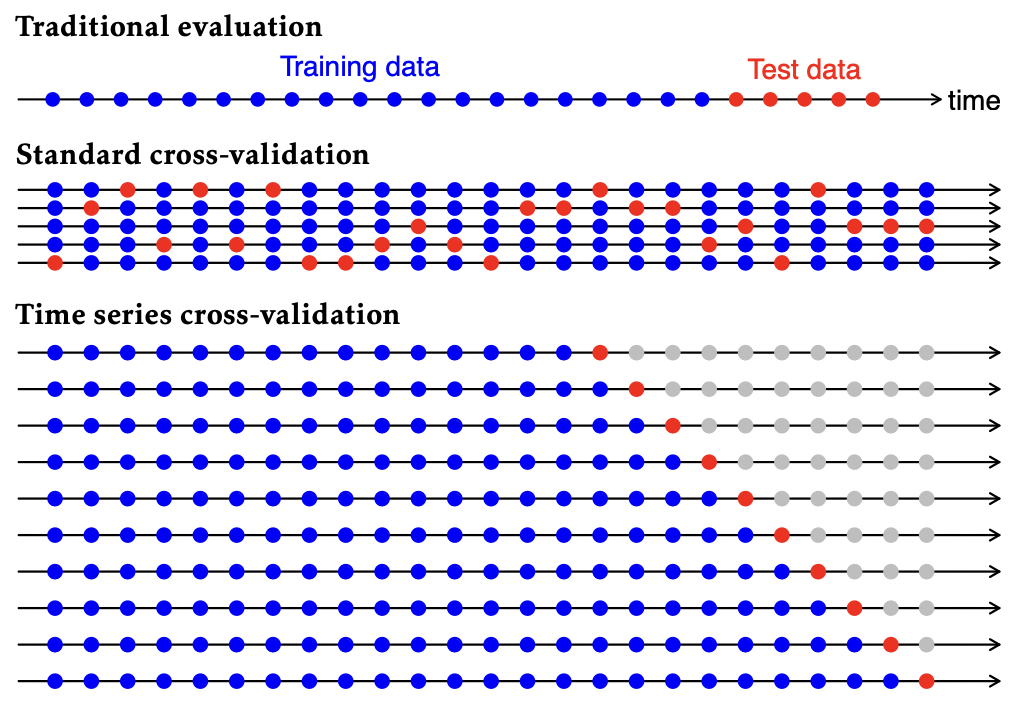}
\caption{Time series cross-validation (backtesting) compared to standard cross-validation from \cite{hyndman2018forecasting}.}
\end{figure}

Backtesting is an alternative to this expensive and time-consuming process where performance on historical simulations functions as a proxy for generalization error. Backtesting involves a series of tests that progress sequentially through time, where every test set consists of a single observation. 
At test $n$, the model trains on the training set consisting of observations prior to the observation in the test set ($i<n$). The forecast is, therefore, not based on future observations, and there is no leakage from the test set to the training set. Then the backtest progresses to the subsequent observation $n+1$, where the training set increases\footnote{The test set can also be a fixed-size FIFO queue.} to include observations $i<n+1$. The backtest progresses until there are no test sets left. The backtesting process is formalized in algorithm \ref{alg:backtesting}.

\begin{algorithm}
\caption{Backtesting}\label{alg:backtesting}
Train the model on the first $k$ observations (of $T$ total observations). 
\begin{algorithmic}
\For{$i=0,1,...,T-k$}
\State Select observation $k+i$ from the test set.
\State Register trade $a_{k+i}$. 
\State Train the model using observations at times $t \leq k+i$. 
\EndFor
\end{algorithmic}    
Measure performance using registered trades $a_k, a_{k+1}, ..., a_T$ and the corresponding prices at time $k, k+1, ..., T$. 
\end{algorithm}

When conducting historical simulations, knowing what information was available during the studied period is critical.
Agents should not have access to data beyond the point at which they are located in the backtest, in order to avoid lookahead bias. 
Lookahead bias is a research bug of inadvertently using future data, whereas real-time production trading is free of such ``feature''.
Data used in forecasting must be stored point-in-time (PIT), which indicates when the information was made available. 
An incorrectly labeled dataset can lead to lookahead bias.

Backtesting is a flawed procedure as it suffers from lookahead bias by design. 
Having experienced the hold-out test sample provides insight into what made markets rise and fall, a luxury not available before the fact. 
Thus, only live trading can be considered genuinely out-of-sample. 
Another form of lookahead bias is repeated and excessive backtest optimization leading to information leakage from the test to the training set. 
A machine learning model will enthusiastically grab any lookahead but fail to generalize to live trading. 
Furthermore, the backtests performed in this thesis rely on assumptions of zero market impact, zero slippage, fractional trading, and sufficient liquidity to execute any trade instantaneously at the quoted price. 
These assumptions do not reflect actual market conditions and will lead to unrealistic high performance in the backtest.

Backtesting should emulate the scientific method, where a hypothesis is developed, and empirical research is conducted to find evidence of inconsistencies with the hypothesis. 
It should be distinct from a research tool for discovering predictive signals. 
It should only be conducted after research has been done. 
A backtest is not an experiment but a simulation to see if the model behaves as expected \cite{de2018advances}. 
Random historical patterns might exhibit excellent performance in a backtest. 
However, it should be viewed cautiously if no ex-ante logical foundation exists to explain the performance \cite{arnott2019backtesting}.

%\afterpage{\newpage}
\newpage
\section{Deep learning}\label{c:DL}

An intelligent agent requires some means of modeling the dynamics of the system in which it operates.
Modeling financial markets is complicated due to low signal-to-noise ratios and non-stationary dynamics. The dynamics are highly nonlinear; thus, several traditional statistical modeling approaches cannot capture the system's complexity.
Moreover, reinforcement learning requires parameterized function approximators, rendering nonparametric learners, e.g., support vector machines and random forests, unsuitable.
Additionally, parametric learners are generally preferred when the predictor data is well-defined\cite{goodfellow2016deep}, such as when forecasting security prices using historical data. 
In terms of nonlinear parametric learners, artificial neural networks comprise the most widely used class. Research suggests that deep neural networks, such as LSTMs and CNNs, effectively model financial data \cite{xiong2015deep, hiransha2018nse, mcnally2018predicting}. Therefore, the algorithmic trading agent proposed in this thesis will utilize deep neural networks to model commodity markets.

This chapter introduces the fundamental concepts of deep learning relevant to this thesis, starting with some foundational concepts related to machine learning \ref{c:ML} and supervised learning \ref{c:SL}. 
Next, section \ref{c:ANN} covers artificial neural networks, central network topologies, and how they are initialized and optimized to achieve satisfactory results. The concepts presented in this chapter are presented in the context of supervised learning but will be extended to the reinforcement learning context in the next chapter (\ref{c:RL}).

\subsection{Machine learning}\label{c:ML}

\acrfull{ml} studies how computers can automatically \textit{learn} from experience without being explicitly programmed. 
A general and comprehensive introduction to machine learning can be found in ``Elements of Statistical Learning'' by Trevor Hastie, Robert Tibshirani, and Jerome Friedman \cite{hastie2009elements}. 
Tom Mitchell defined the general learning problem as ``A computer program is said to learn from experience E with respect to some class of tasks T and performance measure P, if its performance on T, as measured by P, improves with experience E.'' \cite{mitchell1997machine}. 
In essence, ML algorithms extract generalizable predictive patterns from data from some, usually unknown, probability distribution to build a model about the space. It is an optimization problem where performance improves through leveraging data. 
Generalizability relates to a model's predictive performance on independent test data and is a crucial aspect of ML. 
Models should be capable of transferring learned patterns to new, previously unobserved samples while maintaining comparable performance. 
ML is closely related to statistics in transforming raw data into knowledge. 
However, while statistical models are designed for inference, ML models are designed for prediction. 
There are three primary ML paradigms; \textit{supervised} learning, \textit{unsupervised} learning, and \textit{reinforcement} learning.

\subsubsection{No-free-lunch theorem}
The \textit{no-free-lunch} theorem states that there exists no single universally superior ML learning algorithm that applies to all possible datasets. Fortunately, this does not mean ML research is futile. Instead, domain-specific knowledge is required to design successful ML models. The no-free-lunch theorem results only hold when averaged over all possible data-generating distributions. If the types of data-generating distributions are restricted to classes with certain similarities, some ML algorithms perform better on average than others. Instead of trying to develop a universally superior machine learning algorithm, the focus of ML research should be on what ML algorithms perform well on specific data-generating distributions.

\subsubsection{The curse of dimensionality}

The \textit{Hughes phenomenon} states that, for a fixed number of training examples, as the number of features increases, the average predictive power for a classifier will increase before it starts deteriorating. 
Bellman termed this phenomenon \textit{the curse of dimensionality}, which frequently manifests itself in machine learning \cite{sutton2018reinforcement}. 
One manifestation of the curse is that the sampling density is proportional to $N^{1/p}$, where $p$ is the dimension of the input space and $N$ is the sample size. 
If $N = 100$ represents a dense sample for $p=1$, then $N^{1/10} = 100^{10}$ is the required sample size for the same sampling density with $p=10$ inputs. 
Therefore, in high dimensions, the training samples sparsely populate the input space \cite{hastie2009elements}. 
In Euclidean spaces, a non-negative term is added to the distance between points with each new dimension. 
Thus, generalizing from training samples becomes more complex, as it is challenging to say something about a space without relevant training examples.

\subsection{Supervised learning}\label{c:SL}

\acrfull{sl} is the machine learning paradigm where a labeled training set of $N\in \mathbb{N}_+$ observations $\tau = \{ x_i, y_i \}_{i=1}^N$ is used to learn a functional dependence $\mathbf{y} = \hat{f}(\mathbf{x})$ that can predict $\mathbf{y}$ from a previously unobserved $\mathbf{x}$. 
Supervised learning includes regression tasks with numerical targets and classification tasks with categorical targets. 
A supervised learning algorithm adjusts the input/output relationship of $\hat{f}$ in response to the prediction error to the target $y_i -  \hat{f}(x_i)$. 
The hypothesis is that if the training set is representative of the population, the model will generalize to previously unseen examples.

\subsubsection{Function approximation}\label{func-approx}

Function approximation, or function estimation, is an instance of supervised learning that concerns selecting a function among a well-defined class that underlies the predictive relationship between the input vector $\mathbf{x}$ and output variable $\mathbf{y}$. 
In most cases, $\mathbf{x} \in \mathbb{R}^{d}$, where $d \in \mathbb{N}_+$, and $\mathbf{y} \in \mathbb{R}$. 
Function approximation relies on the assumption that there exists a function $f(\mathbf{x})$ that describes the approximate relationship between $(\mathbf{x}, \mathbf{y})$. 
This relationship can be defined as
\begin{equation}
\mathbf{y} = f(\mathbf{x}) + \mathbf{\epsilon}
\end{equation}
where $\epsilon$ is some irreducible error that is independent of $\mathbf{x}$ where $\mathbb{E}[\epsilon] = 0$ and $Var(\epsilon)=\sigma^2_\epsilon$. All departures from a deterministic relationship between $(\mathbf{x}, \mathbf{y})$ are captured via the error $\epsilon$. 
The objective of function approximation is to approximate the function $f$ with a model $\hat{f}$. In reality, this means finding the optimal model parameters $\theta$. Ordinary least squares can estimate linear models' parameters $\theta$.

\paragraph{Bias-variance tradeoff}

Bias and variance are two sources of error in model estimates. 
Bias measures the in-sample expected deviation between the model estimate and the target and is defined as 
\begin{equation}
Bias^2(\hat{f}(\mathbf{x})) = [\mathbb{E}[\hat{f}(\mathbf{x})] - f(\mathbf{x})]^2 
\end{equation}
and is a decreasing function of complexity. 
Variance measures the variability in model estimates and is defined as 
\begin{equation}
Var(\hat{f}(\mathbf{x})) = \mathbb{E}[\hat{f}(\mathbf{x}) - \mathbb{E}[\hat{f}(\mathbf{x})] ]^2 
\end{equation}
and is an increasing function of complexity. 
The out-of-sample mean square error for model estimates is defined as
\begin{equation}
MSE = Bias^2(\hat{f}) + Var(\hat{f}) + Var(\epsilon)
\end{equation}
where the last term $Var(\epsilon)$ is the irreducible noise error due to a target component $\epsilon$ not predictable by $\mathbf{x}$. 
The bias can be made arbitrarily small using a more complex model; however, this will increase the model variance, or generalization error, when switching from in-sample to out-of-sample. 
This is known as the \textit{bias-variance tradeoff}.

\paragraph{Overfitting}

A good ML model minimizes the model error, i.e., the training error (bias) and the generalization error (variance). 
This is achieved at some optimal complexity level, dependent on the data and the model. 
Increasing model complexity, or capacity, can minimize training error. However, such a model is unlikely to generalize well. Therefore, the difference between training error and generalization error will be high. This is known as \textit{overfitting} and happens when the complexity of the ML model exceeds that of the underlying problem. 
Conversely, \textit{underfitting} is when the training error is high because the model's complexity is lower than that of the underlying problem.

\begin{figure}[ht]
\centering
\includegraphics[width=\textwidth]{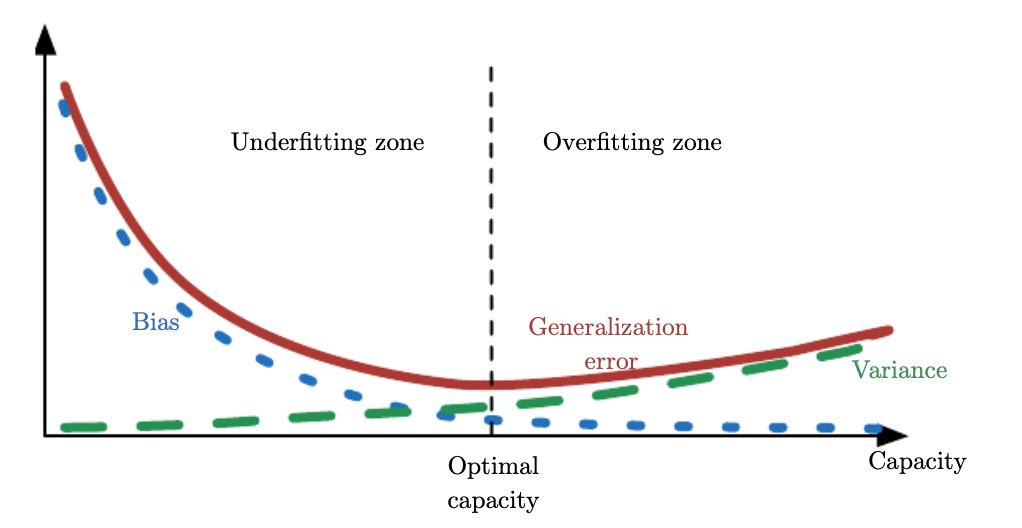}
\caption{An illustration of the relationship between the capacity of a function approximator and the generalization error from \cite{goodfellow2016deep}.}
\end{figure}

To minimize model error, the complexity of the ML model, or its inductive bias, must align with that of the underlying problem. The principle of parsimony, known as \textit{Occam's razor}, states that among competing hypotheses explaining observations equally well, one should pick the simplest one. 
This heuristic is typically applied to ML model selection by selecting the simplest model from models with comparable performance.

Recent empirical evidence has raised questions about the mathematical foundations of machine learning. 
Complex models such as deep neural networks have been shown to decrease both training error and generalization error with growing complexity \cite{zhang2021understanding}. 
Furthermore, the generalization error keeps decreasing past the interpolation limit. 
These surprising results contradict the bias-variance tradeoff that implies that a machine learning model should balance over- and underfitting. 
Belkin et al. \cite{belkin2019reconciling} reconciled these conflicting ideas by introducing a ``double descent'' curve to the bias-variance tradeoff curve. 
This increases performance when increasing the model capacity beyond the interpolation point.

\subsection{Artificial neural networks}\label{c:ANN}

An \acrfull{ann} is a parametric learner fitting nonlinear models. 
The network defines a mapping $h_{\theta} : \mathbb{R}^n \rightarrow \mathbb{R}^m$ where $n$ is the input dimension, $m$ is the output dimension, and $\theta$ are the network weights. 
A neural network has a graph-like topology. It is a collection of nodes organized in layers like a directed and weighted graph. 
The nodes of an ANN are typically separated into layers; the input layer, one or more hidden layers, and the output layer. Their dimensions depend on the function being approximated. 
A multi-layer neural network is called a \acrfull{dnn}.

\subsubsection{Feedforward neural networks}

A \acrfull{ffn}, or fully-connected network, defines the foundational class of neural networks where the connections are a directed acyclic graph that only allows signals to travel forward in the network. 
A feedforward network is a mapping $h_\theta$ that is a composition of multivariate functions $f_1, f_2, ...,f_k, g$, where $k$ is the number of layers in the neural network. It is defined as 
\begin{equation}
h_\theta = g \circ f_k \circ ... \circ f_2 \circ f_1(x)
\end{equation}
The functions $f_j$, $j=1,2,...,k$ represent the network's hidden layers and are composed of multivariate functions. 
The function at layer $j$ is defined as 
\begin{equation}
f_j (x) = \phi_j(\theta_j x + b_j) 
\end{equation}
where $\phi_j$ is the activation function and $b_j$ is the bias at layer $j$. 
The activation function is used to add nonlinearity to the network. 
The network's final output layer function $g$ can be tailored to suit the problem the network is solving, e.g., linear for Gaussian output distribution or Softmax distribution for categorical output distribution.

\begin{figure}[ht]
\centering
\includegraphics[width=\textwidth]{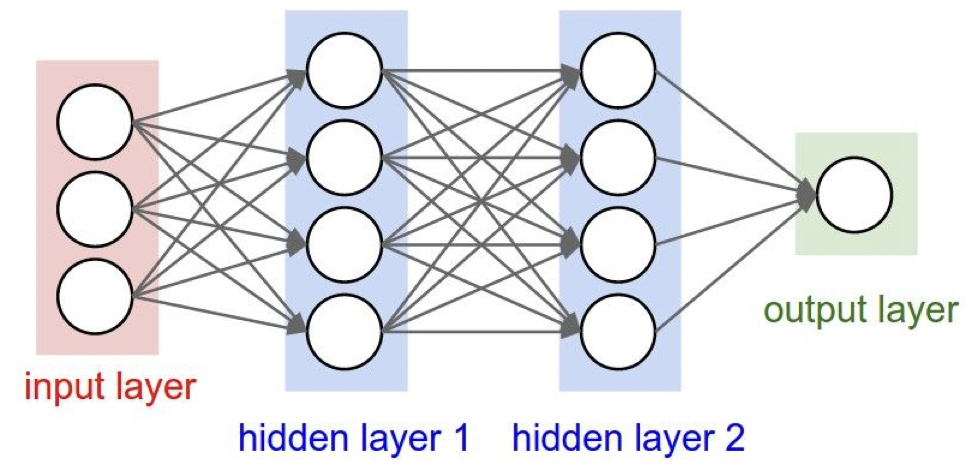}
\caption{Feedforward neural network from \cite{cs231nl4}.}
\end{figure}

\subsubsection{Parameter initialization}

Neural network learning algorithms are iterative and require some starting point from which to begin. 
The initial parameters of the networks can affect the speed and level of convergence or even whether the model converges at all. 
Little is known about weight initialization, a research area that is still in its infancy. 
Further complicating the issue: initial parameters favorable for optimization might be unfavorable for generalization \cite{goodfellow2016deep}. 
Developing heuristics for parameter initialization is, therefore, non-trivial. 
Randomness and asymmetry between the network units are desirable properties for the initial weights \cite{goodfellow2016deep}. 
Weights are usually drawn randomly from a Gaussian or uniform distribution in a neighborhood around zero, while the bias is usually set to some heuristically chosen constant. 
Larger initial weights will prevent signal loss during forward- and backpropagation. However, too large values can result in exploding values, a problem particularly prevalent in recurrent neural networks. The initial scale of the weights is usually set to something like $1/\sqrt{m}$ where $m$ is the number of inputs to the network layer.

Kaiming initialization \cite{he2015delving} is a parameter initialization method that takes the type of activation function (e.g., Leaky-ReLU) used to add nonlinearity to the neural network into account. 
The key idea is that the initialization method should not exponentially reduce or magnify the magnitude of input signals. Therefore, each layer is initialized at separate scales depending on their size. 
Let $m_l \in \mathbb{N}_+$ be the size of the inputs into the layer $l \in \mathbb{N}_+$. Kaiming He et al. recommend initializing weights such that 
$$ \frac{1}{2} m_l \textrm{Var}{[\theta_l]} = 1 $$
Which corresponds to an initialization scheme of 
$$ w_l \sim \mathcal{N}(0, 2/m_l) $$
Biases are initialized at $0$.

\subsubsection{Gradient-based learning}

Neural nets are optimized by adjusting their weights $\theta$ with the help of objective functions. 
Let $J(\theta)$ define the differentiable objective function for a neural network, where $\theta$ are the network weights. 
The choice of the objective function and whether it should be minimized or maximized depends on the problem being solved. 
For regression tasks, the objective is usually to minimize some loss function like mean-squared error (MSE)
\begin{equation}
J(\theta) = \dfrac{1}{n} \sum_{i=1}^n \left(h_{\theta}(x_{i})-y_{i} \right)^2
\end{equation}

Due to neural nets' nonlinearity, most loss functions are non-convex, meaning it is impossible to find an analytical solution to $\nabla J(\theta)=0$. 
Instead, iterative, gradient-based optimization algorithms are used. There are no convergence guarantees, but it often finds a satisfactorily low value of the loss function relatively quickly. 
Gradient descent-based algorithms adjust the weights $\theta$ in the direction that minimizes the MSE loss function. 
The update rule for parameter weights in gradient descent is defined as
\begin{equation}
\theta_{t+1} = \theta_t - \alpha \nabla_\theta J(\theta_t)
\end{equation}
where $\alpha>0$ is the learning rate and the gradient $\nabla J(\theta_t)$ is the partial derivatives of the objective function with respect to each weight. The learning rate defines the rate at which the weights move in the direction suggested by the gradient of the objective function. 
Gradient-based optimization algorithms, also called first-order optimization algorithms, are the most common optimization algorithms for neural networks \cite{goodfellow2016deep}.

\paragraph{Stochastic gradient descent}

Optimization algorithms that process the entire training set simultaneously are known as batch learning algorithms. 
Using the average of the entire training set allows for calculating a more accurate gradient estimate. 
The speed at which batch learning converges to a local minima will be faster than online learning. However, batch learning is not suitable for all problems, e.g., problems with massive datasets due to the high computational costs of calculating the full gradient or problems with dynamic probability distributions.

Instead, \acrfull{sgd} is often used when optimizing neural networks. 
SGD replaces the gradient in conventional gradient descent with a stochastic approximation. Furthermore, the stochastic approximation is only calculated on a subset of the data. This reduces the computational costs of high-dimensional optimization problems. 
However, the loss is not guaranteed to decrease when using a stochastic gradient estimate. 
SGD is often used for problems with continuous streams of new observations rather than a fixed-size training set. 
The update rule for SGD is similar to the one for GD but replaces the true gradient with a stochastic estimate 
\begin{equation}
\theta_{t+1} = \theta_t - \alpha_t \nabla_\theta J^{(j)}(\theta_t)
\end{equation}
where $\nabla_\theta J^{(j)}(\theta)$ is the stochastic estimate of the gradient computed from observation $j$. The total loss is defined as $J(\theta) = \sum_{j=1}^N J^{(j)}(\theta)$ where $N\in\mathbb{N}$ is the total number of observations. 
The learning rate at time $t$ is $\alpha_t>0$. Due to the noise introduced by the SGD gradient estimate, gradually decreasing the learning rate over time is crucial to ensure convergence. Stochastic approximation theory guarantees convergence to a local optima if $\alpha$ satisfies the conditions $\sum \alpha = \infty$ and $\sum \alpha^2 < \infty$. 
It is common to adjust the learning rate using the following update rule $\alpha_t = (1-\beta)\alpha_0 + \beta \alpha_\tau$, where $\beta = \dfrac{t}{\tau}$, and the learning rate is kept constant after $\tau$ iterations, i.e., $\forall t\geq \tau$, $\alpha_t = \alpha_\tau$ \cite{goodfellow2016deep}.

Due to hardware parallelization, simultaneously computing the gradient of $N$ observations will usually be faster than computing each gradient separately \cite{goodfellow2016deep}. 
Neural networks are, therefore, often trained on mini-batches, i.e., sets of more than one but less than all observations. 
Mini-batch learning is an intermediate approach to fully online learning and batch learning where weights are updated simultaneously after accumulating gradient information over a subset of the total observations. 
In addition to providing better estimates of the gradient, mini-batches are more computationally efficient than online learning while still allowing training weights to be adjusted periodically during training. 
Therefore, minibatch learning can be used to learn systems with dynamic probability distributions. 
Samples of the mini-batches should be independent and drawn randomly. 
Drawing ordered batches will result in biased estimates, especially for data with high temporal correlation.

Due to noisy gradient estimates, stochastic gradient descent and mini-batches of small size will exhibit higher variance than conventional gradient descent during training. 
The higher variance can be helpful to escape local minima and find new, better local minima. 
However, high variance can also lead to problems such as overshooting and oscillation that can cause the model to fail to converge. 
Several extensions have been made to stochastic gradient descent to circumvent these problems.

\paragraph{Adaptive gradient algorithm}

The Adaptive Gradient (AdaGrad) is an extension to stochastic gradient descent introduced in 2011 \cite{duchi2011adaptive}. 
It outlines a strategy for adjusting the learning rate to converge quicker and improving the capability of the optimization algorithm. 
A per-parameter learning rate allows AdaGrad to improve performance on problems with sparse gradients. Learning rates are assigned lower for parameters with frequently occurring features and higher for parameters with less frequently occurring features. 
The AdaGrad update rule is given as 
\begin{equation}
\theta_{t+1} = \theta_t - \dfrac{\alpha}{\sqrt{G_t + \epsilon}} g_t
\end{equation}
where $g_t=\nabla_\theta J^{(j)}(\theta_t)$, and $G_t = \sum_{\tau=1}^t {g_t g_t^\top}$, is the outer product of all previous subgradients. $\epsilon > 0$ is a smoothing term to avoid division by zero. 
As training proceeds, the squared gradients in the denominator of the learning rate will continue to grow, resulting in a strictly decreasing learning rate. As a result, the learning rate will eventually become so small that the model cannot acquire new information.

\paragraph{Root mean square propagation}

Root Mean Square Propagation (RMSProp) is an unpublished extension to SGD developed by Geoffrey Hinton. 
RMSProp was developed to resolve the problem of AdaGrad's diminishing learning rate. 
Like AdaGrad, it maintains a per-parameter learning rate. 
To normalize the gradient, it keeps a moving average of squared gradients. This normalization decreases the learning rate for more significant gradients to avoid the exploding gradient problem and increases it for smaller gradients to avoid the vanishing gradient problem. 
The RMSProp update rule is given as
\begin{equation}
\theta_{t+1} = \theta_t - \dfrac{\alpha}{\sqrt{E[g^2]_t + \epsilon}} g_t
\end{equation}
where $E[g^2]_t = \beta E[g^2]_t + (1-\beta)g^2_t$ where $v_t$ is the exponentially decaying average of squared gradients and $\beta>0$ is a second learning rate conventionally set to $\beta=0.9$.

\paragraph{Adam}

The Adam optimization algorithm is an extension of stochastic gradient descent that has recently seen wide adoption in deep learning. It was introduced in 2015 \cite{kingma2014adam} and derives its name from adaptive moment estimation. 
It utilizes the Adaptive Gradient (AdaGrad) Algorithm and Root Mean Square Propagation (RMSProp). 
Adam only requires first-order gradients and little memory but is computationally efficient and works well with high-dimensional parameter spaces. 
As with AdaGrad and RMSProp, Adam utilizes independent per-parameter learning rates separately adapted during training. 
Adam stores a moving average of gradients $E[g]_t=\beta_1 E[g]_t + (1-\beta_1)g_t$ with learning rate $\beta_1>0$. 
Like RMSProp, Adam also stores a moving average of squared gradients $E[g^2]_t$ with learning rate $\beta_2>0$. 
The Adam update rule is given as
\begin{equation}
\theta_{t+1} = \theta_t - \dfrac{\alpha}{\sqrt{\hat{E[g^2]_t}}+\epsilon} \hat{E[g]_t}
\end{equation}
where $\hat{E[g^2]_t}=\dfrac{E[g^2]_t}{1-\beta_2^t}$ and $\hat{E[g]_t}=\dfrac{E[g]_t}{1-\beta_1^t}$. The authors recommend learning rates $\beta_1 = 0.9$, $\beta_2 = 0.999$, as well as $\epsilon = 10^{-8}$. 
Adam has been shown to outperform other optimizers in a wide range of non-convex optimization problems. 
Researchers at Google \cite{andrychowicz2020matters} recommend the Adam optimization algorithm for SGD optimization in reinforcement learning.

\subsubsection{Backpropagation}

Gradient-based optimization requires a method for computing a function's gradient. 
For neural nets, the gradient of the loss function with respect to the weights of the network $\nabla_\theta J(\theta)$ is usually computed using the backpropagation algorithm (backprop) introduced in 1986 \cite{rumelhart1985learning}.
Backprop calculates the gradient of the loss function with respect to each weight in the network. 
This is done by iterating backward through the network layers and repeatedly applying the chain rule. 
The chain rule of calculus is used when calculating derivatives of functions that are compositions of other functions with known derivatives. 
Let $y, z : \mathbb{R} \rightarrow \mathbb{R}$ be functions defined as $y=g(x)$ and $z=f(g(x))=f(y)$. By the chain rule
\begin{equation}
\dfrac{dz}{dx} = \dfrac{dz}{dy} \dfrac{dy}{dx}
\end{equation}
Generalizing further, let $x\in\mathbb{R}^m, y\in\mathbb{R}^n$, and define mappings $g : \mathbb{R}^m \rightarrow \mathbb{R}^n$ and $f : \mathbb{R}^n \rightarrow \mathbb{R}$. If $y=g(x)$ and $z=f(y)$, then the chain rule is
\begin{equation}
\dfrac{\partial z}{\partial x_i} = \sum_j \dfrac{\partial z}{\partial y_j} \dfrac{\partial y_j}{\partial x_i}
\end{equation}
which can be written in vector notation as
\begin{equation}
\nabla_x z = \left( \dfrac{\partial y}{\partial x} \right)^\top \nabla_y z
\end{equation}
where $\dfrac{\partial y}{\partial x}$ is the $n\times m$ Jacobian matrix of $g$. 
Backpropagation is often performed on tensors and not vectors.
However, backpropagation with tensors is performed similarly by multiplying Jacobians by gradients.
Backpropagation with tensors can be performed by flattening a tensor into a vector, performing backprop on the vector, and then reshaping the vector back into a tensor. 
Let $X$ and $Y$ be tensors and $Y=g(X)$ and $z=f(Y)$. The chain rule for tensors is
\begin{equation}
\nabla_x z = \sum_j \left( \nabla_x Y_j \right) \dfrac{\partial z}{\partial Y_j}
\end{equation}

By recursively applying the chain rule, a scalar's gradient can be expressed for any node in the network that produced it. 
This is done recursively, starting from the output layer and going back through the layers of the network to avoid storing subexpressions of the gradient or recomputing them several times.

\subsubsection{Activation function}

The activation function $\phi(\xi)$ adds nonlinearity to a neural net. 
If the activation function in the hidden layer is linear, then the network is equivalent to a network without hidden layers since linear functions of linear functions are themselves linear. 
The activation function must be differentiable to compute the gradient. 
Choosing an appropriate activation function depends on the specific problem. 
Sigmoid functions $\sigma$, like the logistic function, are commonly used, as well as other functions such as the hyperbolic tangent function $tanh$. 
The derivative of the logistic function is close to $0$ except in a small neighborhood around $0$. At each backward step, the $\delta$ is multiplied by the derivative of the activation function. 
The gradient will therefore approach $0$ and thus produce extremely slow learning. 
This is known as the \textit{vanishing gradient} problem. 
For this reason, the \acrfull{relu} is the default recommendation for activation function in modern deep neural nets \cite{goodfellow2016deep}. 
ReLU is a ramp function defined as $ReLU(x) = \max \{0, x\}$.
The derivative of the ReLU function is defined as 
\begin{equation} 
ReLU'(x) = 
 {\begin{cases}
 0 & if\ x<0 \\
 1 & if\ x>0 \\
 \end{cases} }
\end{equation}
The derivative is undefined for $x=0$, but it has subdifferential $[0,1]$, and it conventionally takes the value $ReLU'(0)=0$ in practical implementations. 
Since ReLU is a piecewise linear function, it optimizes well with gradient-based methods.

\begin{figure}[ht]
\centering
\includegraphics{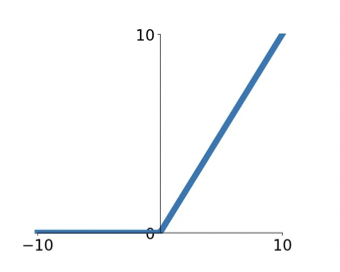}
\caption{ReLU activation function from \cite{cs231nl4}.}
\end{figure}

ReLU suffers from what is known as the \textit{dying ReLU} problem, where a large gradient could cause a node's weights to update such that the node will never output anything but $0$. 
Such nodes will not discriminate against any input and are effectively ``dead". 
This problem can be caused by unfortunate weight initialization or a too-high learning rate. 
Generalizations of the ReLU function, like the Leaky ReLU (LReLU) activation function, has been proposed to combat the dying ReLU problem \cite{goodfellow2016deep}. 
Leaky ReLU allows a small ``leak" for negative values proportional to some slope coefficient $\alpha$, e.g., $\alpha=0.01$, determined before training. 
This allows small gradients to travel through inactive nodes. 
Leaky ReLU will slowly converge even on randomly initialized weights but can also reduce performance in some applications \cite{goodfellow2016deep}.

\begin{figure}[ht]
\centering
\includegraphics{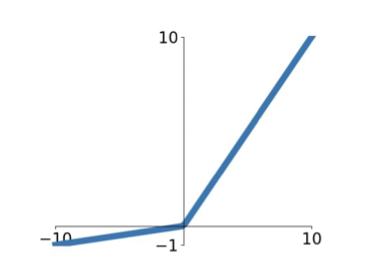}
\caption{Leaky ReLU activation function from \cite{cs231nl4}.}
\end{figure}

\subsubsection{Regularization}\label{c:nn-regularization}

Minimization of generalization error is a central objective in machine learning. 
The representation capacity of large neural networks, expressed by the universal approximation theorem (\ref{universal-approx-the}), comes at the cost of increased overfitting risk. 
Consequently, a critical question in ML is how to design and train neural networks to achieve the lowest generalization error. 
Regularization addresses this question. Regularization is a set of techniques designed to reduce generalization error, possibly at the expense of training error.

Regularization of estimators trades increased bias for reduced variance. If effective, it reduces model variance more than it increases bias. 
\textit{Weight decay} is used to regularize ML loss functions by adding the squared $L^2$ norm of the parameter weights $\Omega(\theta)= \frac{1}{2}||\theta||^2_2$ as a regularization term to the loss function 
\begin{equation}
\tilde{J}(\theta) = J(\theta) + \lambda \Omega(\theta)
\end{equation}
where $\lambda \geq 0$ is a constant weight decay parameter. 
Increasing $\lambda$ punishes larger weights harsher. 
Weight decay creates a tradeoff for the optimization algorithm between minimizing the loss function $J(\theta)$ and the regularization term $\Omega(\theta)$.

\begin{figure}[ht]
\centering
\includegraphics[width=\textwidth]{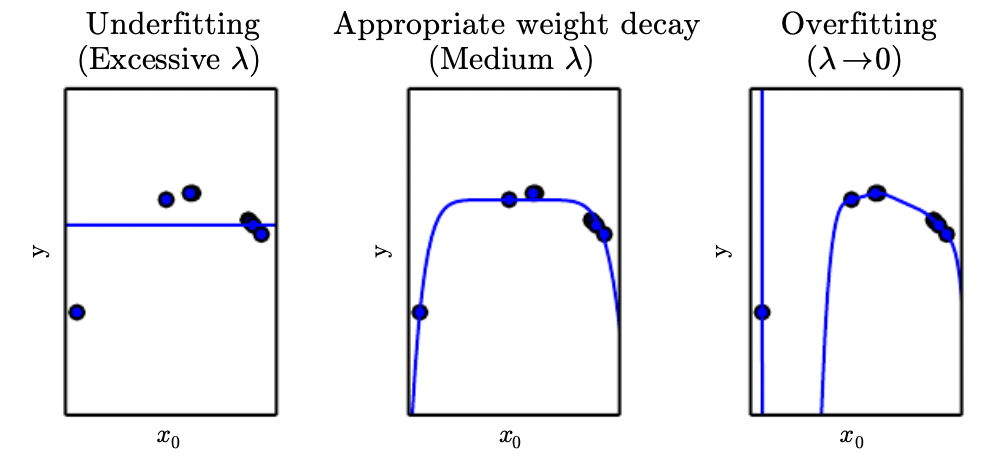}
\caption{An example of the effect of weight decay with parameter $\lambda$ on a high-dimensional polynomial regression model from \cite{goodfellow2016deep}.}
\end{figure}

\textit{Dropout} \cite{srivastava2014dropout} is another regularization strategy that reduces the risk of overfitting by randomly eliminating non-output nodes and their connections during training, preventing units from co-adapting too much. 
Dropout can be considered an ensemble method, where an ensemble of ``thinned'' sub-networks trains the same underlying base network. 
It is computationally inexpensive and only requires setting one parameter $\alpha \in [0,1)$, which is the rate at which nodes are eliminated.

\textit{Early stopping} is a common and effective implicit regularization technique that addresses how many epochs a model should be trained to achieve the lowest generalization error. 
The training data is split into training and validation subsets. 
The model is iteratively trained on the training set, and at predefined intervals in the training cycle, the model is tested on the validation set. 
The error on the validation set is used as a proxy for the generalization error. 
If the performance on the validation set improves, a copy of the model parameters is stored. 
If performance worsens, the learning terminates, and the model parameters are reset to the previous point with the lowest validation set error. 
Testing too frequently on the validation set risks premature termination. 
Temporary dips in performance are prevalent for nonlinear models, especially when trained with reinforcement learning algorithms when the agent explores the state and action space. 
Additionally, frequent testing is computationally expensive. 
On the other hand, infrequent testing increases the risk of not registering the model parameters near their performance peak. 
Early stopping is relatively simple but comes at the cost of sacrificing parts of the training set to the validation set.

\subsubsection{Batch normalization}

Deep neural networks are sensitive to initial random weights and hyperparameters. 
When updating the network, all weights are updated using a loss estimate under the false assumption that weights in the prior layers are fixed. 
In practice, all layers are updated simultaneously. 
Therefore, the optimization step is constantly chasing a moving target. 
The distribution of inputs during training is forever changing. 
This is known as \textit{internal covariate shift}, making the network sensitive to initial weights and slowing down training by requiring lower learning rates.

\textit{Batch normalization} (batch norm) is a method of adaptive reparametrization used to train deep networks. 
It was introduced in 2015 \cite{ioffe2015batch} to help stabilize and speed up training deep neural networks by reducing internal covariate shift. 
Batch norm normalizes the output distribution to be more uniform across dimensions by standardizing the activations of each input variable for each mini-batch. 
Standardization rescales the data to standard Gaussian, i.e., zero-mean unit variance. The following transformation is applied to a mini-batch of activations to standardize it 
\begin{equation}
\hat{x}^{(k)}_{norm} = \dfrac{x^{(k)} - \mathbb{E}[x^{(k)}]}{\sqrt{Var[x^{(k)}] + \epsilon}}
\end{equation}
where the $\epsilon>0$ is a small number such as $10^{-8}$ added to the denominator for numerical stability. 
Normalizing the mean and standard deviation can, however, reduce the expressiveness of the network \cite{goodfellow2016deep}. 
Applying a second transformation step to the mini-batch of normalized activations restores the expressive power of the network
\begin{equation}
\tilde{x}^{(k)} = \gamma \hat{x}^{(k)}_{norm} + \beta 
\end{equation}
where $\beta$ and $\gamma$ are learned parameters that adjust the mean and standard deviation, respectively. This new parameterization is easier to learn with gradient-based methods. 
Batch normalization is usually inserted after fully connected or convolutional layers. 
It is conventionally inserted into the layer before activation functions but may also be inserted after. 
Batchnorm speeds up learning and reduces the strong dependence on initial parameters. 
Additionally, it can have a regularizing effect and sometimes eliminate the need for dropout.

\subsubsection{Universal approximation theorem}\label{universal-approx-the}

In 1989, Cybenko \cite{cybenko1989approximation} proved that a feedforward network of arbitrary width with a sigmoidal activation function and a single hidden layer can approximate any continuous function. 
The theorem asserts that given any $f \in C([0,1]^n)$\footnote{Continuous function on the $n$-dimensional unit cube.}, $\epsilon > 0$ and sigmoidal activation function $\phi$, there is a finite sum of the form 
\begin{equation}
    \hat{f}(x) = \sum^N_{i=1} \alpha_i \phi(\theta_i^\top x + b_i)    
\end{equation} 
where $\alpha_i, b_i \in \mathbb{R}$ and $\theta_i \in \mathbb{R}^2$, for which 
\begin{equation}
    |\hat{f}(x)-f(x)| < \epsilon
\end{equation}
for all $x \in [0,1]^n$. 
Hornik \cite{hornik1991approximation} later generalized to include all squashing activation functions in what is known as the \textit{universal approximation} theorem. 
The theorem establishes that there are no theoretical constraints on the expressivity of neural networks. 
However, it does not guarantee that the training algorithm will be able to learn that function, only that it can be learned for an extensive enough network.

\subsubsection{Deep neural networks}

Although a single-layer network, in theory, can represent any continuous function, it might require the network to be infeasibly large. 
It may be easier or even required to approximate more complex functions using networks of deep topology \cite{sutton2018reinforcement}.
The class of ML algorithms that use neural nets with multiple hidden layers is known as \acrfull{dl}. 
Interestingly, the universal approximation theorem also applies to networks of bounded width and arbitrary depth. 
Lu et al. \cite{lu2017expressive} showed that for any Lebesgue-integrable function $f : \mathbb{R}^n \rightarrow \mathbb{R}$ and any $\epsilon > 0$, there exists a fully-connected ReLU network $A$ with width $(n+4)$, such that the function $F_A$ represented by this network satisfies
\begin{equation}
\int_{\mathbb{R}^n} |f(x) - F_A(x)| dx < \epsilon
\end{equation}
i.e., any continuous multivariate function $f : \mathbb{R}^n \rightarrow \mathbb{R}$ can be approximated by a deep ReLU network with width $d_m \leq n+4$. 

Poggio et al.\cite{poggio2017and} showed that a deep network could have exponentially better approximation properties than a wide shallow network of the same total size. Conversely, a network of deep topology can attain the same expressivity as a larger shallow network. 
They also show that a deep composition of low-dimensional functions has a theoretical guarantee, which shallow networks do not have, that they can resist the curse of dimensionality for a large class of functions. 

Several unique network architectures have been developed for tasks like computer vision, sequential data, and machine translation. As a result, they can significantly outperform larger and more deeply layered feedforward networks. 
The architecture of neural networks carries an inductive bias, i.e., an a priori algorithmic preference. 
A neural network's inductive bias must match that of the problem it is solving to generalize well out-of-sample.

\subsubsection{Convolutional neural networks}\label{c:dl-cnn}

A \acrfull{cnn} is a type of neural network specialized in processing data with a known, grid-like topology such as time-series data (1-dimensional) or images (2-dimensional) \cite{goodfellow2016deep}. Convolutional neural networks have profoundly impacted fields like computer vision \cite{goodfellow2016deep} and are used in several successful deep RL applications \cite{mnih2013playing, hausknecht2015deep, lillicrap2015continuous}. 
A CNN is a neural net that applies convolution instead of general matrix multiplication in at least one layer. 
A convolution is a form of integral transform defined as the integral of the product of two functions after one is reflected about the y-axis and shifted 
\begin{equation}
s(t)=(x*w)(t) = \int x(a)w(t-a)da
\end{equation}
where $x(t)\in\mathbb{R}$ and $w(t)$ is a weighting function.

The convolutional layer takes the input $x$ with its preserved spatial structure. 
The weights $w$ are given as filters that always extend the full depth of the input volume but are smaller than the full input size. 
Convolutional neural nets utilize weight sharing by applying the same filter across the whole input. 
The filter slides across the input and convolves the filter with the image. 
It computes the dot product at every spatial location, which makes up the activation map, i.e., the output. 
This can be done using different filters to produce multiple activation maps. 
The way the filter slides across the input can be modified. 
The stride specifies how many pixels the filter moves every step. 
It is common to zero pad the border if the stride is not compatible with the size of the filter and the input. 

\begin{figure}[ht]
\centering
\includegraphics[width=\textwidth]{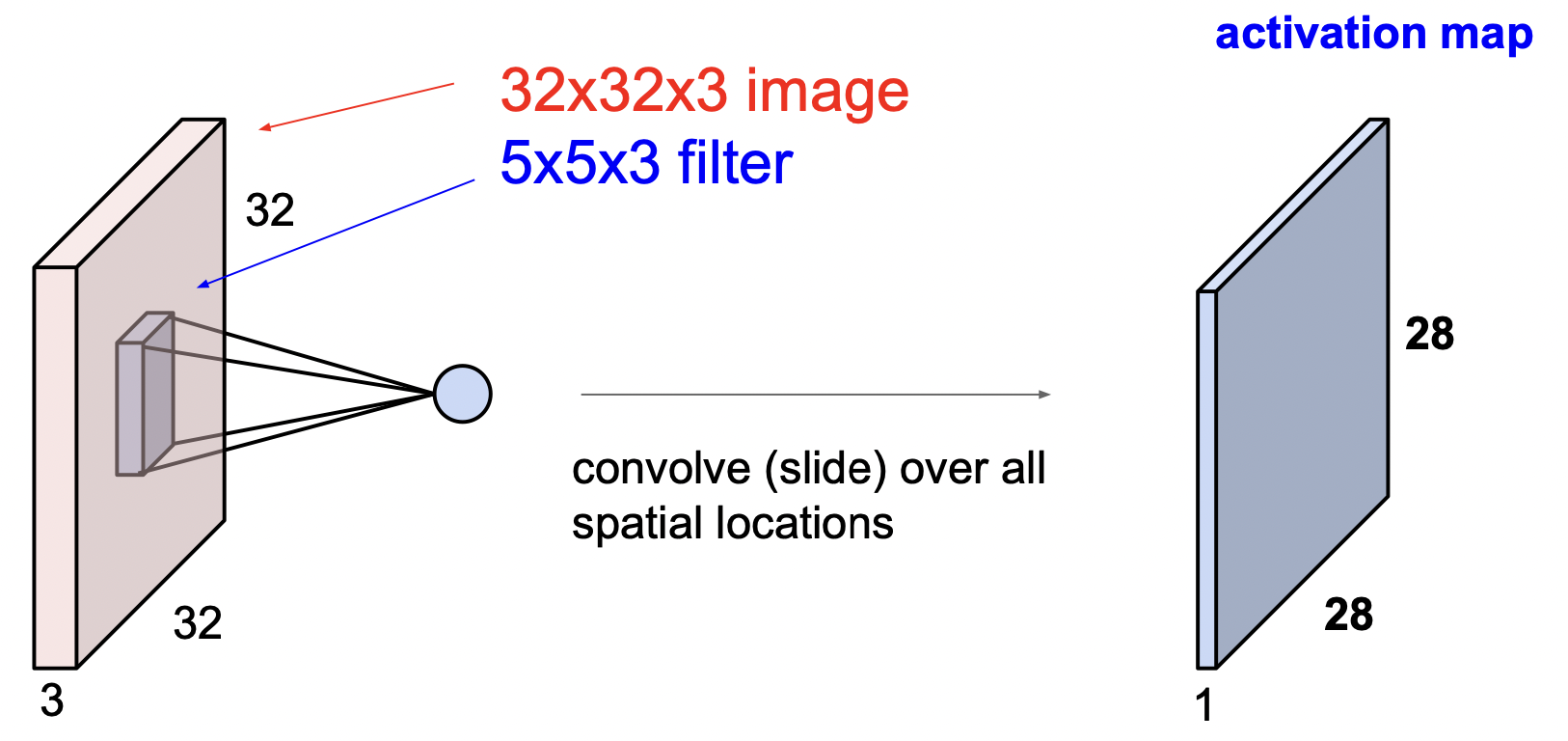}
\caption{3D convolutional layer from \cite{cs231nl5}.}
\end{figure}

After the convolutional layer, a nonlinear activation function is applied to the activation map. 
Convolutional networks may also include pooling layers after the activation function that reduce the dimension of the data. 
Pooling can summarize the feature maps to the subsequent layers by discarding redundant or irrelevant information. 
\textit{Max pooling} is a pooling operation that reports the maximum output within a patch of the feature map. 
Increasing the stride of the convolutional kernel also gives a downsampling effect similar to pooling.

\subsubsection{Recurrent neural networks}

A \acrfull{rnn} is a type of neural network that allows connections between nodes to create cycles so that outputs from one node affect inputs to another. 
The recurrent structure enables networks to exhibit temporal dynamic behavior. 
RNNs scale far better than feedforward networks for longer sequences and are well-suited to processing sequential data. 
However, they can be cumbersome to train as their recurrent structure precludes parallelization. Furthermore, conventional batch norm is incompatible with RNNs, as the recurrent part of the network is not considered when computing the normalization statistic.

\begin{figure}[ht]
\centering
\includegraphics[width=\textwidth]{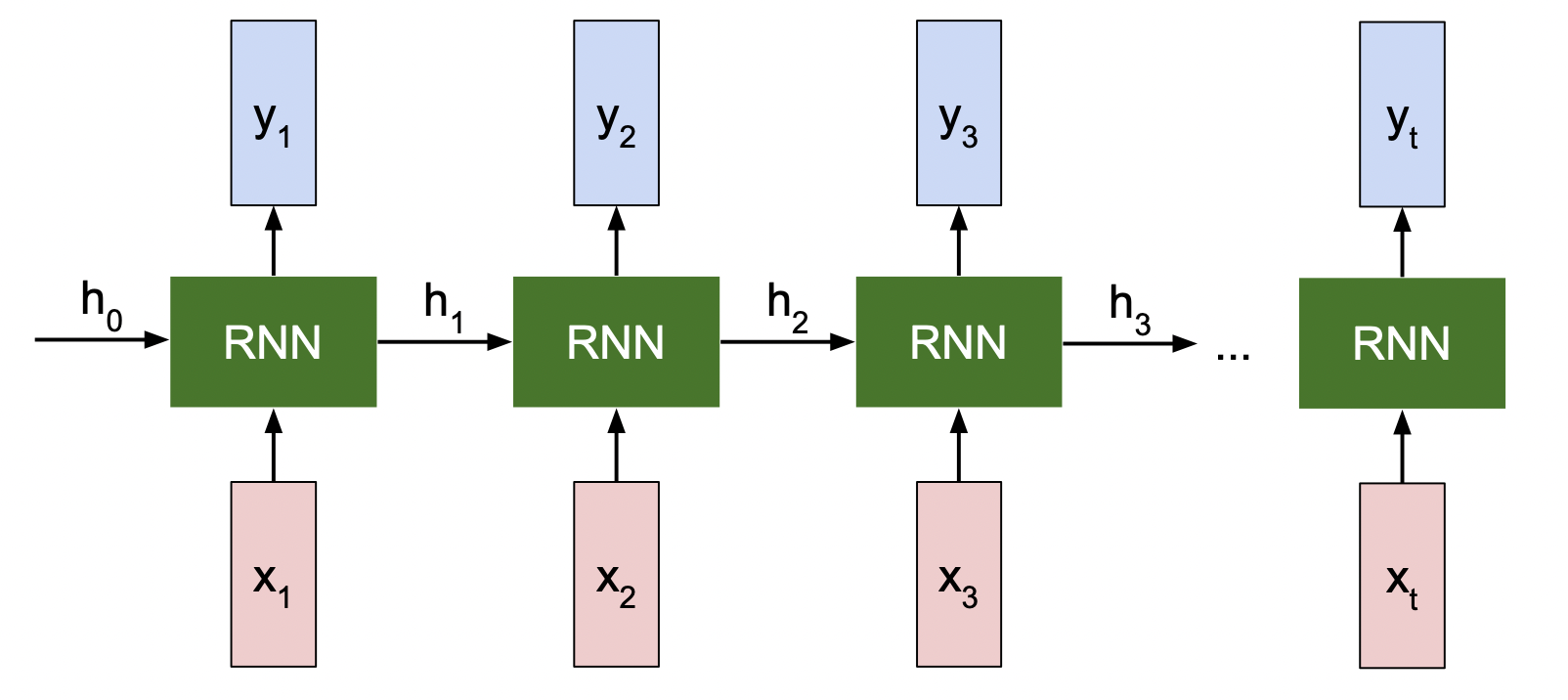}
\caption{Recurrent neural network from \cite{cs231nl10}.}
\end{figure}

RNNs generate a sequence of hidden states $h_t$. 
The hidden states enable weight sharing that allows the model to generalize over examples of various lengths. 
Recurrent neural networks are functions of the previous hidden state $h_{t-1}$ and the input $x_t$ at time $t$. 
The hidden units in a recurrent neural network are often defined as a dynamic system $h^{(t)}$ driven by an external signal $x^{(t)}$
\begin{equation}
h^{(t)}=f(h^{(t-1)}, x^{(t)}; \theta)
\end{equation}

Hidden states $h_t$ are utilized by RNNs to summarize problem-relevant aspects of the past sequence of inputs up to $t$ when forecasting future states based on previous states. 
Since the hidden state is a fixed-length vector, it will be a lossly summary. 
The forward pass is sequential and cannot be parallelized. 
Backprop uses the states computed in the forward pass to calculate the gradient. 
The backprop algorithm used on unrolled RNNs is called \textit{backpropagation through time} (BPTT). 
All nodes that contribute to an error should be adjusted. In addition, for an unrolled RNN, nodes far back in the calculations should also be adjusted. 
\textit{Truncated} backpropagation through time that only backpropagates for a few backward steps can be used to save computational resources at the cost of introducing bias.

Every time the gradient backpropagates through a vanilla RNN cell, it is multiplied by the transpose of the weights. A sequence of vanilla RNN cells will therefore multiply the gradient with the same factor multiple times. 
If $x>1$ then $\lim_{n\rightarrow \infty}x^n = \infty$, and if $x<1$ then $\lim_{n\rightarrow \infty}x^n = 0$. 
If the largest singular value of the weight matrix is $>1$, the gradient will exponentially increase as it backpropagates through the RNN cells. Conversely, if the largest singular value is $<1$, the opposite happens, where the gradient will shrink exponentially. 
For the gradient of RNNs, this will result in either exploding or vanishing gradients. 
This is why vanilla RNNs trained with gradient-based methods do not perform well, especially when dealing with long-term dependencies. Bengio et al. \cite{bengio1994learning} present theoretical and experimental evidence supporting this conclusion. 
Exploding gradients lead to large updates that can have a detrimental effect on model performance. 
The standard solution is to clip the parameter gradients above a certain threshold. 
Gradient clipping can be done element-wise or by the norm over all parameter gradients. Clipping the gradient norm has an intuitive appeal over elementwise clipping. 
Since all gradients are normalized jointly with the same scaling factor, the gradient still points in the same direction, which is not necessarily the case for element-wise gradient clipping \cite{goodfellow2016deep}. 
Let $\lVert \mathbf{g} \rVert$ be the norm of the gradient $\mathbf{g}$ and $v>0$ be the norm threshold. If the norm crosses over the threshold $\lVert \mathbf{g} \rVert > v$, the gradient is clipped to 
\begin{equation}
    \mathbf{g} \leftarrow \frac{\mathbf{g}^v}{\lVert \mathbf{g} \rVert}
\end{equation}
Gradient clipping solves the exploding gradient problem and can improve performance for reinforcement learning with nonlinear function approximation \cite{andrychowicz2020matters}. 
For vanishing gradients, however, the whole architecture of the recurrent network needs to be changed. This is currently a hot topic of research \cite{goodfellow2016deep}.

\paragraph{Long short-term memory}\label{c:dl-lstm}

\acrfull{lstm} is a form of gated RNN designed to have better gradient flow properties to solve the problem of vanishing and exploding gradients. 
LSTMs were introduced in 1997 \cite{hochreiter1997long} and are traditionally used in natural language processing \cite{goodfellow2016deep}. 
Recently, LSMT networks have been successfully applied to financial time series forecasting \cite{siami2018forecasting}. 
Although new architectures like transformers have impressive natural language processing and computer vision performance, LSTMs are still considered state-of-the-art for time series forecasting.

\begin{figure}[ht]
\centering
\includegraphics[width=\textwidth]{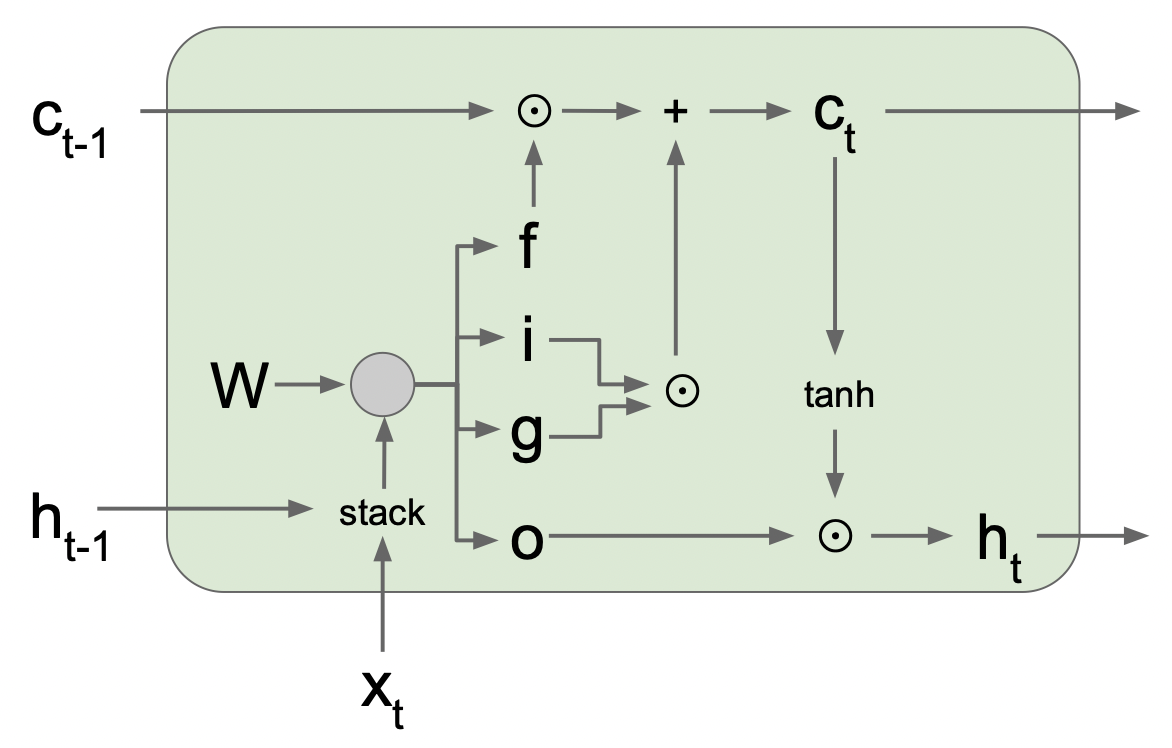}
\caption{LSTM cell from \cite{cs231nl10}.}
\end{figure}

The LSTM is parameterized by the weight $W\in \mathbb{R}^n$, which is optimized using gradient-based methods. 
While vanilla RNNs only have one hidden state, LSTMs maintain two hidden states at every time step. One is $h_t$, similar to the hidden state of vanilla RNNs, and the second is $c_t$, the cell state that gets kept inside the network. 
The cell state runs through the LSTM cell with only minor linear interactions. 
LSTMs are composed of a cell and four gates which regulate the flow of information to and from the cell state and hidden state
\begin{itemize}
    \item Input gate $i$; decides which values in the cell state to update
    \item Forget gate $f$; decides what to erase from the cell state
    \item Output gate $o$; decides how much to output to the hidden state
    \item Gate gate $g$; how much to write to cell, decides how much to write to the cell state
\end{itemize}

The output from the gates is defined as 
\begin{equation}
   \begin{vmatrix} 
   i \\
   f \\
   o \\
   g \\
   \end{vmatrix} 
= 
   \begin{vmatrix} 
   \sigma \\
   \sigma \\
   \sigma \\
   \tanh \\
   \end{vmatrix} 
W
   \begin{vmatrix} 
   h_{t-1} \\
   x_t \\
   \end{vmatrix} 
\end{equation}
where $\sigma$ is the sigmoid activation function.
The cell state $c_t$ and hidden state $h_t$ are updated according to the following rules
\begin{equation}
c_t = f \odot c_{t-1} + i \odot g
\end{equation}
\begin{equation}
h_t = o \odot \tanh{(c_t)}
\end{equation}

When the gradient flows backward in the LSTM, it backpropagates from $c_t$ to $c_{t-1}$, and there is only elementwise multiplication by the $f$ gate and no multiplication with the weights. 
Since the LSTMs backpropagate from the last hidden state through the cell states backward, it is only exposed to one $tanh$ nonlinear activation function. Otherwise, the gradient is relatively unimpeded. 
Therefore, LSTMs handle long-term dependencies without the problem of exploding or vanishing gradients.

%\afterpage{\newpage}
\newpage
\section{Reinforcement learning}\label{c:RL}

An algorithmic trading agent maps observations of some predictor data to market positions. This mapping is non-trivial, and as noted by Moody et al. \cite{moody1998performance}, accounting for factors such as risk and transaction costs is difficult in a supervised learning setting. Fortunately, reinforcement learning provides a convenient framework for optimizing risk- and transaction-cost-sensitive algorithmic trading agents. 

The purpose of this chapter is to introduce the fundamental concepts of reinforcement learning relevant to this thesis. 
A more general and comprehensive introduction to reinforcement learning can be found in ``Reinforcement Learning: An Introduction'' by Richard Sutton and Andrew Barto \cite{sutton2018reinforcement}. An overview of deep reinforcement learning may be found in ``Deep Reinforcement Learning'' by Aske Plaat \cite{plaat2022deep}.
This chapter begins by introducing reinforcement learning \ref{c:RL-intro} and the Markov decision process framework \ref{c:RL-mdp}, and some foundational reinforcement learning concepts \ref{c:RL-reward}, \ref{c:RL-vfuncpolicy}. 
Section \ref{c:RL-funcapprox} discusses how the concepts introduced in the previous chapter (\ref{c:DL}) can be combined with reinforcement learning to generalize over high-dimensional state spaces. 
Finally, section \ref{c:RL-pgm} introduces policy gradient methods, which allow an agent to optimize a parameterized policy directly.

\subsection{Introduction}\label{c:RL-intro}

\acrfull{rl} is the machine learning paradigm that studies how an intelligent agent can learn to make optimal sequential decisions in a time series environment under stochastic or delayed feedback. 
It is based on the concept of learning optimal behavior to solve complex problems by training in an environment that incorporates the structure of the problem. 
The agent optimizes a policy that maps states to actions through reinforcement signals from the environment in the form of numerical rewards. 
The goal of using RL to adjust the parameters of an agent is to maximize the expected reward generated due to the agent's actions. 
This goal is accomplished through trial and error exploration of the environment.
A key challenge of RL is balancing exploring uncharted territory and exploiting current knowledge, known as the \textit{exploration-exploitation} tradeoff. 
Although it has been studied for many years, the exploration-exploitation tradeoff remains unsolved. 
Each action must be tried multiple times in stochastic environments to get a reliable estimate of its expected reward. 
For environments with non-stationary dynamics, the agent must continuously explore to learn how the distribution changes over time. The agent-environment interaction in RL is often modeled as a Markov decision process.

\subsection{Markov decision process}\label{c:RL-mdp}

A \acrfull{mdp} is a stochastic control process and a classical formalization of sequential decision-making. 
A MPD is a tuple $(\mathcal{S}, \mathcal{A}, p, \mathcal{R}, \gamma)$, where 
\begin{itemize}
    \item $\mathcal{S}$ is a countable non-empty set of states (state space). 
    \item $\mathcal{A}$ is a countable non-empty set of actions (action space)
    \item $p(s' | s,a)=Pr(s_{t+1}=s' | s_t=s, a_t=a)$ is the transition probability matrix. 
    \item $\mathcal{R} \subset \mathbb{R}$ is the set of all possible rewards. 
    \item $\gamma \in [0,1]$ is the discount rate. 
\end{itemize}

The agent interacts with the environment at discrete time steps $t=0,1,2,3,...$, which are not necessarily fixed intervals of real-time. 
At each step $t$, the agent receives a representation of the state of the environment $s_t \in \mathcal{S}$, where $s_0 \in \mathcal{S}$ is the initial state drawn from some initial state distribution $p_0 \in \Delta(\mathcal{S})$. 
Based on the state $s_t=s$, the agent chooses one of the available actions in the current state $a_t\in A(s)$. 
After performing the action $a_t$, the agent receives an immediate numerical reward $r_{t+1} \in \mathcal{R}$ and the subsequent state representation $s_{t+1} \in \mathcal{S}$. This interaction with a Markov decision process produces a sequence known as a \textit{trajectory}: $s_0, a_0, r_1, s_1, a_1, r_2, s_2, a_2, r_3, ...$. This sequence is finite for episodic tasks (with the termination time usually labeled $T$); for continuing tasks, it is infinite.

\begin{figure}[ht]
\centering
\includegraphics[width=\textwidth]{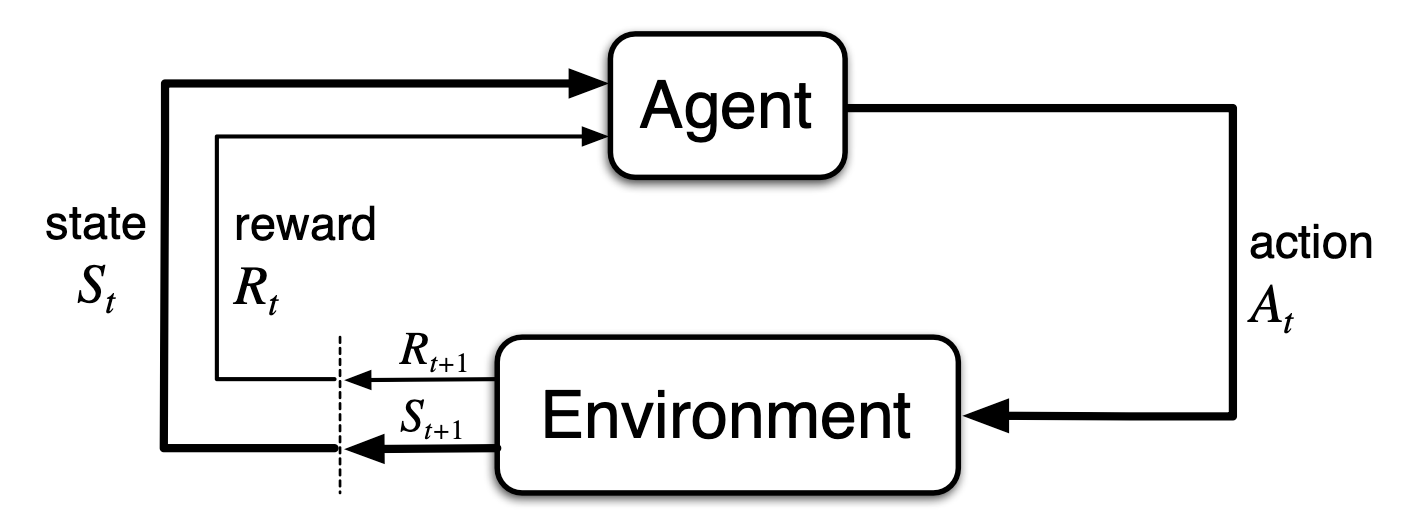}
\caption{Agent-environment interaction from \cite{sutton2018reinforcement}.}
\end{figure}

The dynamics of the system can be completely described by the one-step transition function $p : \mathcal{S} \times \mathcal{R} \times \mathcal{S} \times \mathcal{A} \rightarrow [0,1]$ that is defined as
\begin{equation}
p(s', r | s, a) = Pr\{s_t=s', r_t=r | s_{t-1}=s, a_{t-1}=a\}
\end{equation}
for all $s, s' \in \mathcal{S}$, $r \in \mathcal{R}$, and $a \in \mathcal{A}(s)$. 
It defines a probability distribution such that 
\begin{equation}
\sum_{s \in \mathcal{S}} \sum_{a\in \mathcal{A}(s)} p(s', r | s, a) = 1
\end{equation}
for all $s\in \mathcal{S}$, and $a \in \mathcal{A}(s)$. 
Note that the one-step transition function depends only on the current state $s$ and not previous states, i.e., the state has the Markov property. 
Essentially, MDPs are Markov chains with actions and rewards. 
The transition probabilities $p : \mathcal{S} \times \mathcal{S} \times \mathcal{A} \rightarrow [0,1]$ are defined through the dynamics function $p(s',r|s,a)$, as
\begin{equation}
p(s' |s,a) = Pr(s_t =s' | s_{t-1}=s, a_{t-1}=a) = \sum_{r \in \mathcal{R}} p(s', r | s,a)
\end{equation}
The reward generating function $r : \mathcal{S} \times \mathcal{A} \rightarrow \mathbb{R}$, is defined through the dynamics function $p(s',r|s,a)$, as
\begin{equation}
r(s,a) = \mathbb{E}[r_{t+1} | s_t=s, a_t=a] = \sum_{r \in \mathcal{R}} r \sum_{s' \in \mathcal{S}} p(s', r | s,a)
\end{equation}
The reward-generating function determines the expected reward from performing an action $a$ in a state $s$. 
In practice, the dynamics of the system $p(s', r | s, a)$ are seldom known a priori but learned through interacting with the environment.

\subsubsection{Infinite Markov decision process}

A \textit{finite} MDP is a Markov decision process with countably finite state space $|\mathcal{S}|<\infty$, action space $|\mathcal{A}|<\infty$. 
Finite MDPs can be described as tabular and solved by dynamic programming algorithms with convergence guarantees if the state and action space dimensions $|\mathcal{S} \times \mathcal{A}|$ are not too large. Unfortunately, the applicability of these methods is severely limited by the assumption that state-action spaces are countable. 
These assumptions must be relaxed for MDPs to have significant real-world applications. 
Fortunately, the same theory for finite MDPs also applies to continuous and countably infinite state-action spaces under function approximation. 
The system dynamics are then described by a transition probability function $\mathcal{P}$ instead of a matrix.

\subsubsection{Partially observable Markov decision process}

Let $s_t$ be the environment state and $s^a_t$ be the agent state. A Markov decision process assumes full observability of the environment, i.e., that $s_t = s^a_t$. 
A \acrfull{pomdp} relaxes this assumption and allows for optimal decision-making in environments that are only partially observable to the agent. Since they are generalizations of MDPs, all algorithms used for MDPs are compatible with POMDPs. 
A POMDP is a tuple $(\mathcal{S}, \mathcal{A}, \mathcal{O}, \mathcal{P}, \mathcal{R}, \mathcal{Z}, \gamma)$ that extends a MDP with two additional elements
\begin{itemize}
    \item $\mathcal{O}$ is the observation space. 
    \item $\mathcal{Z}$ is the observation function, $\mathcal{Z} = Pr(o_{t+1} = o | s_{t+1}=s', a_t = a)$. 
\end{itemize}

An agent in a POMDP cannot directly observe the state; instead, it receives an observation $o \in \mathcal{O}$ determined by the observation function $\mathcal{Z}$. 
The agent state approximates the environment state $s^a_t \approx s_t$. 
However, a single observation $o$ is not a Markovian state signal. 
Direct mapping between observation and action is insufficient for optimal behavior, and a memory of past observations is required. 
The history of a POMDP is a sequence of actions and observations $h_t = \{ o_1, a_1, ..., o_t, a_t \}$. 
The agent state can be defined as the history $s^a_t = h_t$. 
However, storing and processing the complete history of every action scales linearly with time, both in memory and computation.
A more scalable alternative is a stateful sequential model like a recurrent neural network (RNN). 
In this model, the agent state is represented by the network $s^a_t = f_\theta(s^a_{t-1}, o_t)$.

A state can be split into an agent's internal state and the environment's external state. Anything that cannot be changed arbitrarily by the agent is considered external and, thus, part of the external environment. 
On the other hand, the internal data structures of the agent that the agent can change are part of the internal environment.

\subsection{Rewards}\label{c:RL-reward}

The goal of a reinforcement learning agent is to maximize the expected return $\mathbb{E}[G_t]$, where the return $G_t$ is defined as the sum of rewards
\begin{equation}
G_t = r_{t+1} + r_{t+2} + ... + r_{T} 
\end{equation}
In an episodic setting, where $t = 0,...,T$, this goal is trivial to define as the sequence of rewards is finite. 
However, some problems, like financial trading, do not naturally break into subsequences and are known as continuing problems. 
For continuing problems, where $T=\infty$ and there is no terminal state, it is clear that the sum of rewards $G_t$ could diverge. 
Discounting was introduced to solve the problem of returns growing to infinity. 
Discounted returns are defined as 
\begin{equation}
G_t = \sum_{k=0}^\infty \gamma^k r_{t+k+1} = r_{t+1} + \gamma G_{t+1}
\end{equation}
where $\gamma \in [0,1]$ is the discount rate used to scale future rewards. 
Setting $\gamma = 0$ suggests that the agent is myopic, i.e., only cares about immediate rewards. 
As long as $\gamma< 1$ and the reward sequence is bounded, the discounted return $G_t$ is finite. 
Discounting allows reinforcement learning to be used in continuing problems.

Reinforcement signals $r_{t+1}$ from the environment can be immediate or delayed. 
Games and robot control are typical examples of delayed reward environments, where an action affects not only the immediate reward but also the next state and, through that, all subsequent rewards. 
An example of delayed reward is when chess players occasionally sacrifice a piece to gain a positional advantage later in the game. 
Although sacrificing a piece in isolation is poor, it can still be optimal long-term. Consequently, temporal credit assignment is a fundamental challenge in delayed reward environments. 
AlphaZero \cite{silver2017mastering} surpassed human-level play in chess in just 24 hours, starting from random play, using reinforcement learning. 
Interestingly, AlphaZero seems unusually (by human standards) open to material sacrifices for long-term positional compensation, suggesting that the RL algorithm estimates delayed reward better than human players. 
Throughout this thesis, financial trading is modeled as a stochastic immediate reward environment. This choice is justified in chapter \ref{part:problemsetting}. 
Therefore, the problem reduces to an associative reinforcement learning problem, a specific instance of the full reinforcement learning problem. 
It requires generalization and trial-and-error exploration but not temporal credit assignment. 
The methods presented in this chapter will only be those relevant in an immediate reward environment. 
Unless otherwise stated, the discount rate $\gamma$, a tradeoff between immediate and delayed rewards, is assumed to be zero, making the agent myopic. 
As a result, the return $G_t$ in an immediate reward environment is defined as the immediate reward 
\begin{equation}
G_t = \sum_{k=0}^{\infty} \gamma^k r_{t+k+1} = r_{t+1}
\end{equation}

\subsection{Value function and policy}\label{c:RL-vfuncpolicy}

A stochastic policy is a mapping from states to a probability distribution over the action space and is defined as
\begin{equation}
\pi :  \mathcal{S} \rightarrow \Delta(\mathcal{A})
\end{equation}
The stochastic policy is a probability measure $\pi (a | s) = Pr\{ a_t = a | s_t = s \}$, which is the probability that the agent performs an action $a$, given that the current state is $s$. 
Stochastic policies can be advantageous in problems with perceptual aliasing. Furthermore, it handles the exploration-exploitation tradeoff without hard coding it. 
A deterministic policy maps states $\mathcal{S}$ to actions $\mathcal{A}$, and is defined as
\begin{equation}
\mu : \mathcal{S} \rightarrow \mathcal{A}
\end{equation}
RL algorithms determine how policies are adjusted through experience, where the goal is for the agent to learn an optimal or near-optimal policy that maximizes returns.

Value-based RL algorithms, like Q-learning and SARSA, estimate a state-value function or an action-value function and extract the policy from it. 
The state-value function $V_\pi(s)$ is the expected return when starting in state $s$ and following policy $\pi$. It is defined $\forall s \in \mathcal{S}$ as 
\begin{equation}
V_\pi(s)= \mathbb{E}_\pi[G_t | s_t = s] 
\end{equation}
The action-value function $Q_\pi(s,a)$ is the expected return when performing action $a$ in state $s$ and then following the policy $\pi$. It is defined $\forall s \in \mathcal{S}, a \in \mathcal{A}(s)$ as 
\begin{equation}
Q_\pi(s,a)= \mathbb{E}_\pi[G_t | s_t = s, a_t = a]
\end{equation}
An example of a value-based policy is the $\epsilon$-greedy policy, defined as
\begin{equation}
 \pi(a|s, \epsilon) =
 \begin{cases}
 \argmax_a{Q_\pi(s, a)} & \text{with probability $1-\epsilon$}\\
 \text{sample random action } a \sim \mathcal{A}(s) & \text{with probability $\epsilon$}
 \end{cases} 
\end{equation}
where $\epsilon \in [0,1]$ is the exploration rate. 

Reinforcement learning algorithms are divided into \textit{on}-policy and \textit{off}-policy algorithms. 
The same policy that generates the trajectories is being optimized for on-policy algorithms. In contrast, for off-policy algorithms, the policy generating trajectories differs from the one being optimized. 
For off-policy learning, the exploration can be delegated to an explorative behavioral policy while the agent optimizes a greedy target policy.

\subsection{Function approximation}\label{c:RL-funcapprox}

Tabular reinforcement learning methods include model-based methods like dynamic programming as well as model-free methods like Monte-Carlo and temporal-difference learning methods (e.g., Q-learning and SARSA). 
Unfortunately, tabular methods require discrete state and action spaces, and due to the curse of dimensionality, these spaces must be relatively small. Thus, their applicability to real-world problems is limited. Complex environments like financial trading cannot be represented in discrete states. 
Instead, feature vectors represent states in environments where the state space is too large to enumerate. 
As most states will probably never be visited and visiting the same states twice is unlikely, it is necessary to generalize from previous encounters to states with similar characteristics. 
This is where the concepts of function approximation discussed in the previous chapter (\ref{func-approx}) come in. 
Using function approximation, samples from the desired function, e.g., the value function, are generalized to approximate the entire function.

Value-based reinforcement learning algorithms such as \acrfull{dqn} \cite{mnih2013playing} and \acrfull{drqn} \cite{hausknecht2015deep} use deep neural networks as function approximators to generalize over a continuous space that is optimized using the Q-learning algorithm. 
DQN and DRQN achieved superhuman performance in playing Atari games using raw pixels as input into a convolutional neural network that outputs action-value estimates of future returns. 
These value-based algorithms are still limited by discrete action space and the curse of dimensionality as it has to calculate the Q-value of every single action. 
In the games DQN and DRQN are tested on, the agent is limited to a small discrete set of actions (between 4 and 18). 
However, for many applications, a discrete action space is severely limiting. 
Furthermore, these algorithms use the naive exploration heuristics $\epsilon$-greedy, which is not feasible in critical domains. 
Fortunately, policy gradient methods bypass these problems entirely.

\subsection{Policy gradient methods}\label{c:RL-pgm}

While value-based reinforcement learning algorithms extract a policy from action-value estimates, \acrfull{pg} methods learn a parameterized policy and optimize it directly. 
The policy’s parameter vector is $\theta\in \mathbb{R}^{d'}$, with the policy defined as 
\begin{equation}
   \pi_\theta(a|s)=Pr\{a_t=a|s_t=s,\theta_t=\theta\} 
\end{equation}

Continuous action space is modeled by learning the statistics of the probability of the action space. 
A natural policy parameterization in continuous action spaces is the Gaussian distribution $a \sim \mathcal{N}(\mu_\theta(s), \sigma_\theta(s)^2)$ defined as
\begin{equation}
\pi_\theta(a|s)= \dfrac{1}{\sigma_\theta(s)\sqrt{2\pi}} e^{-\dfrac{(a-\mu_\theta(s))^2}{2\sigma_\theta(s)^2}}
\end{equation}
where $\mu_\theta(s)\in\mathbb{R}$ and $\sigma_\theta(s)\in\mathbb{R_+}$ are parametric function approximations of the mean and standard deviation, respectively. 
The mean decides the space where the agent will favor actions, while the standard deviation decides the degree of exploration. 
It is important to note that this gives a probability density, not a probability distribution like the softmax distribution.

For policy gradient methods in the continuous time setting, the goal of optimizing the policy $\pi_\theta$ is to find the parameters $\theta$ that maximize the average rate of return per time step \cite{sutton2018reinforcement}. 
The performance measure $J$ for the policy $\pi_\theta$ in the continuing setting is defined in terms of the average rate of reward per time step as
\begin{align*}
 J(\pi_\theta) &= \int_{\mathcal{S}} d^\pi(s) \int_{\mathcal{A}} r(s,a) \pi_\theta(a|s) da ds && \\
 &= \mathbb{E}_{s \sim d^\pi, a \sim \pi_\theta} [r(s,a)]\numberthis \label{perfmeasure1}  
\end{align*}
where $d^\pi(s) = \lim_{t \rightarrow \infty} Pr\{ s_t = s | a_{0:t} \sim \pi_\theta \}$ is the steady-state distribution under the policy $\pi_\theta$. 

Policy optimization aims to find the parameters $\theta$ that maximize the performance measure $J$. 
Gradient ascent is used as the optimization algorithm for the policy. 
The policy parameter $\theta$ is moved in the direction suggested by the gradient of $J$ to maximize the return, yielding the following gradient ascent update
\begin{equation}
\theta_{t+1}=\theta_t+\alpha \widehat{\nabla_\theta J(\pi_{\theta_t})}
\end{equation}
where $\alpha$ is the step-size and $\widehat{\nabla_\theta J(\pi_{\theta_t})}$ is a stochastic estimate whose expectation approximates the gradient of $J$ with respect to $\theta$ \cite{sutton2018reinforcement}.

The policy gradient theorem\footnote{For the full proof see chapter 13.6 in \cite{sutton2018reinforcement}} for the continuing case provides the following expression for the gradient 
\begin{align*}
\nabla_\theta J(\pi_\theta) &=  \int_{\mathcal{S}} d^\pi(s) \int_\mathcal{A} Q_{\pi} (s,a) \nabla_\theta \pi_\theta(a|s) dads && \\
 &= \mathbb{E}_{s \sim d^\pi, a \sim \pi_\theta} [Q_\pi (s,a) \nabla_\theta \log{\pi_\theta(a|s)} ] \numberthis \label{perfmeasure2}  
\end{align*}
Even though the steady-state distribution $d^\pi$ depends on the policy parameters $\theta$, the gradient of the performance measure does not involve the gradient of $d^\pi$, allowing the agent to simulate paths and update the policy parameter at every step \cite{sutton2018reinforcement}.

\subsubsection{REINFORCE}

REINFORCE is an on-policy direct policy optimization algorithm derived using the policy gradient theorem \cite{sutton2018reinforcement}. 
The algorithm is on-policy. Consequently, the agent will encounter the states in the proportions specified by the steady-state distribution. 
Using the policy gradient theorem, the calculation of the policy gradient reduces to a simple expectation. 
The only problem is estimating the action-value function $Q_\pi (s,a)$. 
REINFORCE solves this problem by using the sampled return $G_t$ as an unbiased estimate of the action-value function $Q_\pi (s_t,a_t)$. 
Observing that the state-value is equal to the expectation of the sampled return, i.e., $\mathds{E}_\pi[G_t|s_t,a_t]=Q_\pi(s_t,a_t)$, the following expression for the policy gradient can be defined
\begin{align*}
 \nabla_\theta J(\pi_\theta) &= \mathbb{E}_{s \sim d^\pi, a \sim \pi_\theta} [Q_\pi (s,a) \nabla_\theta \log{\pi_\theta(a|s)} ] && \\
 &= \mathbb{E}_{s \sim d^\pi, a \sim \pi_\theta} [G_t \nabla_\theta \log{\pi_\theta(a|s)} ] \numberthis \label{reinforcegrad}  
\end{align*}
This expression can be sampled on each time step t, and its expectation equals the gradient. 
The gradient ascent policy parameter update for REINFORCE is defined as
\begin{equation}
\theta_{t+1}=\theta_t+\alpha G_t \nabla_\theta \log{\pi_{\theta_t}(a_t|s_t)} 
\end{equation}
where $\alpha$ is the step size. 
The direction of the gradient is in the parameter space that increases the probability of repeating action $a_t$ on visits to $s_t$ in the future the most \cite{sutton2018reinforcement}. The higher the return, the more the agent wants to repeat that action. The update is inversely proportional to the action probability to adjust for different frequencies of visits to states, i.e., some states might be visited often and have an advantage over less visited states.

While REINFORCE is unbiased and only requires estimating the policy, it might exhibit high variance due to the high variability of sampled returns (if the trajectory space is large). 
High variance leads to unstable learning updates and slower convergence. 
Furthermore, the stochastic policy used to estimate the gradient can be disadvantageous in critical domains such as health care or finance. 
Thankfully, both these problems can be solved by a class of policy gradient methods called actor-critic methods.

\subsubsection{Actor-critic}\label{c:rl-pgm-ac}

Policy-based reinforcement learning is effective in high-dimensional and continuous action space, while value-based RL is more sample efficient and more convenient for online learning. 
\acrfull{ac} methods seek to combine the best of both worlds where a policy-based actor chooses actions, and the value-based critic critique those actions. 
The actor optimizes the policy parameters using stochastic gradient ascent in the direction suggested by the critic.
The critic's value function is optimized using stochastic gradient descent to minimize the loss to the target. 
This use of a critic introduces bias since the critique is an approximation of the return and not actual observed returns like in actor-based algorithms like REINFORCE. 
There are numerous actor-critic algorithms like advantage actor-critic (A2C) \cite{sutton2018reinforcement}, asynchronous advantage actor-critic (A3C) \cite{mnih2016asynchronous}, and proximal policy optimization (PPO) \cite{schulman2017proximal}, that have exhibited impressive performance in a variety of applications. 
These methods rely on stochastic policies and computing the advantage function. 
For critical domains such as finance, a deterministic policy directly optimized by a learned action-value function might be more appropriate. 
Fortunately, the policy gradient framework can be extended to deterministic policies \cite{silver2014deterministic, lillicrap2015continuous}. 

The idea behind deterministic actor-critic algorithms is based on Q-learning, where a network $Q(s,a)$ approximates the return. 
Q-learning can be extended to high-dimensional state spaces by defining the Q-network as a function approximator $Q_{\phi}(s,a) : \mathcal{S} \times \mathcal{A} \rightarrow \mathbb{R}$, parameterized by $\phi \in \mathbb{R}^{b'}$. 
If the Q-network is optimal ($Q_\phi^*$), finding the optimal action ($a^*$) in a small discrete action space is trivial; $a^*(s) = \argmax_a{Q_\phi^*(s,a)}$. 
However, the exhaustive computations required for this process are not feasible in high-dimensional or continuous action spaces due to the curse of dimensionality. 
This problem can be bypassed by learning a deterministic policy $\mu_\theta(s) : \mathcal{S} \rightarrow \mathcal{A}$, parameterized by $\theta \in \mathbb{R}^{d'}$, as an approximator to $a(s)$, such that $\max_a{Q_\phi(s,a)} \approx Q_\phi(s, \mu(s))$.

\paragraph{Deterministic policy gradient}

Let $\mu_\theta : \mathcal{S} \rightarrow \mathcal{A}$ be the deterministic policy parameterized by $\theta \in \mathbb{R}^{d'}$. 
The performance measure $J$ for the deterministic policy $\mu_\theta$ in the continuous time average reward setting is defined as
\begin{align*}
 J(\mu_\theta) &= \int_{\mathcal{S}} d^\mu(s) r(s,\mu_\theta(s)) ds && \\
 &= \mathbb{E}_{s \sim d^\mu} [r(s,\mu_\theta(s))]\numberthis \label{perfmeasuredeterm}  
\end{align*}
Initially, there was a belief that the deterministic policy gradient did not exist; however, it was proven by Silver et al. \cite{silver2014deterministic}, which provides the following expression for the gradient
\begin{align*}
 \nabla_\theta J(\mu_\theta) &= \int_{\mathcal{S}} d^\mu(s) \nabla_\theta \mu_\theta(s) \nabla_a Q^\mu(s,a) |_{a=\mu_\theta(s)} ds && \\
 &= \mathbb{E}_{s \sim d^\mu} [\nabla_\theta \mu_\theta(s) \nabla_a Q^\mu(s,a) |_{a=\mu_\theta(s)}]\numberthis \label{graddeterm}  
\end{align*}
The deterministic policy gradient theorem holds for both on-policy and off-policy methods. 
Deterministic policies only require integrating over the state space and not both the state and action space like stochastic policies. 
The true action-value can be approximated by a parameterized critic, i.e., $Q_\phi(s,a) \approx Q^\mu(s,a)$.

\paragraph{Off-policy learning}

Learning a deterministic policy in continuous action spaces on-policy will generally not ensure sufficient exploration and can lead to sub-optimal solutions. 
To solve this problem, the deterministic actor-critic algorithm learns off-policy by introducing an exploration policy $\mu'_\theta$ defined as 
\begin{equation}
\mu'_\theta (s) = \mu_\theta (s) + \mathcal{W}
\end{equation}
where $\mathcal{W}$ is sampled noise from a noise-generating function. 
The exploration policy $\mu'_\theta$ explores the environment and generates trajectories that optimize the target policy $\mu_\theta$ and Q-network $Q_\phi$.

\paragraph{Q-network optimization}

Let $Q_{\phi}(s,a) : \mathcal{S} \times \mathcal{A} \rightarrow \mathbb{R}$ be the Q-network parameterized by $\phi \in \mathbb{R}^{b'}$. 
The Q-network is iteratively updated to fit a target defined by the recursive relationship $y = r + \gamma \max_{a'} Q(s',a')$ known as the Bellman equation \cite{sutton2018reinforcement}. The Bellman equation reduces to the immediate reward in an immediate reward environment, where $\gamma=0$. 
The goal is to find the weights $\phi$ that minimize the loss (usually MSE) to the target
\begin{equation}
L(Q_\phi) = \mathbb{E}_{s \sim d^{\mu'} ,a \sim \mu', r\sim E} [(Q_\phi (s,a) - y)^2]
\end{equation}
where $d^{\mu'}$ is the steady-state distribution under the exploration policy $\mu'_\theta$, and $E$ is the environment. 
The gradient of the loss function with respect to the Q-network parameter weights $\phi$ is defined as
\begin{equation}
 \nabla_\phi L(Q_\phi) = \mathbb{E}_{s \sim d^{\mu'} ,a \sim \mu', r\sim E} [(Q_\phi (s,a) - y) \nabla_\phi Q_\phi (s,a)]
\end{equation}
and is used to calculate the backward pass in the Q-network's stochastic gradient descent optimization algorithm.

\paragraph{Replay memory}

Learning policies and Q-networks with large nonlinear function approximators is generally considered difficult and unstable and do not come with convergence guarantees. Another challenge of combining deep neural networks with reinforcement learning is that most ML optimization algorithms assume that samples are independent and identically distributed (IID). 
The IID assumption is rarely valid for RL agents sequentially exploring the state space. 
Furthermore, minibatch learning is advantageous as it efficiently utilizes hardware optimization. 
The introduction of replay memory \cite{mnih2013playing} addresses these problems and trains large nonlinear function approximators stably and robustly. 
A replay memory $\mathcal{D} = \{ \tau_{t-k+1}, \tau_{t-k+2}, ..., \tau_{t} \}$ is a finite cache storing the past $k$ transitions $\tau_t = (s_t, a_t, r_{t})$. 
A minibatch $\mathcal{B} \subseteq \mathcal{D}$ of $|\mathcal{B}|>0$ transitions are randomly sampled from the replay memory and used to update both the policy and Q-network.

Randomly sampled batches are ineffective for training recurrent neural networks, which carry forward hidden states through the mini-batch. 
Deep Recurrent Q-Network (DRQN) \cite{hausknecht2015deep} is an extension of DQN for recurrent neural networks. 
DRQN uses experience replay like DQN; however, the sampled batches are in sequential order. 
The randomly sampled batch $\mathcal{B} \subseteq \mathcal{D}$ consists of the transitions $\mathcal{B} = \{ \tau_i, \tau_{i+1}, ..., \tau_{i+|\mathcal{B}|-2}, \tau_{i+|\mathcal{B}|-1} \}$, where $i$ is some random starting point for the batch.
The RNNs initial hidden state is zeroed at the start of the mini-batch update but then carries forward through the mini-batch.

\afterpage{\blankpage}
\newpage
\part{Methodology}\label{part:methodology}

%\afterpage{\newpage}
\newpage
\section{Problem Setting}\label{part:problemsetting}

In reinforcement learning, the agent learns through interaction with the environment. Thus, developing a model of the environment, in this case, the commodity market, is necessary to optimize an algorithmic trading agent through reinforcement signals.
Commodities trading involves sequential decision-making in a stochastic and nonstationary environment to achieve some objective outlined by the stakeholder. This chapter describes a discrete-time Markov decision process that models this environment \footnote{Although this thesis focuses on commodities, the model's general concepts apply to other financial markets.}. 
Neither the strong assumption of countable state-actions nor the assumption of full environment observability can be satisfied. 
Thus, based on the previously proposed financial markets dynamical system \cite{zhang2020deep, huang2018financial, liu2018practical}, this chapter presents an infinite partially observable MDP for commodities trading.

\subsection{Assumptions}\label{part:prob-assumptions}

Since the model will be tested ex-post by backtesting, described in section \ref{c:at-backtest}, it is necessary to make a couple of simplifying assumptions about the markets the agent operates in: 
\begin{enumerate}
 \item No slippage, i.e., there is sufficient liquidity in the market to fill any orders placed by the agent, regardless of size, at the quoted price. In other words, someone is always willing to take the opposite side of the agent's trade. This assumption relates to external factors that may affect the price between the time the agent is quoted the price and the time the order is filled\footnote{In reality, prices may significantly change between receiving a quote and placing an order.}.
 \item No market impact, i.e., the money invested by the agent is not significant enough to move the market. This assumption relates to the agent's own trades' impact on the market. The reasonability of this assumption depends on the depth of the market. 
\end{enumerate}

\subsection{Time Discretization}\label{c:ps-td}

Financial trading involves continuously reallocating capital in one or more financial assets. 
The problem does not naturally break into sub-sequences with terminal states. 
Therefore, this MDP is in the continuous-time setting.
A discretization operation is applied to the continuous timeline to study the reinforcement learning-based algorithmic trading described in this thesis, discretizing the timeline into steps $t = 0,1,2, ...$. 
As described in section \ref{c:at-subsampl}, sampling at fixed time intervals is unsatisfactory in markets where activity varies throughout the day and exhibits undesirable statistical properties like non-normality of returns and heteroskedasticity. 
Instead of the traditional time-based constant duration $\Delta t$, the observations are sampled as a function of dollar volume based on the ideas from Mandelbrot and Taylor \cite{mandelbrot1997variation, mandelbrot1967distribution}, and Clark \cite{clark1973subordinated} presented in section \ref{c:at-subsampl}. Dollar volume-based sampling provides better statistical properties for the agent and can, without human supervision, adapt to changes in market activity.

In practice, observations are sampled by sequentially summing the product of the volume $v_i$ and price $p_i$ of every trade in the market and then sampling a new observation once this sum breaches a predefined threshold $\delta > 0$ before starting the process over again. 
Define the sum of the total transacted dollar volume from the past sampled point $k$ to point $i$ as
\begin{equation}
\chi_i = \sum_{j=k+1}^{i} v_j \cdot p_j
\end{equation}
where $i \geq k+1$. 
Once $\chi_i$ breaches the threshold, i.e., $\chi_i > \delta$, the sub-sampling scheme samples the trade at time $i$ as a new observation, $k=i+1$, and resets the sum of dollar volume $\chi_{i+1} = 0$.

Due to the increasing volume in the energy futures markets in recent years, defining an appropriate threshold $\delta$ is complicated. 
On the one hand, the purpose of using this sampling scheme is that the sampling frequency will deviate throughout the day and weeks depending on the transacted dollar volume. 
However, if structural changes in the market significantly alter the transacted dollar volume over long periods, e.g., three months, it would be advantageous for the threshold to adjust to that change. 
A constant threshold will therefore be unsatisfactory as it would not be reactive to these structural changes over long periods.
A more robust alternative is a threshold that adjusts itself without human supervision. 
Therefore, the threshold $\delta$ is defined using a simple moving average over the daily dollar volume of the past 90 days, avoiding lookahead bugs. 
The threshold is tuned using one parameter, the target number of samples per day, defined as $tgt \in \mathbb{R}_+$. 
The threshold $\delta$ is defined as 
\begin{equation}
\delta = \frac{SMA_{90d}(v \cdot p)}{tgt} 
\end{equation}
which is the threshold needed to achieve the target number of samples per day in the past 90 days. 
The threshold continuously updates as trades occur in the market. 
There is no guarantee that the threshold will lead to the desired amount of samples per day, as it is computed from historical data. 
Nonetheless, it does achieve satisfactory results, even in unstable markets.

The time discretization scheme presented in this section represents progress in the research area from fixed time-interval-based sampling, providing better statistical properties while being more robust and adaptive to changing market environments.

\subsection{State Space}\label{c:ps-ss}

The universe of possible investments is limited to one instrument.
The state space of a financial market includes all market participants and their attitudes toward the financial instrument. 
Thus, the state space $\mathcal{S}$ is continuous and partially observable. 
Representing the environment state $\mathbf{s}_t$ to an algorithmic trading agent is impossible, so it needs to be approximated by an agent state, i.e., $\mathbf{s}^a_t \approx \mathbf{s}_t$. 
This thesis adopts the philosophy of technical traders described in section \ref{c:at-forecasting}. It uses past trades, specifically their price and volume, as observations $\mathbf{o}_t$ of the environment. 
Let $k \in \mathbb{R}_+$ be the number of trades for the instrument during the period $(t-1, t]$. 
An observation $\mathbf{o}_t$ at time $t$ is defined as 
\begin{equation}
\mathbf{o}_t=[\mathbf{p}_t, \mathbf{v}_t]
\end{equation}
where
\begin{itemize}
 \item $\mathbf{p}_t \in \mathbb{R}^{k}$ are the prices of all $k$ trades during the period $(-1, t]$. The opening price is denoted $p_t$. 
 \item $\mathbf{v}_t \in \mathbb{R}^{k}$ are the volumes of all $k$ trades during the period $(-1, t]$. 
\end{itemize}

A single observation $\mathbf{o}_t$ is not a Markovian state signal, and the agent state can be defined by the entire history $\mathbf{h}_t$. 
However, this alternative is not scalable. 
Section \ref{c:ps-td} introduced the time discretization scheme for this environment, which is a form of sub-sampling. 
However, the computational and memory requirements still grow linearly with the number of samples, so a history cut-off is also employed. 
In other words, the agent will only have access to the past $n \in \mathbb{N}_+$ observations $\mathbf{o}_{t-n+1:t}$. 
In addition, the recursive mechanism of considering the past action as a part of the internal state of the environment introduced by Moody et al. \cite{moody1998performance} is adopted to consider transaction costs. 
The agent state is formed by concatenating the external state consisting of stacking the $n$ most recent observations with the internal state consisting of the past action $a_{t-1}$, i.e., 
\begin{equation}
\mathbf{s}^a_t = \{ \mathbf{o}_{t-n+1:t}, a_{t-1} \}
\end{equation}
The dimension of the agent state vector is $\dim{(\mathbf{s}^a_t)} = 2kn+1$.

\subsection{Action Space}

At every time step $t$, the agent can buy or sell the instrument on the market. 
The opening price of a period $p_t$, the price the agent can buy or sell the instrument for at time $t$, is the last observed price, i.e., the closing price of the previous period $(t-1, t]$.
The no slippage assumption from section \ref{part:prob-assumptions} implies that the instrument can be bought or sold in any quantity for the time $t$ at the opening price of that period $p_{t}$.

Some trading environments allow the agent to output the trade directly; e.g., $a_t = -5$ corresponds to selling five contracts, or $a_t = +10$ corresponds to purchasing ten contracts. However, despite its intuitive nature, it can be problematic because the agent must maintain a continuous record of the number of contracts it holds and the amount of available balance at all times in order to avoid making irrational decisions such as selling contracts they do not own or purchasing more contracts than they can afford. Adding this layer of complexity complicates the learning process.
Instead, a more straightforward approach is to have the agent output its desired position weight. In this case, a trade is not directly outputted but inferred from the difference between the agent's current position and its chosen next position.

At every step $t$, the agent performs an action $a_t \in [-1,1]$, representing the agent's position weight during the period $(t, t+1]$. 
The weight represents the type and size of the position the agent has selected, where 
\begin{itemize}
 \item $a_t > 0$ indicates a long position, where the agent bets the price will rise from time $t$ to time $t+1$. The position is proportional to the size of the weight, where $a_t=1$ indicates that the agent is maximally long. 
 \item $a_t = 0$ indicates no position. 
 \item$a_t < 0$ indicates a short position, where the agent bets the price will fall. $a_t = -1$ indicates that the agent is maximally short. This thesis assumes that there is no additional cost or restriction on short-selling. 
\end{itemize}
The trading episode starts and ends (if it ends) with no position, i.e., ${a}_{0} = {a}_{T} = 0$.

The weight $a_t$ represents a fraction of the total capital available to the agent at any time. 
For this problem formulation, it is irrelevant if $a_t = 1$ represents $\$1$ or $\$100$ million. 
However, this requires that any fraction of the financial instrument can be bought and sold. 
E.g., if the agent has $\$100$ to trade and wants to take the position $a_t = 0.5$, i.e., a long position worth $\$50$, the price might not be a factor of $50$, meaning that the agent would not get the exact position it selected. 
The fractional trading assumption is less reasonable the smaller the amount of capital available to the agent. 
On the other hand, the assumptions made in section \ref{part:prob-assumptions} are less reasonable the higher the amount of capital.

\subsection{Reward Function}

As noted in section \ref{c:at-mapping2pos}, the goal of an algorithmic trading agent should not be to minimize forecast loss but to maximize returns, as it is more in line with the ultimate goal of the trader. 
Transaction costs represent a non-trivial expense that must be accounted for to generalize to real-world markets. 
Moreover, section \ref{c:at-mpt} introduced the philosophy of modern portfolio theory, which advocates maximizing risk-adjusted returns. 
An advantage of reinforcement learning is that the trading agent can be directly optimized to maximize returns while considering transaction costs and risk. 
This section introduces a reward function sensitive to transaction costs and risk.

The reward $r_t$ is realized at the end of the period $(t-1, t]$ and includes the return of the position ${a}_{t-1}$ held during that interval.
The objective of financial trading is generally to maximize future returns, or in more vernacular terms; to buy when the price is low and sell when the price is high. 
The multiplicative return of a financial instrument at time $t$ is defined as the relative change in price from time $t-1$ to $t$
\begin{equation}\label{eq:y}
y_t = \frac{p_t}{p_{t-1}} -1
\end{equation}
Multiplicative returns, unlike additive returns, have the advantage that they are insensitive to the size of the capital traded.
Logarithmic returns $\log{(y_t + 1)}$ are typically used in algorithmic trading for their symmetric properties \cite{jiang2017deep, huang2018financial, zhang2020cost}. 
The gross log return realized at time $t$ is
\begin{equation}
{r}^{gross}_t = \log{\left( y_t +1 \right)} {a}_{t-1} 
\end{equation}

At the end of the period $(t-1,t]$, due to price movements $y_t$ in the market, the weight ${a}_{t-1}$ evolve into 
\begin{equation}
{a}'_t = \frac{a_{t-1} \frac{p_t}{p_{t-1}}}{a_{t-1} y_t + 1}
\end{equation}
where $a'_t \in \mathbb{R}$. 
At the start of the next period $t$, the agent must rebalance the portfolio from its current weight ${a}'_t$ to its chosen weight ${a}_t$. 
As noted in section \ref{c:at-commarkets}, the subsequent trades resulting from this rebalancing are subject to transaction costs. The size of the required rebalancing at time $t$ is represented by $|| {a}_{t} - {a}'_{t} ||$. 
The log-return net of transaction costs at time $t$ is defined as
\begin{equation}\label{eq:net-ret}
{r}^{net}_t = {r}^{gross}_t - \lambda_c || {a}_{t-1} - {a}'_{t-1} || 
\end{equation}
where $ \lambda_\eta \in [0,1]$ is the transaction cost fraction that is assumed to be identical for buying and selling.

The log-return net of transaction costs assumes that the trader is risk-neutral, which is rarely true. 
The Sharpe ratio is the most common measure of risk-adjusted return; however, as noted in section \ref{c:at-mapping2pos}, directly optimizing the Sharpe ratio might not be optimal. 
Instead, this thesis adopts the variance over returns \cite{zhang2020cost} as a risk term 
\begin{equation}\label{eq:risk-term}
\sigma^2(r^{net}_i | i=t-L+1,...,t)=\sigma^2_{L}(r^{net}_t)
\end{equation}
where $L \in \mathbb{N}_+$ is the lookback window to calculate the variance of returns. In this thesis, the lookback window is $L=60$. 
In conclusion, subtracting the risk term defined in equation \ref{eq:risk-term} from the net returns defined in equation \ref{eq:net-ret} gives the risk-adjusted log-return net of transaction costs $r_t$, defined as
\begin{equation}\label{eq:final-ret}
r_t = {r}^{net}_t - \lambda_\sigma \sigma^2_{L}({r}^{net}_t)
\end{equation}
where $\lambda_\sigma \geq 0$ is a risk-sensitivity term that can be considered a trade-off hyperparameter for the stochastic gradient descent optimizer. 
If $\lambda_\sigma = 0$, the agent is risk-neutral. 
The reinforcement learning agents are optimized using the reward function defined in equation \ref{eq:final-ret}.

%\afterpage{\newpage}
\newpage
\section{Reinforcement learning algorithm}\label{c:m-RL-algo}

This chapter presents two model-free reinforcement learning algorithms that solve the trading MDP defined in chapter \ref{part:problemsetting}. 
There are three types of reinforcement learning algorithms: critic-based, actor-based, and actor-critic-based.
Despite the popularity of critic-based algorithms, such as Q-learning, they are unsuitable in this context due to their inability to handle high-dimensional or continuous action spaces.
Actor-based and actor-critic-based methods, known as policy gradient methods (\ref{c:RL-pgm}), are appropriate since they can handle continuous action and state spaces.
Furthermore, policy gradient methods are suitable for continuing tasks like trading. 
As both actor-based and actor-critic-based methods have advantages and disadvantages, it remains to be determined which methodology is most appropriate for this problem. 
Actor-based RL methods like REINFORCE are generally successful in stochastic continuous action spaces and have been applied to both single instrument trading and portfolio optimization \cite{jiang2017deep, zhang2020deep}. 
However, actor-based RL suffers high variance in learning and tends to be unstable and inconvenient in online learning. 
Actor-critic methods like Deep Deterministic Policy Gradient (DDPG) \cite{lillicrap2015continuous} have become popular lately and have been applied to several RL trading and portfolio optimization problems \cite{liu2018practical, yang2020deep}. 
Deterministic policies can be appropriate for financial trading, and off-policy learning combined with replay memory can be practical for online learning. 
However, training two neural networks is generally deemed to be unstable.
Thus, the selection of a reinforcement learning algorithm is non-trivial. 
This chapter presents an actor-based algorithm (\ref{algo:dp}) and an actor-critic-based algorithm (\ref{algo:ac}) for solving the trading MDP.

\subsection{Immediate reward environment}

The zero market impact assumption in chapter \ref{part:prob-assumptions} implies that the agent's participation in the market will not affect future prices $\mathbf{p}$. 
In other words, the zero market assumption implies that the agent's actions will not affect the future external state of the environment. 
However, actions performed at the start of period $t$ affect the transaction costs paid by the agent at the start of the subsequent period $t+1$. 
The reward $r_{t+1}$ depends on transaction costs incurred at time $t$, and thus the agent's previous action $\mathbf{a}_{t-1}$ will affect the following action. 
In this framework, this influence is encapsulated by adopting the recursive mechanism introduced by Moody et al. \cite{moody1998performance} of considering the past action as a part of the internal state of the environment. 
Consequently, large position changes are discouraged.

The goal of the policy gradient agent is to find the policy parameters $\theta$ that maximize the average rate of reward per time step. 
All rewards are equally important to the final return through commutativity. 
Since the agent does not affect the subsequent state of the environment, the goal is to maximize the expected immediate reward $\mathbb{E}[r_{t+1}]$,  exactly expressed in equation \ref{eq:final-ret} as the expected cumulative logarithmic return net of transaction costs and the risk-sensitivity term. 
Therefore, the action-value of the action ${a}_t$ is its immediate reward $r_{t+1}$, i.e., 
\begin{equation}
Q(s_t,a_t)=r_{t+1}
\end{equation}
$\forall s_t \in \mathcal{S}, a_t \in \mathcal{A}(s_t)$. 
As an immediate reward process, the reward function can be directly optimized by the policy gradient from rewards.

The actor-based direct policy gradient method introduced in section \ref{algo:dp} optimizes the policy by using the immediate reward directly. In contrast, the actor-critic method introduced in section \ref{algo:ac} optimizes the policy using critique from a Q-network optimized to minimize the loss to the immediate reward.

\subsection{Direct policy gradient}\label{algo:dp}

The first actor-based reinforcement learning algorithm is a direct policy gradient method inspired by the REINFORCE algorithm. 
Instead of computing learned probabilities for each action, the direct policy gradient method stochastically samples actions from a Gaussian distribution. 
Let $\pi_{\theta, \epsilon} : \mathcal{S} \rightarrow \Delta (\mathcal{A})$ be the stochastic policy parameterized by the weights $\theta \in \mathbb{R}^{d'}$. 
The policy is defined as a normal probability density over a real-valued scalar action
\begin{equation}
\pi_{\theta, \epsilon} ({a}| \mathbf{s}) = \frac{1}{\epsilon \sqrt{2\pi}} e^{\left(-\frac{\left({a} - \mu_\theta(\mathbf{s}) \right)^2}{2 \epsilon^2}\right)}
\end{equation}
where the mean is given by a parametric function approximator $\mu_\theta(\mathbf{s}) : \mathbb{R}^{|\mathbf{s}|} \rightarrow [-1,1]$ that depends on the state and outputs an independent mean for the Gaussian distribution. 
The standard deviation is given as an exploration rate $\epsilon$. 
The exploration rate $\epsilon \in \mathbb{R}$ is positive and decays at $\lambda_\epsilon \in [0,1]$ to encourage exploration of the action space in early learning epochs. The rate has a minimum $\epsilon_{min} \geq 0$ such that $\epsilon \geq \epsilon_{min}$, $\forall t$. 
After each episode, the exploration rate updates according to the following update rule
\begin{equation}
\epsilon \leftarrow \max{(\lambda_\epsilon \epsilon, \epsilon_{min})}
\end{equation}
At every step $t$, the agent samples an action $a_t \sim \pi_\theta$ from the policy and clips the action to the interval $[-1,1]$.

The novel idea of using the exploration rate $\epsilon$ as a controlled, decaying standard deviation of the stochastic policy represents progress in the research area. 
As $\epsilon$ approaches $0$, the policy becomes more or less deterministic to the mean given by the parametric function approximation $\mu_\theta$, which is advantageous in critical domains such as financial trading. 
However, being a stochastic policy, the stochastic sampling required for the REINFORCE update is still available, blending the best of both worlds for an algorithmic trading agent in an immediate reward environment.

\paragraph{Optimization}

As the model should be compatible with pre-trade training and online learning, optimization is defined in an online stochastic batch learning scheme. 
Trajectories are divided into mini-batches $[t_{s}, t_{e}]$, where $t_{s} < t_{e}$. 
The policy's performance measure on a mini-batch is defined as 
\begin{equation}
J(\pi_{\theta, \epsilon})_{[t_{s},t_{e}]} = \mathbb{E}_{\pi_{\theta, \epsilon}} \left[ \sum_{t=t_s + 1}^{t_e} r_t \right]
\end{equation}
i.e., the expected sum of immediate rewards during the mini-batch $[t_s, t_e]$ when following the policy $\pi_{\theta, \epsilon}$. 
Using the policy gradient theorem, the gradient of the performance measure $J$ with respect to the parameter weights $\theta$ is defined as
\begin{equation}
\nabla_\theta J(\pi_{\theta, \epsilon})_{[t_{s},t_{e}]} = \mathbb{E}_{\pi_{\theta, \epsilon}} \left[ \sum_{t=t_s + 1}^{t_e} r_t \nabla_\theta \log \pi_{\theta, \epsilon}(a_t | s_t) \right]
\end{equation}
This expectation is empirically estimated from rollouts under $\pi_{\theta, \epsilon}$. 
The parameter weights are updated using a stochastic gradient ascent pass 
\begin{equation}
\theta \leftarrow \theta + \alpha \nabla_\theta J(\pi_{\theta, \epsilon})_{[t_s, t_e]}
\end{equation}

\paragraph{Pseudocode}
The pseudocode for the actor-based algorithm is given in Algorithm \ref{alg:dir}. 

\begin{algorithm}
\caption{Actor-Based Algorithm for Trading}\label{alg:dir}
Input: a differentiable stochastic policy parameterization $\pi_{\theta, \epsilon}(a | s)$ 

Algorithm parameters: learning rate $\alpha^\theta>0$, mini-batch size $b > 0$, initial exploration rate $\epsilon \geq 0$, exploration decay rate $\lambda_\epsilon \in [0,1]$, exploration minimum $\epsilon_{min} \geq 0$

Initialize: empty list $\mathcal{B}$ of size $b$

\begin{algorithmic}
\Repeat
 \State Receive initial state of the environment $s_0 \in\mathcal{S}$
 \Repeat
 \For{t = 0,1,...,T-1}
  \State Sample action $a_t \sim \pi_{\theta, \epsilon} (\cdot |s_t)$
  \State Execute action $a_t$ in the environment and observe $r_t$ and $s_{t+1}$
  \State Store pair of reward $r_t$ and log-probabilities $\nabla_\theta \ln \pi_{\theta, \epsilon} (a_t |s_t)$ in $\mathcal{B}$
  \If{$|\mathcal{B}| == b$ or $s_t$ is terminal}
    \State Update the policy $\pi_{\theta, \epsilon}$ by one step of gradient ascent using: $$ \nabla_\theta J(\pi_{\theta, \epsilon}) \approx \sum_{\mathcal{B}} r_t \nabla_\theta \ln{\pi_{\theta, \epsilon} (a_t | \mathbf{s}_t)} $$
   \State Reset $\mathcal{B}$ to empty list
  \EndIf
  \EndFor
\Until terminal state 
\State Update the exploration rate $\epsilon = \max{(\epsilon \lambda_\epsilon, \epsilon_{min})}$
\Until convergence
\end{algorithmic}
\end{algorithm}

\subsection{Deterministic actor-critic}\label{algo:ac}

There are several different actor-critic algorithms available, like asynchronous advantage actor-critic (A3C) \cite{mnih2016asynchronous} and proximal policy optimization (PPO) \cite{schulman2017proximal}. 
However, due to the critical nature of the problem, actor-critic algorithms that utilize deterministic policies are of interest, as stochasticity may negatively affect the model's performance.
Thus, the second RL algorithm is an off-policy deterministic actor-critic algorithm inspired by the deep deterministic policy gradient algorithm \cite{lillicrap2015continuous}.

Let $\mu_\theta : \mathcal{A} \rightarrow \mathcal{S}$ be the deterministic policy parameterized by $\theta \in \mathbb{R}^{d'}$. 
This is the same function approximator used to generate the mean for the Gaussian action selection in the direct policy gradient algorithm \ref{algo:dp}. 
However, the function approximator is now used to map states to actions directly and not as the mean in a probability distribution over the actions. 
Instead of optimizing the policy using the log probabilities of the sampled action scaled by the reward, the deterministic policy is optimized using a learned action-value critic relying on the deterministic policy gradient theorem. 
Let $Q_{\phi}(s,a) : \mathcal{S} \times \mathcal{A} \rightarrow \mathbb{R}$ be the Q-network critic parameterized by $\phi \in \mathbb{R}^{b'}$. 
Although exploring the state space is unnecessary in this problem, as the external environment is not affected by the agent's actions, it can be advantageous to explore the action space in the early stages of learning to provide the agent with training examples from the entire action space. 
To ensure the sufficient exploration of the action space, the algorithm is trained off-policy with an exploration policy $\mu'_\theta$ defined as 
\begin{equation}
\mu'_\theta (s) = \mu_\theta (s) + \epsilon \mathcal{W}
\end{equation}
where $\mathcal{W} \sim \mathcal{U}_{[-1,1)}$ is sampled noise from an uniform distribution. 
The exploration parameters $\epsilon, \epsilon_{min}, \lambda_\epsilon$ are defined similarly for the direct policy gradient algorithm \ref{algo:dp}. 
Clipping agents' actions to the interval $[-1,1]$ prevents them from taking larger positions than their available capital.

\paragraph{Optimization}

Both the actor and critic networks are updated using randomly sampled mini-batches $\mathcal{B}$ from a replay memory $\mathcal{D}$. 
The replay memory provides random batches in sequential order for stateful RNNs, and random batches not in sequential order that minimize correlation between samples for non-stateful DNNs. 
The exploration policy $\mu'_\theta$ explores the environment and generates transitions $\tau$ stored in the replay memory $\mathcal{D}$. 

The objective function $J$ for the deterministic policy $\mu_\theta$ is defined as 
\begin{equation}
J(\mu_\theta) = \mathbb{E}_{s \sim \mathcal{B}} [Q_\phi (s, \mu_\theta (s))]
\end{equation}
and its gradient is given as
\begin{equation}
\nabla_\theta J(\mu_\theta) = \mathbb{E}_{s \sim \mathcal{B}} [\nabla_\theta \mu_\theta(s) \nabla_a Q_\phi (s,a) |_{a=\mu_\theta (s)}]
\end{equation}

Since the environment is an immediate reward environment, the target for the Q-network updates is the immediate reward, i.e., $y=r$. 
MSE loss is used as a loss function as the outliers are of critical importance to the success of the trading agent. 
The loss function $L(\phi)$ for the Q-network $Q_\phi$ is defined as 
\begin{equation}
L(Q_\phi) = \mathbb{E}_{s,a,r \sim \mathcal{B}} [(Q_\phi (s,a) - r)^2]
\end{equation}
and its gradient is given as
\begin{equation}
\nabla_\phi L(Q_\phi) = \mathbb{E}_{s,a,r \sim \mathcal{B}} [(Q_\phi (s,a) - r) \nabla_\phi Q_\phi (s,a)]
\end{equation}

\paragraph{Pseudocode}

The pseudocode for the deterministic actor-critic algorithm is given in Algorithm \ref{alg:det}. 

\begin{algorithm}
\caption{Actor-Critic Algorithm for Trading}\label{alg:det}
Input: a differentiable deterministic policy parameterization $\mu_\theta(s)$ 

Input: a differentiable state-action value function parameterization $Q_\phi(s,a)$ 

Algorithm parameters: learning rates $\alpha^\theta>0$, $\alpha^\phi>0$, mini-batch size $b > 0$, replay memory size $d\geq b$, initial exploration rate $\epsilon \geq 0$, exploration decay rate $\lambda_\epsilon \in [0,1]$, exploration minimum $\epsilon_{min} \geq 0$

Initialize empty replay memory cache $\mathcal{D}$

\begin{algorithmic}
\Repeat
\State Receive initial state of the environment $s_0 \in\mathcal{S}$
\For{t = 1,...,T}
 \State Select action $a_t = \mu_\theta (s_t) + \epsilon \mathcal{W}$ from the exploration policy
 \State Execute $a_t$ in the environment and observe $r_t$ and $s_{t+1}$
 \State Store transition $\tau_t = (s_t, a_t, r_t)$ in the replay memory $\mathcal{D}$
 \State Sample a random mini-batch $\mathcal{B}$ of $|\mathcal{B}|$ transitions $\tau$ from $\mathcal{D}$
 \State Update the Q-network by one step of gradient descent using $$ \nabla_\phi \frac{1}{|\mathcal{B}|} \sum_{(s,a,r) \in \mathcal{B}} (Q_\phi (s,a) - r)^2 $$
 \State Update the policy by one step of gradient ascent using $$ \nabla_\theta \frac{1}{|\mathcal{B}|} \sum_{s \in \mathcal{B}} Q_\phi (s,\mu_\theta(s)) $$
 \EndFor
\State Update the exploration rate $\epsilon = \max{(\epsilon \cdot \lambda_\epsilon, \epsilon_{min})}$
\Until convergence
\end{algorithmic}
\end{algorithm}

%\afterpage{\newpage}
\newpage
\section{Network topology}\label{c:network-topology}

The reinforcement learning algorithms introduced in chapter \ref{c:m-RL-algo} utilize function approximation to generalize over a continuous state and action space. 
Section \ref{c:at-forecasting} introduced function approximators for extracting predictive patterns from financial data, where empirical research suggested the superiority of deep learning methods.
Thus, the function approximators introduced in this chapter rely on deep learning techniques introduced in \ref{c:DL}. 
In the research presented in section \ref{c:at-forecasting}, the function approximators based on convolutional neural networks (\ref{c:dl-cnn}) and those based on the LSTM (\ref{c:dl-lstm}) consistently performed well. 
Thus, this section introduces two function approximators based on CNNs and LSTMs, respectively. The sequential information layer, presented in section \ref{c:nt-sil}, leverages these techniques to extract predictive patterns from the data. 
Furthermore, the decision-making layer that maps forecasts to market positions, presented in section \ref{c:nt-dml}, employs the recursive mechanism introduced by Moody et al. \cite{moody1998performance}, enabling the agent to consider transaction costs.

The direct policy gradient algorithm presented in chapter \ref{algo:dp} is an actor-based RL algorithm that only uses a parameterized policy network. The deterministic actor-critic algorithm presented in chapter \ref{algo:ac} uses a parameterized policy network and a parameterized critic network. 
This chapter outlines these function approximators, which fortunately consist of many of the same components. 
Section \ref{c:nt-pn} describes the policy network, while section \ref{c:nt-qn} describes the Q-network. The last section \ref{c:nt-opt} describes the optimization and regularization of the networks.

\subsection{Network input}\label{c:nt-network-input}

The first step is to specify the input into the networks. 
Section \ref{c:ps-ss} defined the agent state $\mathbf{s}_t^a$ of the partially observable environment. 
This section describes the modified version of the agent state $\mathbf{s}^{a'}_t$, which both the policy and critic agents receive as input. 
The modified agent state applies two forms of processing to the network input; the first is extracting the relevant observations from the agent state, which ensures that the network input is of fixed size, and the second is normalizing the network input, which is advantageous for the non-linear function approximators introduced in this chapter. 
This thesis adopts the philosophy of technical traders of the price reflecting all necessary information, and therefore the past price is used to represent the agent state. 
The primary reason for selecting the price series alone as the state representation is to examine the ability of a general deep reinforcement learning model to extract patterns from raw, relatively unprocessed data. Although additional data sources could aid the agent in discovering predictive patterns, that is beyond the scope of this thesis.

Adopting the ideas of Jiang et al. \cite{jiang2017deep}, the agent state is down-sampled by extracting the three most relevant prices from a period; the closing price, the highest price, and the lowest price. 
Thus, the price tensor used to represent the external agent state at time $t$ is defined as 
\begin{equation}
\hat{\mathbf{p}}_t = \left[ {p}_t, {p}_t^{high}, {p}_t^{low} \right]
\end{equation}

Normalizing input data for neural networks speeds up learning \cite{goodfellow2016deep} and is beneficial for reinforcement learning as well \cite{andrychowicz2020matters}. 
However, normalizing the whole time series ex-ante is a form of lookahead. 
The normalization scheme can only use data up to time $\leq t$ for the observation $\mathbf{p}_t$ $\forall t$. 
The choice of instrument weights depends on relative log returns rather than absolute price changes. The price tensor $\hat{\mathbf{p}}_t$ is normalized using the log-returns from the previous closing price $p_{t-1}$. 
Additionally, adopting the ideas from Zhang et al. \cite{zhang2020deep}, the input is further normalized by dividing by a volatility term defined as
\begin{equation}
\sigma^2 \left( \log{\left( \frac{p_i}{p_{i-1}} \right)} | i=t-L+1, ..., t \right) \sqrt{L} = \sigma^2_{L, t} \sqrt{L}
\end{equation}
where $L \in \mathbb{N}_+$ is the lookback window to calculate the volatility of the closing price, which is set to $L=60$ similarly as \cite{zhang2020deep}. 
The normalized price tensor at time $t$ is thus defined as 
\begin{equation}
\bar{\mathbf{p}}_t = \log{ \left( \hat{\mathbf{p}}_t \oslash p_{t-1} \right) } \oslash \sigma^2_{L, t} \sqrt{L}
\end{equation}

As a precaution against outliers in volatile markets, which can be detrimental to the performance of DNNs, the normalized price tensor $\bar{\mathbf{p}}_t$ is clipped to the interval $[-1,1]$. 

Stacking the past $n$ observations produces the approximated environment state. 
Thus, the final price tensor is adjusted to contain the $n$ most recent observations $\bar{\mathbf{p}}_{t-n+1:t} \in \mathbb{R}^{3 \times n}$. 
The stacked price tensor is considered the external agent state and defined as $\mathbf{x}^S_t = \bar{\mathbf{p}}_{t-n+1:t}$
The networks also adopt the recursive mechanism introduced by Moody et al. \cite{moody1998performance} of considering the past action as a part of the internal environment, allowing the agent to take the effects of transaction costs into account. 
The instrument weight from the previous period $a_{t-1}$ is inserted into the final decision-making layer after extracting the sequential features in the sequential information layer. 
The modified agent state thus approximates the state of the environment
\begin{equation}
\mathbf{s}^{a'}_t = (\mathbf{x}^S_t, a_{t-1})
\end{equation}

The policy networks and Q-network receive this modified agent state $\mathbf{s}^{a'}_t$ as input. As an action-value function, the Q-network also takes the current action $a_t$ as input.

\subsection{Policy network}\label{c:nt-pn}

The deterministic policy \ref{algo:ac} and the mean-generating parametric function approximator in the stochastic policy \ref{algo:dp} are the same function approximator $\mu_\theta : \mathbb{R}^{|\mathcal{S}|} \rightarrow [-1,1]$ parameterized by $\theta \in \mathbb{R}^{d'}$, and will in this chapter be referred to as the policy network. 
The policy network consists of a sequential information layer, a decision-making layer, and a $\tanh$ function. 
The input to the policy network is the modified agent state $\mathbf{s}^{a'}_t$. 
The external part of the agent state $\mathbf{x}^S_t$, i.e., the price tensor of stacked observations, is input into the sequential information layer. 
The sequential information layer output is concatenated with the previous action $a_{t-1}$ to produce input into the decision-making layer. The output from the decision-making layer maps to a $\tanh$ function that produces the action constrained to the action space $[-1,1]$.

\begin{figure}[ht]
\centering
\includegraphics[width=\textwidth]{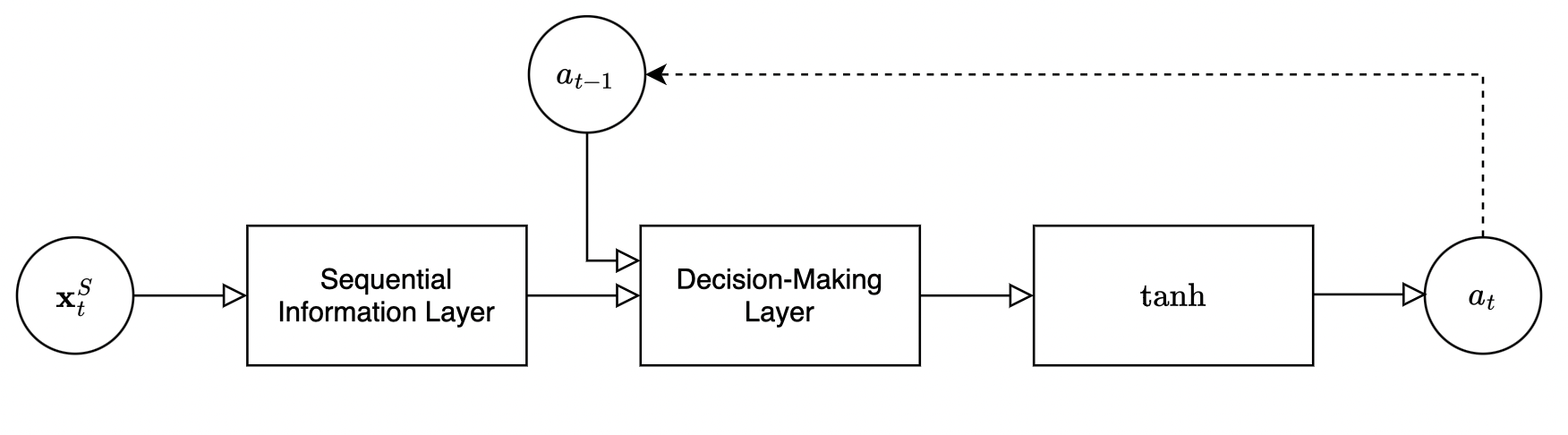}
\caption{Policy network architecture}
\end{figure}

\subsection{Q-network}\label{c:nt-qn}

The Q-network $Q_\phi : \mathbb{R}^{|\mathcal{S}|} \times \mathbb{R}^{|\mathcal{A}|} \rightarrow \mathbb{R}$ is a function approximator parameterized by $\phi \in \mathbb{R}^{b'}$. 
It is an action-value function that assigns the value of performing a specific action in a specific state and thus takes two arguments, the modified agent state $\mathbf{s}^{a'}_t$ and the action $a_t$. 
Other than that, there are two differences between the critic and policy networks. 
Firstly, the Q-network has an additional layer before the sequential information net that concatenates the agent state $\mathbf{s}^a_t$ and the current action $a_t$ and maps it through a fully-connected layer into a leaky-ReLU activation function with negative slope $0.01$ and dropout with probability $0.2$. 
The second difference is that the output after the decision-making layer does not map to a $\tanh$ function since the Q-network outputs action-values, which are not constrained to any specific interval.

\begin{figure}[ht]
\centering
\includegraphics[width=\textwidth]{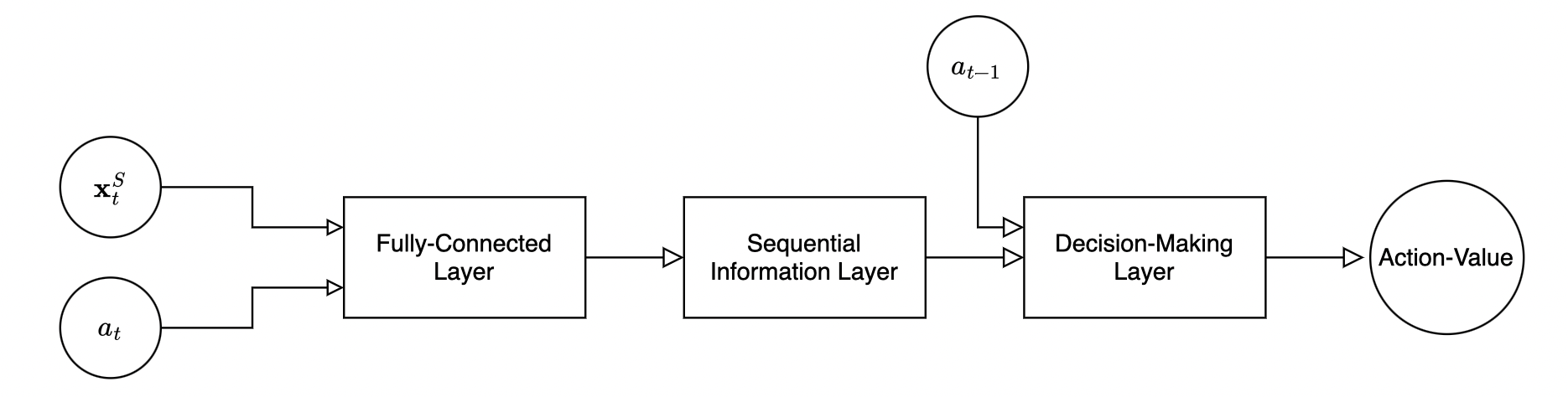}
\caption{Q-network architecture}
\end{figure}

\subsection{Sequential information layer}\label{c:nt-sil}

In essence, an algorithmic trading agent places bets on the relative price change, or returns, of financial instruments. The agent's success ultimately depends on its ability to predict the future. However, doing so in highly competitive and efficient markets is non-trivial. To remain competitive in continuously evolving markets, the agent must \textit{learn} to recognize patterns and generate rules based on past experiences. 
The sequential information layer extracts the sequential features from the input data and is arguably the most integral part of the model. 
Let $\mathbf{x}^I_t$ be the input into the sequential information net (for the policy network $\mathbf{x}^I_t=\mathbf{x}^S_t$). 
The sequential information layer is a parametric function approximator that takes the input $\mathbf{x}^I_t$ and outputs a feature vector $\mathbf{g}_t$, defined as
\begin{equation}
f^{S}(\mathbf{x}^I_t)=\mathbf{g}_t
\end{equation}

The choice of the appropriate function approximator for this task is non-trivial. The inductive bias of the model must align with that of the problem for the model to generalize effectively. Therefore, selecting a model that captures the problem's underlying structure while also being efficient and scalable is imperative. 
Research on financial time series forecasting found that deep learning models, specifically those based on the CNN and LSTM architecture, consistently outperformed traditional time series forecasting methods such as the ARIMA and GARCH \cite{xiong2015deep, mcnally2018predicting, siami2018forecasting, sezer2020financial}. 
The universal approximation theorem (\ref{universal-approx-the}) establishes that there are no theoretical constraints on feedforward networks'\footnote{Of arbitrary width or height.} expressivity. 
However, feedforward networks are not as naturally well-suited to processing sequential data as CNNs and LSTMs. Therefore, they may not achieve the same level of performance, even though it is theoretically possible. Additionally, feedforward networks may require significantly more computing power and memory to achieve the same performance as CNNs or LSTMs on sequential data. 
Transformers were also considered due to their effectiveness in forecasting time series \cite{makridakis2022m5}. Transformers employ an encoder-decoder architecture and rely on attention mechanisms to capture long-term dependencies. 
Thus, they do not require a hidden state, like RNNs, and are relatively easy to parallelize, enabling efficient training on large datasets. 
A variant called decision transformers \cite{chen2021decision} has been applied to offline reinforcement learning. However, it is unclear how to apply the transformer in its conventional encoder-decoder topology to online reinforcement learning. Therefore, the transformer is, for the moment, unsuitable for this problem. 
The \textit{gated recurrent unit} (GRU) is a newer version of the recurrent neural network that is less computationally complex than the LSTM. However, LSTMs are generally considered superior for forecasting financial data \cite{sezer2020financial}.

This section defines two distinct DNN topologies for the sequential information layer; the first is based on convolutional neural networks, while the second is based on recurrent neural networks, specifically the LSTM. The two sequential information topologies both consist of two hidden layers, which is enough for the vast majority of problems. Performance is usually not improved by adding additional layers.

\subsubsection{Convolutional neural network}

The CNN-based sequential information layer topology includes two 1-dimensional convolutional layers. 
In the absence of established heuristics, determining the parameters for a CNN can be challenging. Thus, the parameters chosen for these layers are partly informed by research on CNNs in financial time series forecasting \cite{sezer2020financial} and partly determined through experimentation. 
The first convolutional layer has kernel size $3$ and stride $1$ and processes each of the $3$ columns in the input $\mathbf{x}^I_t$ as separate channels of size $1\times n$, where $n$ is the number of stacked observations. 
It outputs $32$ feature maps of size $1\times n-2$. 
The second convolutional layer has kernel size $3$ and stride $1$ and outputs $32$ feature maps of size $1 \times n-4$. 
Batch norm is used after both convolutional layers on the feature maps to stabilize and speed up learning. 
The CNN uses the Leaky-ReLU activation function with a negative slope of $0.01$ after the batch norm layers to generate the activation maps. Dropout with probability $p=0.2$ is used between the layers. 
Max pooling with kernel size $2$ and stride $2$ is applied after the final convolutional layer to down-sample the output before all activation maps are concatenated into one big activation map.

\begin{figure}[ht]
\centering
\includegraphics[width=\textwidth]{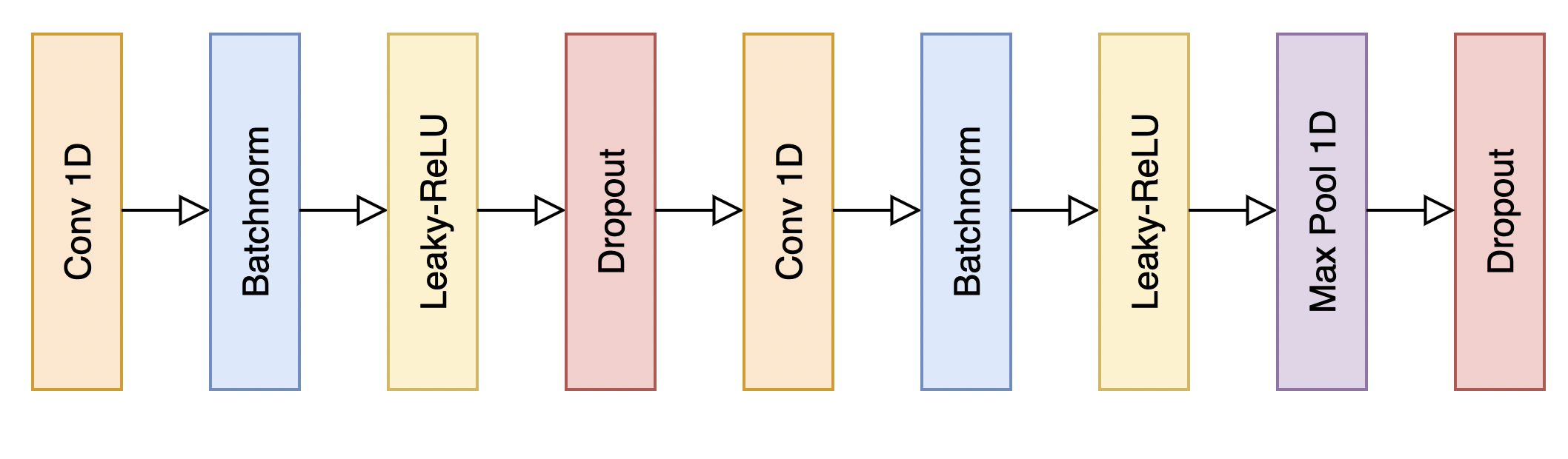}
\caption{Convolutional sequential information layer architecture}
\end{figure}

\subsubsection{Long Short-Term Memory}

The second sequential information network topology introduces memory through a recurrent neural network to solve the partially observable MDP. 
Long Short-term Memory (LSTM) is the go-to solution for environments where memory is required and are good at modeling noisy nonstationary data. 
The LSTM sequential information net architecture consists of two stacked LSTM layers. 
Both LSTM layers have $128$ units in the hidden state, which was chosen experimentally. Following both LSTM layers, the network employs dropout with dropout-probability $p=0.2$. 
The LSTM cell contains three sigmoid functions and one hyperbolic tangent function, so inserting an activation function after the LSTM layer is superfluous. 
Batchnorm is incompatible with RNNs, as the recurrent part of the network is not considered when computing the normalization statistic and is, therefore, not used.

\begin{figure}[ht]
\centering
\includegraphics[width=0.45\textwidth]{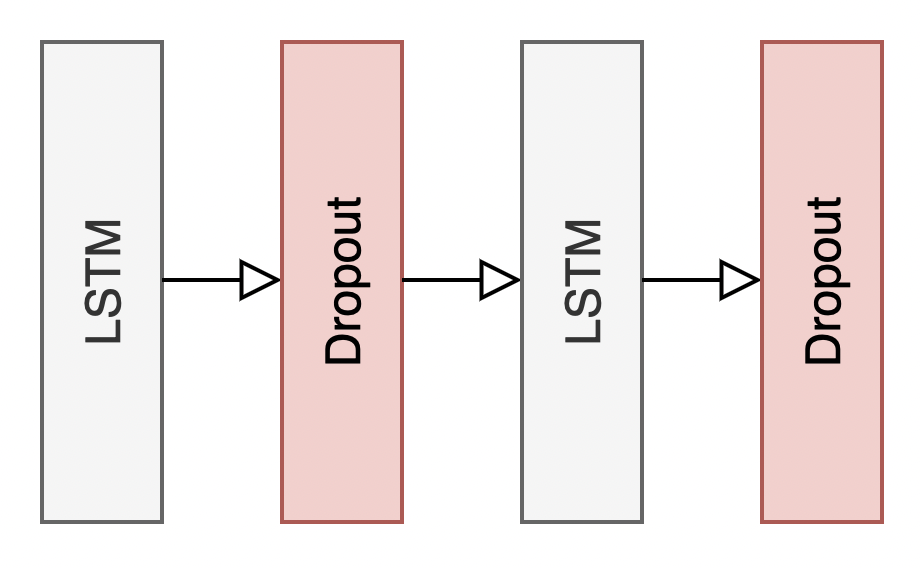}
\caption{LSTM sequential information layer architecture}
\end{figure}

\subsection{Decision-making layer}\label{c:nt-dml}

In the decision-making layer, the previous action $a_{t-1}$ is concatenated into the features $\mathbf{g}_t$, i.e., the output from the sequential information layer, and mapped to a single output value. 
The previous action weight $a_{t-1}$ allows the agent to consider transaction costs when making trading decisions (policy network) or when assigning value to actions in states (Q-network). 
The mapping comprises a fully-connected layer between the features and the previous action to a single output value. 
This output value is the action (or Gaussian mean) for the policy (after mapping it to a $\tanh$ function) or the action-value for the Q-network. 
The input to the decision-making layer is defined as
\begin{equation}
\mathbf{x}_t^{D} = (\mathbf{g}_t, a_{t-1})
\end{equation}
The decision-making layer is a dot product between a weight vector $\mathbf{w}^D \in \mathbb{R}^{|\mathbf{x}_t^{D}|}$ and the input $\mathbf{x}_t^{D}$, defined as
\begin{equation}
f_{D}(\mathbf{x}_t^{D})= (\mathbf{w}^{D})^{\top} \mathbf{x}^{D}_t 
\end{equation}

\subsection{Network optimization}\label{c:nt-opt}

The weights $\theta$ and $\phi$ of the policy network $\mu_\theta$ and Q-network $Q_\phi$ are initialized using Kaiming initialization which ensures that the initial weights of the network are not too large and have a small variance, which helps to prevent the network from getting stuck in local minima and allows the network to generalize better. 
Additionally, Kaiming initialization considers the type of activation function used in the network, contrary to conventional initialization schemes. 
The weight initialization scheme centers the initial output distribution of the networks around $0$ with a small standard deviation regardless of the input. 
The weights are updated using the Adam stochastic gradient descent algorithm on mini-batches, allowing the network to update the weights more efficiently and accurately than other SGD algorithms.
The gradient norm for each mini-batch is clipped to $1$ to prevent exploding gradients. 
There are many potential activation functions for neural networks, including the default recommendation, the ReLU. To combat the ``dying ReLU problem'', the leaky-ReLU activation function is used in the networks. The negative slope, or the ``leak'', is set to the standard value of $0.01$.

\subsubsection{Regularization}

Machine learning research generally focuses on problems with complex structures and high signal-to-noise ratios, such as image classification.
For these problems, complicated non-linear models like neural nets have demonstrated their effectiveness.
However, in a high-noise environment such as financial forecasting, where the R-squared is often of order $10^{-4}$ \cite{isichenko2021quantitative}, anything beyond linear regression poses a significant overfitting risk. 
An overfitted network will likely perform well on the training set but generalize poorly out-of-sample. 
An algorithmic trading agent that performs well on the training set is of little use, and it is imperative to reduce the generalization error. 
Therefore, regularization is needed to mitigate the risk of overfitting and reduce the generalization error. 
A description of the regularization techniques used for these networks is provided in section \ref{c:nn-regularization}.

For ML models to be generalizable, they must learn data patterns rather than individual data points to identify a bigger picture agnostic of noisy details. Regularization techniques such as weight decay limit the capacity of the networks by adding a parameter norm penalty to the objective function. Weight decay uses the $L^2$ norm; other norms, such as the $L^1$ norm, can also be used. The $L^2$ norm is appropriate since it punishes outliers harsher and is easier to optimize with gradient-based methods.
The parameter $\lambda_{wd}$ controls the degree of penalization, balancing the tradeoff between increased bias and decreased variance. 
The network optimizer introduced in this section uses weight decay with the constant parameter $\lambda_{wd} = 0.001$ to mitigate the network's overfitting risk. 
Experimentally, this value delivered the optimal balance for the bias-variance tradeoff. 
Although increasing the weight decay penalty could further reduce overfitting risk, this was too restrictive for the networks.

It is important to note that weight decay reduces, but does not eliminate, the risk of overfitting. 
Dropout is another explicit regularization technique almost universally used in deep neural networks. 
Dropout forces the network to learn multiple independent data representations, resulting in a more robust model. 
When training networks on noisy financial data, dropout effectively ensures the network ignores the noise.
Similarly to weight decay, the dropout rate is a tradeoff.
There is no established heuristic for choosing the dropout rate; instead, it is usually chosen through experimentation. 
In this case, a dropout rate of $0.2$ provided a suitable regularizing effect where the model generalized well and produced accurate predictions. Dropout is used between all hidden layers in the networks.

Although explicit regularizers such as weight decay and dropout reduce overfitting risk, it remains tricky to determine the optimal training duration. This challenge is addressed with early stopping, which functions as an implicit regularizer. The networks are trained in an early stopping scheme, with testing on the validation set every $10$th epoch. 
As reinforcement learning involves random exploration, the models are tested slightly less frequently than conventional to prevent premature stopping.

\afterpage{\blankpage}
\newpage
\part{Experiments}\label{part:experiments}

%\afterpage{\newpage}
\newpage
\section{Experiment and Results}\label{sec:expandres}

%\subsection{Introduction/Aim}\label{exp:intro}
%What did you do and why? 
%What did you do? 
%Why did you do it? 

Experiments play a vital role in science and provide the basis for scientific knowledge. This chapter presents the experiments and results where the methods presented in part \ref{part:methodology} are tested on historical market data using the backtesting framework described in section \ref{c:at-backtest}. 
The backtest requires simplifying market assumptions, specified in chapter \ref{part:problemsetting}. 
Section \ref{exp:methods} details the experiment setting. 
The results of the experiment are presented and discussed in sections \ref{exp:results} and \ref{exp:discussion-of-res}. 
Finally, the overall approach is discussed in section \ref{exp:discussion-of-mod}. 
The experiment aims to answer the research questions posed at the start of this thesis. 
\begin{enumerate}
    \item Can the risk of algorithmic trading agents operating in volatile markets be controlled?
    \item What reinforcement learning algorithms are suitable for optimizing an algorithmic training agent in an online, continuous time setting?
    \item What deep learning architectures are suitable for modeling noisy, non-stationary financial data? 
\end{enumerate}

\subsection{Materials and Methods}\label{exp:methods}

Chapter \ref{c:m-RL-algo} described two reinforcement learning algorithms to solve the commodity trading problem; the direct policy gradient (PG) and the deterministic actor-critic (AC). 
Chapter \ref{c:network-topology} described two sequential information layers, one based on CNN architecture and the other LSTM-based.
Both the actor and critic in the actor-critic algorithm are modeled using the same architecture. 
In total, that leaves four combinations which are specified below with their respective abbreviations
\begin{itemize}
    \item \textbf{PG-CNN:} Direct policy gradient algorithm where the policy network is modeled using the CNN-based sequential information layer. 
    \item \textbf{PG-LSTM:} Direct policy gradient algorithm where the policy network is modeled using the LSTM-based sequential information layer. 
    \item \textbf{AC-CNN:} Deterministic actor-critic algorithm where the policy and Q-network are modeled using the CNN-based sequential information layer. 
    \item \textbf{AC-LSTM:} Deterministic actor-critic algorithm where the policy and Q-network are modeled using the LSTM-based sequential information layer. 
\end{itemize}

\subsubsection{Baselines}\label{baselines}

Defining a baseline can be helpful when evaluating the performance of the methods presented in part \ref{part:methodology}. 
A challenge with testing algorithmic trading agents is the lack of established baselines. 
However, the by far most common alternative is the \textit{buy-and-hold} baseline \cite{moody1998performance}\cite{zhang2020deep}\cite{zhang2020cost}. 
The buy-and-hold baseline consists of buying and holding an instrument throughout the experiment, i.e., $a_t=1, \forall t$. 
Compared to a naive buy-and-hold baseline, an intelligent agent actively trading a market should be able to extract excess value and reduce risk.

\subsubsection{Hyperparameters}

Table \ref{hyperparameters} shows the hyperparameters used in this experiment.
The learning rates for the policy network and Q-network are denoted $\alpha_{actor}$ and $\alpha_{critic}$, respectively, and were tuned experimentally. 
$|\mathcal{B}|$ is the batch size, and $|\mathcal{D}|$ is the replay memory size. 
Large batch sizes are necessary to obtain reliable gradient estimates. However, large batch sizes also result in less frequent updates to the agent and updates that may contain outdated information. As a result of this tradeoff, the batch and replay memory sizes used in this experiment were selected as appropriate values.
The transaction cost fraction $\lambda_c$ is set to a reasonable value that reflects actual market conditions. 
The initial exploration rate is denoted $\epsilon$, with decay rate $\lambda_\epsilon$ and minimum $\epsilon_{min}$. 
The number of stacked past observations is given by $n$, considered a reasonable value for the agent to use for short-term market prediction.

\begin{center}
 \begin{tabular}{ |p{1cm}||p{0.8cm}|p{0.75cm}|p{0.6cm}||p{0.6cm}||p{0.85cm}||p{0.35cm}|p{0.5cm}|p{0.5cm}|p{0.4cm}|  }
 \hline
 Model & $\mathbf{\alpha_{actor}}$ & $\mathbf{\alpha_{critic}}$ & $\mathbf{|\mathcal{B}|}$ &  $\mathbf{|\mathcal{D}|}$ & $\mathbf{\lambda_c}$ & $\mathbf{\epsilon}$ & $\mathbf{\lambda_\epsilon}$ & $\mathbf{\epsilon_{min}}$ & $n$ \\
 \hline
 \textbf{PG}   & $0.0001$    & - &   128 & - & $0.0002$ & $1$ & $0.9$ & $0.01$ & $20$ \\
 \textbf{AC} & $0.0001$  & $0.001$   &128 & $1 000$ & $0.0002$ & $1$ & $0.9$ & $0.01$ & $20$\\
 \hline
\end{tabular}
\captionof{table}{Hyperparameters}\label{hyperparameters}
\end{center}

\subsubsection{Training scheme}\label{exp:trainingscheme}

The dataset is split into three parts; a training set, a validation set, and a test set, in fractions of $1/4$, $1/4$, and $1/2$, respectively. 
The RL agents train on the training set, the first $1/4$ of the dataset, and then validate on the validation set, the next $1/4$. 
Early stopping is used, with testing every $10$th epoch. 
The early stopping frequency is low because the RL agents exhibit randomness and stochasticity, especially in the early epochs. 
Setting a high early stopping frequency can cause premature convergence. 
The weight initialization scheme, described in section \ref{c:nt-opt}, causes the initial action distribution of the policy to be centered around $0$ with a small standard deviation. 
However, the agents learn faster when exploring the edge values of the state space in the early stages. 
Exploration of the action space is controlled, for both RL algorithms, by the three exploration parameters $\epsilon, \lambda_\epsilon, \epsilon_{min}$. 
During training, the exploration rate starts at $\epsilon=1$ and decays at $\lambda_\epsilon=0.9$ per episode to a minimum exploration rate of $\epsilon_{min}=0.01$.

When the agent has finished training, it tests once out-of-sample on the test set, the last $1/2$ of the dataset. 
Leaving half of the dataset for final testing ensures the test set is sufficiently large to evaluate the trading agents.
Exploring the action space is no longer necessary after initial training. Therefore, $\epsilon=0$ for the out-of-sample test. 
According to the optimization strategies specified in their respective pseudocodes \ref{alg:dir} and \ref{alg:det}, the RL agents continuously refit themselves as they observe transitions. 
The results section (\ref{exp:results}) presents results from these backtests.

\subsubsection{Performance metrics}\label{performance-metrics}

The objective of the algorithmic trading agent is described by modern portfolio theory (\ref{c:at-mpt}) of maximizing risk-adjusted returns, usually represented by the Sharpe ratio. 
Thus, the Sharpe ratio defined in equation \ref{eq:sharpe} will be the primary performance metric for the backtest. 
The reward function defined in equation \ref{eq:final-ret} is not a comparable performance measure to related work. 
Instead, the standard method for assessing performance by linear returns net of transaction costs is adopted. In a backtest, the agent interacts with the environment and generates a sequence of actions $\{ a_0, a_1, ..., a_{T-1}, a_T \}$. 
The linear net return after $T\in \mathbb{N}_+$ trades is defined as
\begin{equation}\label{eq:tradereturnsperformance}
R_{T} = \prod_{t=1}^T \left( y_t \cdot a_{t-1} \right)+1 - \lambda_c || a'_{t-1} - a_{t-1} || 
\end{equation}
where $y_t, a_t, a'_t, \lambda_c$ are defined in chapter \ref{part:problemsetting}. 
The return $R_{T}$ is used to calculate the Sharpe ratio. 
As there is randomness in the models, either through stochastic action selection or random mini-batch sampling, along with random weight initialization, performance is averaged over $10$ runs. 
In addition to the Sharpe ratio, additional performance metrics can help paint a clearer picture of the performance of an algorithmic trading agent. 
This thesis adopts some of the performance metrics most frequently found in related work \cite{jiang2017deep, zhang2020cost, zhang2020deep}. 
The performance metrics used in this thesis are defined as
\begin{enumerate}
    \item $\mathbb{E}[R]$: the annualized expected rate of linear trade returns. 
    \item $Std(R)$: the annualized standard deviation of linear trade returns. 
    \item Sharpe: a measure of risk-adjusted returns defined in equation \ref{eq:sharpe}. The risk-free rate is assumed to be zero, and the annualized Sharpe ratio is thus $\mathbb{E}[R]/Std(R)$. 
    \item MDD: \acrfull{mdd}, the maximum observed loss from any peak. 
    \item Hit: the rate of positive trade returns. 
\end{enumerate}

\subsubsection{Dataset}

The dataset consists of the front-month \textit{TTF Natural Gas Futures} contracts from 2011 to 2022. 
The observations are sampled according to the transacted euro volume on the exchange, defined in section \ref{c:ps-td}. 
Larger sample sizes are desirable to ensure statistical significance, especially for highly overparameterized approximators such as neural networks. In addition, predictability is generally higher over shorter horizons \cite{isichenko2021quantitative}. 
However, as sampling frequency (and therefore trading frequency) increases, simplifying assumptions, such as no impact and perfect liquidity, become increasingly consequential. 
Thus, an appropriate target number of samples per day is $tgt=5$, which provides a little over $20 \; 000$ total samples. 
The data processing is limited to what is described in section \ref{c:nt-network-input}.

\begin{figure}[ht]
\centering
\includegraphics[width=\textwidth]{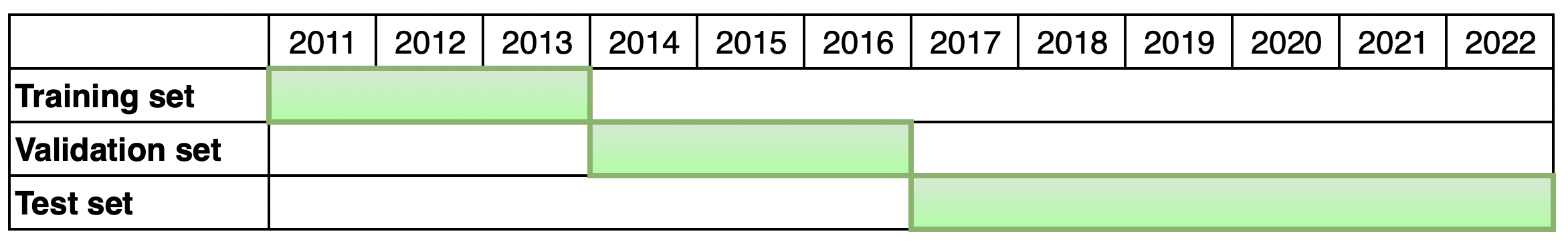}
\caption{The training-validation-test split}
\label{tvs-split}
\end{figure}

The first quarter of the dataset consisting of trades from 01/01/2011 to 01/01/2014 makes up the training set. 
The validation set is the second quarter of the dataset from 01/01/2014 to 01/01/2017. 
Finally, the test set is the second half of the dataset from 01/01/2017 to 01/01/2023.  
Figure \ref{tvs-split} illustrates the training-validation-test split.

\subsection{Results}\label{exp:results}

This section presents the results of the experiments described in the previous section. The models are tested using four different values of the risk-sensitivity term $\lambda_\sigma$ ($0$, $0.01$, $0.1$, and $0.2$), and the results of all four values are presented. 
The results are visualized using three tools; a table and two types of plots, and they are briefly described below
\begin{itemize}
    \item The table consists of the performance metrics (described in section \ref{performance-metrics}) of each model (described in section \ref{exp:methods}) from the backtests. 
    \item A standard line plot illustrates the performance of the models against the baseline indexed in time of the cumulative product of logarithmic trade returns, where the trade returns are defined in equation \ref{eq:tradereturnsperformance}. 
    \item A boxplot illustrates the distribution of the monthly logarithmic returns\footnote{Again, trade returns are defined in equation \ref{eq:tradereturnsperformance} and resampled to produce monthly values. The logarithmic monthly returns are then calculated based on these values.} of each model and the baseline. Boxplots summarize a distribution by its sampled median, the first quantile ($Q_1$), and the third quantile ($Q_3$), represented by the box. The upper whisker extends to the largest observed value within $Q_3 + \frac{3}{2}IQR$, and the lower whisker extends to the smallest observed value within $Q_1 -  \frac{3}{2}IQR$, where the interquartile range (IQR) is $Q_3 - Q_1$. Dots represent all values outside of the whiskers (outliers). 
\end{itemize}
The plots display the performance of all models and the baseline and are grouped by risk-sensitivity terms.

Table \ref{experimentres-wrefit} below shows the results of the backtests averaged over $10$ runs. The variation between runs was small enough to warrant the level of precision of the results given in the table.

\begin{center}
 \begin{tabular}{ |p{2cm}||p{1cm}||p{1cm}||p{1cm}||p{1cm}||p{1cm}|  }
 \hline
 & $\mathbb{E}[R]$ & $Std(R)$ & Sharpe &  MDD & Hit \\
 \hline
 Buy \& Hold   & $0.271$ & $0.721$ & $0.376$ & $0.877$ & $0.524$ \\
  \hline
 \multicolumn{6}{|c|}{$\lambda_\sigma = 0$} \\
 \hline
 PG-CNN   & $\mathbf{0.403}$ & $0.558$ & $\mathbf{0.722}$ & $0.753$ & $0.529$ \\
 PG-LSTM   & $0.297$ & $\mathbf{0.502}$ & $0.591$ & $0.726$ & $0.527$ \\
 AC-CNN   & $0.302$ & $0.610$ & $0.495$ & $0.724$ & $0.538$ \\
 AC-LSTM   & $0.226$ & $0.694$ & $0.325$ & $\mathbf{0.637}$ & $\mathbf{0.541}$ \\
  \hline
  Average   & $0.307$ & $0.591$ & $0.533$ & $0.710$ & $0.534$ \\
 \hline
 \multicolumn{6}{|c|}{$\lambda_\sigma = 0.01$} \\
 \hline
 PG-CNN   & $\mathbf{0.401}$ & $0.437$ & $\mathbf{0.918}$ & $0.665$ & $0.537$ \\
 PG-LSTM   & $0.258$ & $0.326$ & $0.791$ & $0.540$ & $0.526$ \\
 AC-CNN   & $0.346$ & $0.471$ & $0.735$ & $0.601$ & $\mathbf{0.545}$ \\
 AC-LSTM   & $0.251$ & $\mathbf{0.300}$ & $0.837$ & $\mathbf{0.443}$ & $0.535$ \\
  \hline
  Average   & $0.314$ & $0.383$ & $0.820$ & $0.562$ & $0.536$ \\
 \hline
 \multicolumn{6}{|c|}{$\lambda_\sigma = 0.1$} \\
 \hline
 PG-CNN   & $\mathbf{0.371}$ & $0.356$ & $\mathbf{1.042}$ & $0.591$ & $0.537$ \\
 PG-LSTM   & $0.235$ & $0.264$ & $0.890$ & $0.373$ & $0.524$ \\
 AC-CNN   & $0.091$ & $0.239$ & $0.380$ & $0.392$ & $\mathbf{0.539}$ \\
 AC-LSTM   & $0.110$ & $\mathbf{0.190}$ & $0.579$ & $\mathbf{0.261}$ & $0.525$ \\
 \hline
  Average   & $0.202$ & $0.262$ & $0.723$ & $0.404$ & $0.531$ \\
  \hline
   \multicolumn{6}{|c|}{$\lambda_\sigma = 0.2$} \\
 \hline
 PG-CNN   & $\mathbf{0.243}$ & $0.298$ & $\mathbf{0.815}$ & $0.410$ & $0.533$ \\
 PG-LSTM   & $0.179$ & $0.247$ & $0.725$ & $0.373$ & $\mathbf{0.537}$ \\
 AC-CNN   & $0.136$ & $\mathbf{0.198}$ & $0.687$ & $0.454$ & $0.531$ \\
 AC-LSTM   & $0.114$ & $0.229$ & $0.498$ & $\mathbf{0.341}$ & $0.522$ \\
 \hline
  Average   & $0.168$ & $0.243$ & $0.681$ & $0.394$ & $0.531$ \\
  \hline
\end{tabular}
\captionof{table}{Backtest results}\label{experimentres-wrefit}
\end{center}

A pair of plots (line plot and boxplot) are grouped by risk-term values ($0$, $0.01$, $0.1$, and $0.2$, respectively).

\begin{figure}[H]
\centering
\includegraphics[width=\textwidth]{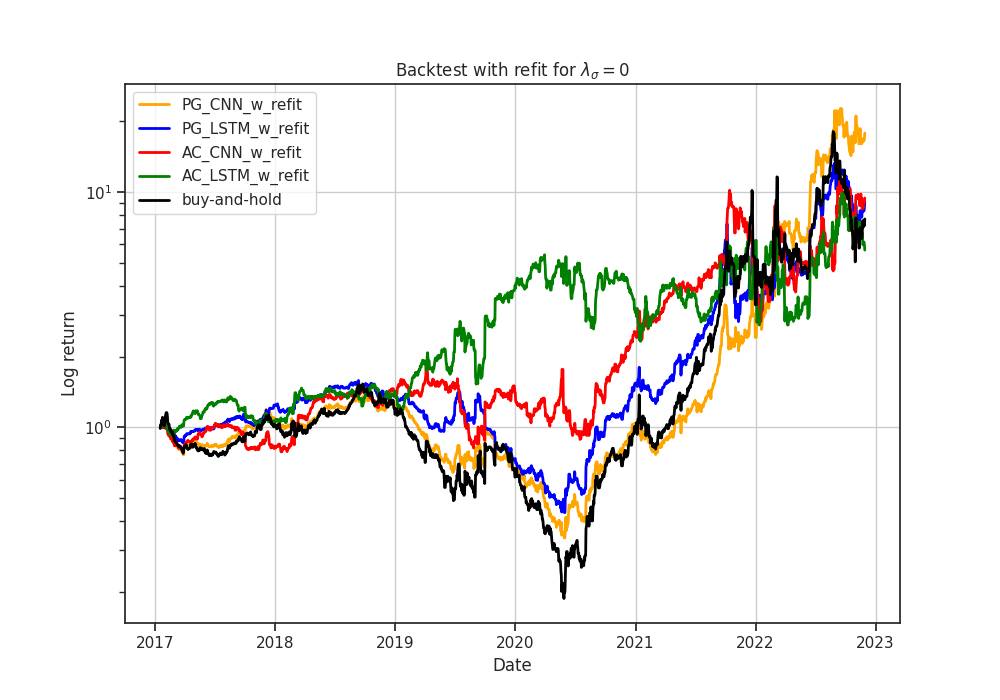}
\caption{Cumulative logarithmic trade returns for $\lambda_\sigma=0$}
\label{lineplot0}
\end{figure}

\begin{figure}[H]
\centering
\includegraphics[width=0.8\textwidth]{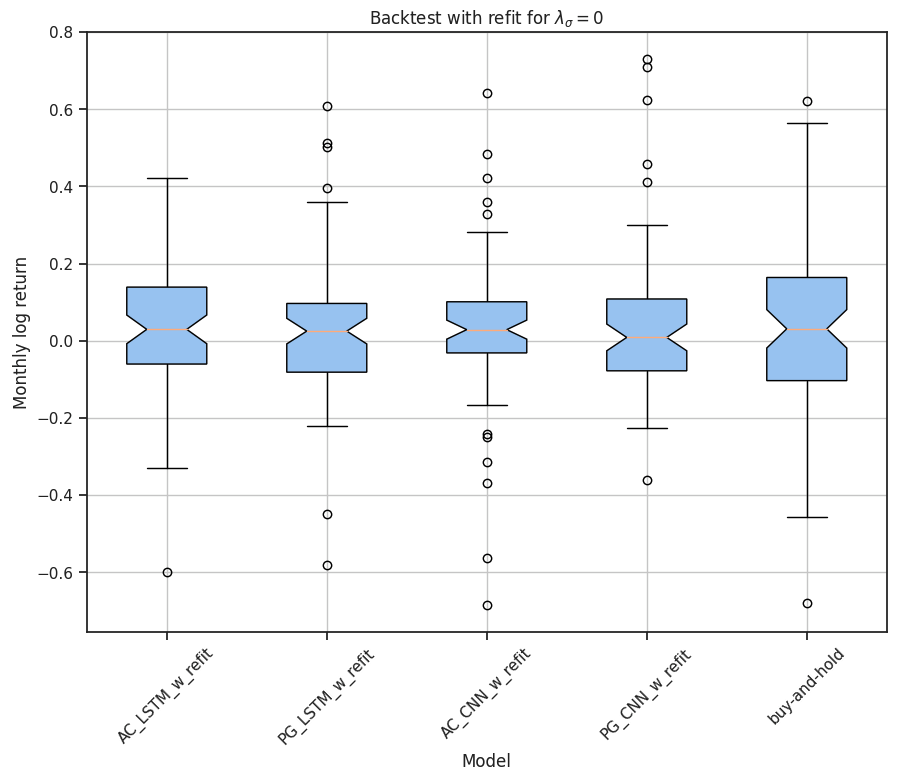}
\caption{Boxplot of monthly logarithmic trade returns for $\lambda_\sigma=0$}
\label{boxplot0}
\end{figure}

\begin{figure}[H]
\centering
\includegraphics[width=\textwidth]{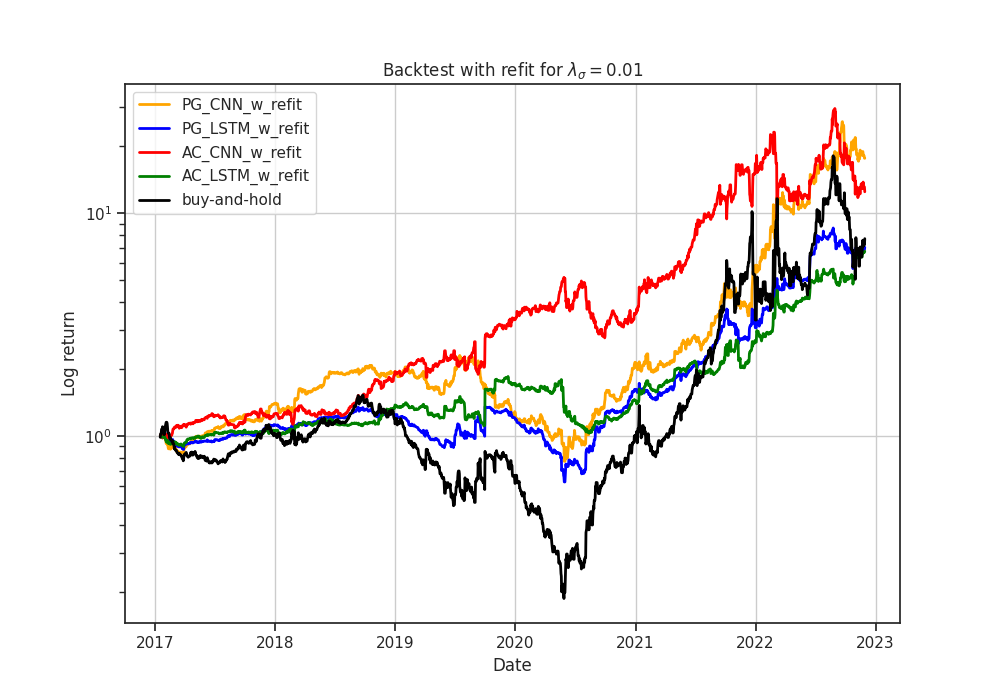}
\caption{Cumulative logarithmic trade returns for $\lambda_\sigma=0.01$}
\label{lineplot001}
\end{figure}

\begin{figure}[H]
\centering
\includegraphics[width=0.8\textwidth]{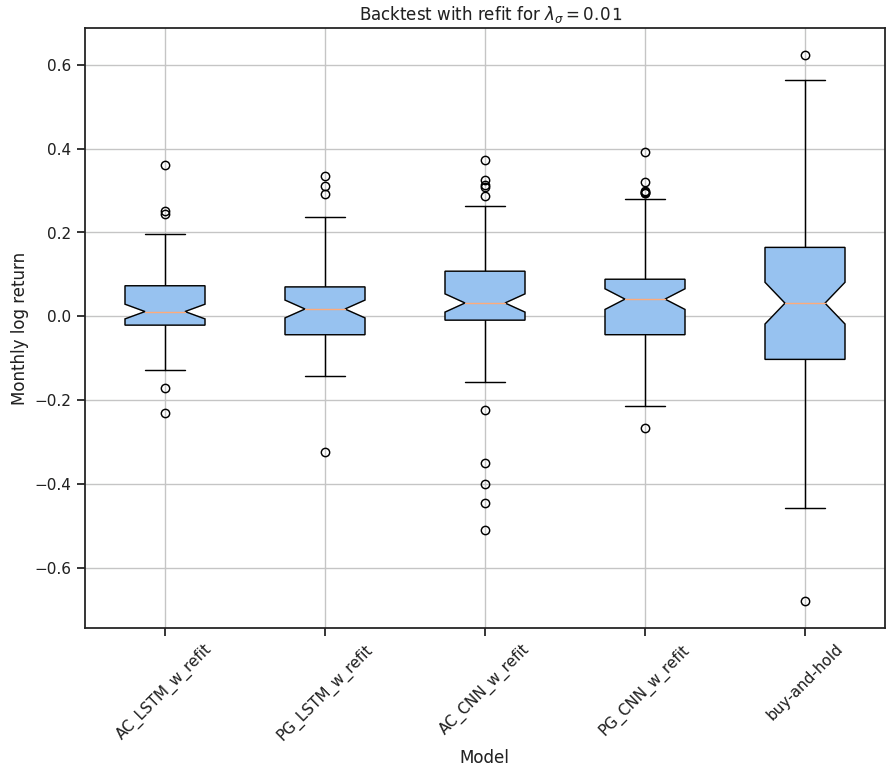}
\caption{Boxplot of monthly logarithmic trade returns for $\lambda_\sigma=0.01$}
\label{boxplot001}
\end{figure}

\begin{figure}[H]
\centering
\includegraphics[width=\textwidth]{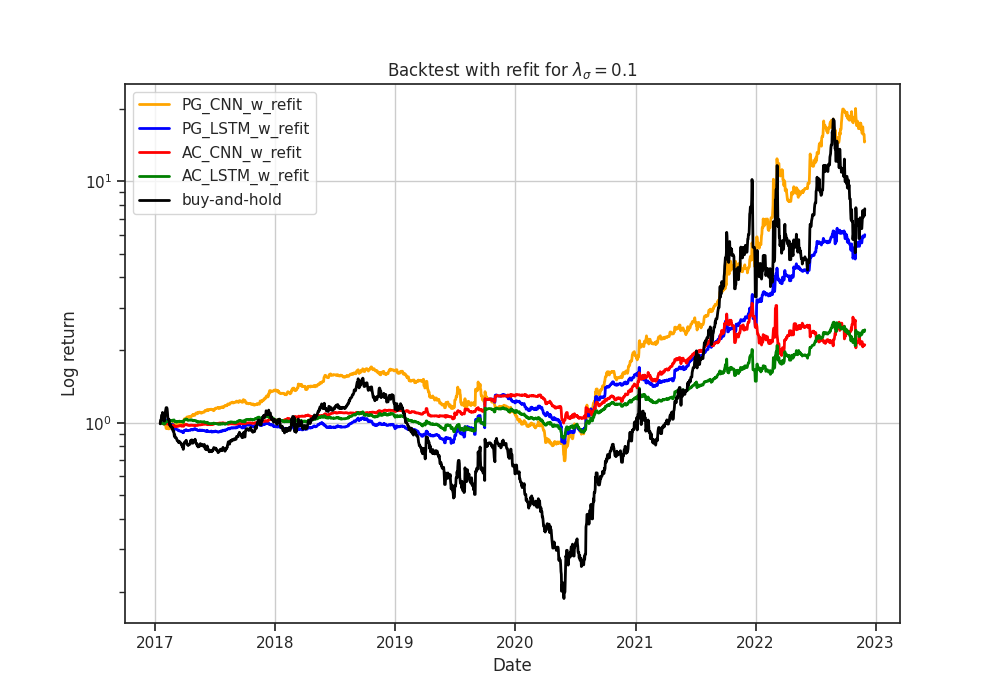}
\caption{Cumulative logarithmic trade returns for $\lambda_\sigma=0.1$}
\label{lineplot01}
\end{figure}

\begin{figure}[H]
\centering
\includegraphics[width=0.8\textwidth]{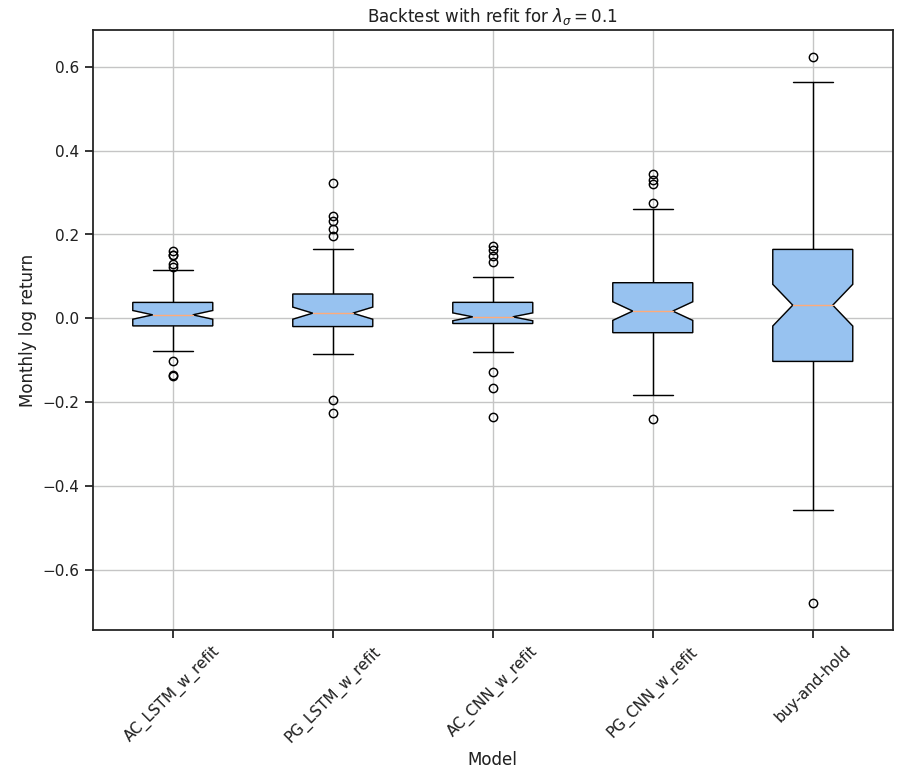}
\caption{Boxplot of monthly logarithmic trade returns for $\lambda_\sigma=0.1$}
\label{boxplot01}
\end{figure}

\begin{figure}[H]
\centering
\includegraphics[width=\textwidth]{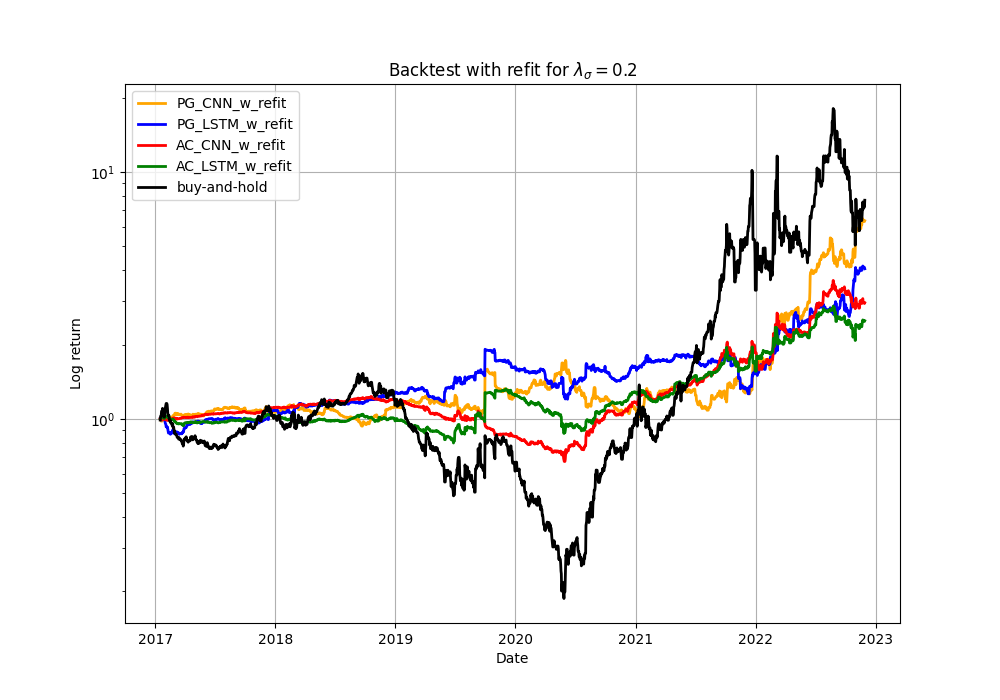}
\caption{Cumulative logarithmic trade returns for $\lambda_\sigma=0.2$}
\label{lineplot02}
\end{figure}

\begin{figure}[H]
\centering
\includegraphics[width=0.8\textwidth]{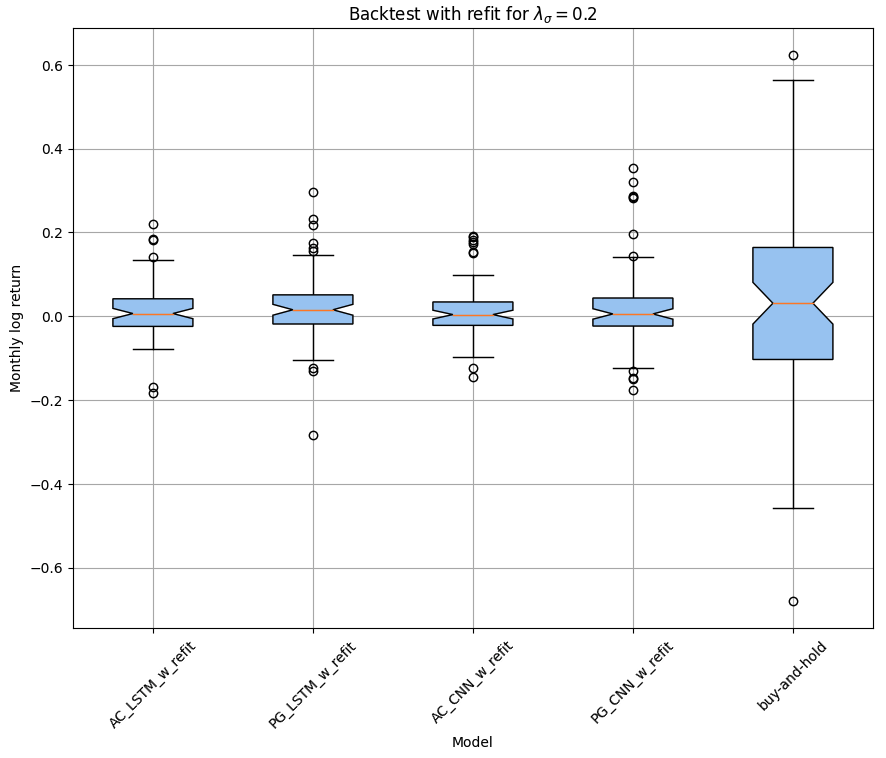}
\caption{Boxplot of monthly logarithmic trade returns for $\lambda_\sigma=0.2$}
\label{boxplot02}
\end{figure}

\subsection{Discussion of results}\label{exp:discussion-of-res}

This section will discuss the experiment results and how they relate to the three research questions posed at the start of this thesis. 

% Compare against baseline - risk
\subsubsection{Risk/reward}
The first research question posed at the start of this thesis was: 
\begin{quote}
    Can the risk of algorithmic trading agents operating in volatile markets be controlled? 
\end{quote}
Note some general observations about the buy-and-hold baseline against which the models are compared.
The baseline mirrors the direction of the natural gas market during the out-of-sample test. 
Due to the volatility and upwards price pressure stemming from the energy crisis in 2021-2022, the buy-and-hold baseline has a high annualized expected return but also a high annualized standard deviation of returns. 
As a result, the Sharpe ratio, the primary performance metric in this experiment, is relatively low. 
The primary goal of the deep reinforcement learning models is to increase the Sharpe ratio. 
Increasing the Sharpe ratio is achieved by increasing the expected returns or decreasing the standard deviation. 
The reward function \ref{eq:final-ret} defines this trade-off, where the risk-sensitivity term $\lambda_\sigma$ functions as a trade-off hyperparameter.
Low values of $\lambda_\sigma$ make the agent more risk-neutral, i.e., more concerned with increasing expected returns and less concerned with decreasing the standard deviation of returns. 
Conversely, high values of $\lambda_\sigma$ make the agent more risk-averse, i.e., more concerned with decreasing the standard deviation of returns and less concerned with increasing the expected returns. 
The experiments in this thesis use four risk-sensitivity terms; $0$, $0.01$, $0.1$, and $0.2$. 
For $\lambda_\sigma=0$, the agent is risk-neutral and only concerned with maximizing the expected return. 
For values exceeding $0.1$, the agent becomes so risk-averse that it hardly participates in the market. 
This trade-off is evident in the results where the annualized expected return and the standard deviation are on average $83\%$ and $143\%$ higher, respectively, for $\lambda_\sigma=0$ compared to $\lambda_\sigma=0.2$. 
The boxplots (figures \ref{boxplot0}, \ref{boxplot001}, \ref{boxplot01}, \ref{boxplot02}) illustrate how the monthly logarithmic returns are more concentrated as $\lambda_\sigma$ increase. 
This phenomenon is also observed in the line plots (figures \ref{lineplot0}, \ref{lineplot001}, \ref{lineplot01}, \ref{lineplot02}), where it is apparent that the variability of returns decreases as $\lambda_\sigma$ increases. 
The agent's action, i.e., the position size (and direction), is its only means to affect the reward. 
By definition, a risk-averse agent will choose outcomes with low uncertainty, so a preference for smaller positions is a natural consequence of increasing risk sensitivity. Consequently, the risk-averse deep RL agents have a significantly lower maximum drawdown than the baseline, but they do not fully capitalize on the increasing prices from mid-2020 to 2023.

Averaged over all the deep RL models for all risk-sensitivity terms, they produce $83 \%$ higher Sharpe than the baseline. 
This gain comes from reducing the standard deviation of returns, which is reduced by $49\%$, whereas the return is only slightly reduced by $8\%$. 
Looking at the various risk-sensitivity terms, the risk-neutral agents (i.e., those where $\lambda_\sigma=0$), on average, increase the returns by $13\%$ compared to the baseline. 
Although they have no risk punishment, they also decrease the standard deviation of returns by $18\%$. 
This last point is surprising but could be a byproduct of an intelligent agent trying to maximize returns. 
For $\lambda_\sigma=0.01$, the deep RL agents, on average, produce $16\%$ increased returns and $47\%$ reduced standard deviation of returns compared to the baseline. 
This combination results in a $118\%$ higher Sharpe ratio. 
For $\lambda_\sigma=0.1$, the agents on average produce $25\%$ lower returns; however, the standard deviation of returns is reduced more by $64\%$. Thus, the Sharpe is increased by $92\%$ compared to the baseline. 
The most risk-averse agents (i.e., those where $\lambda_\sigma=0.2$) on average produce $38\%$ lower returns with $66\%$ lower standard deviation of returns, yielding an $83\%$ increase in Sharpe compared to the baseline. 
The risk-sensitivity term $\lambda_\sigma=0.01$ produces the highest Sharpe ratio on average. 
Thus, the backtests suggest that of the four risk-sensitivity options tested in this thesis, $\lambda_\sigma=0.01$ strikes the best risk/reward balance.

\subsubsection{RL models}

The second research question posed at the start of this thesis was: 
\begin{quote}
    What reinforcement learning algorithms are suitable for optimizing an algorithmic training agent in an online, continuous time setting? 
\end{quote}
A curious result from the experiment is that, for three out of four risk-sensitivity terms\footnote{$\lambda_\sigma=0$, $0.01$, and $0.1$}, the model with the highest hit-rate has the lowest Sharpe. 
In other words, the model making the highest rate of profitable trades also produces the lowest risk-adjusted returns. 
This result illustrates the complexity of trading financial markets and justifies the methods chosen in this thesis. 
Firstly, there is no guarantee that a higher percentage of correct directional calls will result in higher returns. Therefore, a forecast-based supervised learning approach optimized for making correct directional calls may not align with the stakeholder's ultimate goal of maximizing risk-adjusted returns. 
For an algorithmic trading agent to achieve the desired results, e.g., making trades that maximize the Sharpe ratio, it should be optimized directly for that objective. 
However, doing so in a supervised learning setting is not straightforward. Reinforcement learning, on the other hand, provides a convenient framework for learning optimal sequential behavior under uncertainty. 
Furthermore, discrete position sizing, a drawback of value-based reinforcement learning, can expose the agent to high risks. However, the agent can size positions based on confidence through the continuous action space offered by policy gradient methods, allowing for more effective risk management.
Section \ref{c:at-mapping2pos} presented research arguing that, for algorithmic trading, reinforcement learning is superior to supervised learning and policy gradient methods are superior to value-based methods, and this result supports those arguments.

Although previous research supported policy gradient methods, there was no consensus on which one was superior in this context. 
Chapter \ref{c:m-RL-algo} presented two policy gradient methods: one based on an actor-only framework and the other based on an actor-critic framework, and discussed their respective advantages and disadvantages. The previous section (\ref{exp:results}) presented results from the backtests where both algorithms were tested out-of-sample. 
For all risk-sensitivity terms, the direct policy gradient algorithm, on average, produces a $49\%$ higher Sharpe ratio than the deterministic actor-critic algorithm. 
Comparing the two algorithms using the same network architecture and risk sensitivity term reveals that the actor-based algorithm outperforms the actor-critic-based algorithm in 7 out of 8 combinations. 
The only case where the actor-critic-based algorithm performs better\footnote{PG-LSTM vs. AC-LSTM for $\lambda_\sigma=0.01$} is the case with the smallest performance gap. 
Furthermore, the actor-only direct policy gradient method strictly increases the Sharpe ratio for both network architectures as the risk-sensitivity parameter $\lambda_\sigma$ increases to a maximum at $\lambda_\sigma=0.1$. 
The actor-critic method does not follow this pattern, suggesting it fails to achieve its optimization objective.

% Observed unbiased reward vs. estimated reward. 
The performance gap between the actor-based and actor-critic-based algorithms is significant enough to warrant a discussion. 
An explanation for the performance gap could be that the actor-critic-based algorithm optimizes the policy using a biased Q-network reward estimate instead of the observed unbiased reward. 
As a data-generating process, the commodity market is complex and non-stationary. 
If the Q-network closely models the data-generating distribution, using reward estimates from sampled experience from a replay memory is an efficient method for optimizing the policy. 
On the other hand, it is also clear that a policy that is optimized using Q-network reward estimates that are inaccurate will adversely affect performance. 
The direct policy gradient algorithm optimizes the policy using the observed unbiased reward and avoids this problem altogether. 
Given that the reward function is exactly expressed, optimizing it directly, as the direct policy gradient method does, is the most efficient approach.
Many typical RL tasks work well with the actor-critic framework, but the backtests indicate that financial trading is not one of them.

\subsubsection{Networks}

% Compare networks against each other
The third and final research question posed at the start of this thesis was: 
\begin{quote}
    What deep learning architectures are suitable for modeling noisy, non-stationary financial data? 
\end{quote}
In the research presented in section \ref{c:at-forecasting}, two types of deep learning architectures stood out; the long short-term memory and the convolutional neural network. 
Chapter \ref{c:network-topology} presented two types of parametric function approximators based on the CNN- and LSTM-architecture, respectively. 
The previous section (\ref{exp:results}) presented results from the backtests where both these function approximators are tested out-of-sample. 
On average, the CNN-based models produce over $5\%$ higher Sharpe than those based on the LSTM, which is surprising, as LSTMs are generally viewed as superior in sequence modeling and, due to their memory, are the preferred option when modeling POMDPs. 
In contrast to the CNN, the LSTM can handle long-term dependencies, but it seems the lookback window provides enough historical information for the CNN to make trade decisions.  
However, the performance gap is not big enough to say anything conclusive, and the LSTM outperforms the CNN in some tests, so it is unclear which is most suitable.

One interesting observation is that the CNN-based models produce higher returns and standard deviation of returns compared to the LSTM.
On average, the CNN-based models produce $37\%$ higher returns and $15\%$ higher standard deviation of returns. 
From the line plots in figures \ref{lineplot0}, \ref{lineplot001}, \ref{lineplot01}, and \ref{lineplot02}, it looks like a possible explanation for this is that the LSTM-based models prefer smaller position sizes compared to the CNN-based models. 
One potential reason for this phenomenon involves the difference in how the CNN and LSTM are optimized. 
Generally speaking, the CNN-based model is far easier and quicker to optimize than the LSTM-based model, partly due to batch norm, which in its conventional form is incompatible with RNNs. 
Another reason is that when the LSTM is trained for long sequences, the problem of vanishing gradients makes back-propagating the error difficult and slow. Increasing the learning rate leads to exploding gradients. 
The CNN-based model with batch norm quickly and effectively adjusts its parameters to take full advantage of newly observed information during out-of-sample tests. The LSTM-based model, on the other hand, adjusts its parameters much slower. 
As a result, the actions it selects often end up someplace in the middle of the action space causing smaller position sizes, lower returns, and lower standard deviation of returns. 
For that reason, the author of this thesis theorizes that the performance gap between the CNN-based and LSTM-based models would increase with time.

\subsection{Discussion of model}\label{exp:discussion-of-mod}

Following the discussion of the results, it is interesting to take a step back and have a more general discussion of the model. This includes discussing the strengths and weaknesses of the model, as well as potential applications and limitations.

\subsubsection{Environment}

Solving complex real-world problems with reinforcement learning generally requires creating a simplified version of the problem that lends itself to analytical tractability. Usually, this involves removing some of the frictions and constraints of the real-world problem. In the context of financial trading, the environment described in chapter \ref{part:problemsetting} makes several simplifying assumptions about the environment, including no market impact, no slippage, the ability to purchase or sell any number of contracts at the exact quoted price, no additional costs or restrictions on short-selling, and fractional trading. It is imperative to note that these assumptions do not necessarily reflect real-world conditions. As such, it is crucial to know the problem formulation's limitations and how it will negatively affect the model's generalizability to the real-world problem. 
Poorly designed environments, where agents learn to exploit design flaws rather than the actual problem, are a frequent problem in reinforcement learning \cite{sutton2018reinforcement}. At the same time, these simplifying assumptions allow for a clean theoretical analysis of the problem. 
Furthermore, the environment introduces some friction through transaction costs, an improvement over many existing models.

Lookahead bias in the input data is avoided by using the price series alone as input, as described in section \ref{c:nt-network-input}. 
The price series of a financial instrument is generally the most reliable predictor of future prices. 
However, price series only provide a limited view of the market and do not consider the broader economic context and the potential impact of external factors. As a result, an agent relying solely on price series may miss out on meaningful predictive signals.
Furthermore, since the model learns online, an effective data governance strategy is required to ensure the quality and integrity of the real-time input data stream, as data quality issues can harm the model's performance.
The dollar bars sampling scheme described in section \ref{c:ps-td} has solid theoretical foundations for improving the statistical properties of the sub-sampled price series compared to traditional time-based sampling. When using this sampling scheme, however, the agent cannot be certain of the prediction horizon, which complicates forecasting.

Commodity trading firms often conduct asset-backed trading in addition to paper trading, which incurs additional costs for booking pipeline capacity, storage capacity, or LNG tankers. The model does not currently include these costs, but the environment could be adjusted to include them.

\subsubsection{Optimization}

Statistical learning relies on an underlying joint feature-target distribution $F(x,y)$, with non-vanishing mutual information. 
The algorithmic trading agent approximates this function by learning the distribution through historical data. 
As financial markets are nonstationary, statistical distributions constantly change over time, partly because market participants learn the market dynamics and adjust their trading accordingly.
In order to remain relevant for the near future, the model must be continuously refitted using only data from the immediate past, at the expense of statistical significance. 
On the other hand, training a complex model using only a relatively small set of recent data is challenging in a high-noise setting such as financial forecasting, often resulting in poor predictive power.
This tradeoff between timeliness and statistical significance is known as the \textit{timeliness-significance} tradeoff \cite{isichenko2021quantitative}. 
The timeliness-significance tradeoff highlights a central challenge in optimizing algorithmic trading models.

This thesis investigates the use of reinforcement learning in algorithmic trading, a field traditionally dominated by supervised learning-based approaches. 
Supervised learning is a straightforward method for easily labeled tasks, such as forecasting financial prices. Reinforcement learning, on the other hand, is better suited to complex problems, such as sizing positions, managing risk and transaction costs. In fact, with only minor modifications, the model outlined in this thesis can optimize an agent trading a portfolio of arbitrary size. 
For this reason, reinforcement learning was chosen as the algorithmic trading agents' optimization framework. 
Temporal credit assignment is one of the main strengths of reinforcement learning, making it ideal for game playing and robotics, involving long-term planning and delayed reward. In this problem, however, temporal credit assignment does not apply since trading is an immediate reward process. 
Furthermore, the complexity of reinforcement learning compared to supervised learning comes at a price.
As well as requiring a model of the environment in which the agent interacts and learning to associate context with reward-maximizing actions, reinforcement learning introduces added complexity by introducing the exploration-exploitation tradeoff. 
With more moving parts, reinforcement learning can be significantly more challenging to implement and tune and is generally less sample efficient than supervised learning. 
The learning process is more convoluted as the agent learns through reinforcement signals generated by interaction with the environment and involves stochasticity. Consequently, the model can display unstable behavior where the policy diverges, or the agent overfits to noise. 
E.g., if the market experiences a sustained downward trend, the agent can be deceived into believing that the market will continue to decline indefinitely. 
As a result, the agent may adjust its policy to always short the market, which will have disastrous effects once the market reverses. 
The phenomenon is caused by the temporal correlation of sequential interactions between RL agents and the market, and that reinforcement learning is sample inefficient, making it difficult to obtain good gradient estimates.
Replay memory can ensure that gradient estimates are derived from a wide variety of market conditions.
However, replay memory introduces biased gradient estimates, which, according to backtests, is a poor tradeoff. 
The timeliness-significance tradeoff further complicates this problem of obtaining suitable gradient estimates. 
A supervised learning framework is more straightforward and avoids much of the complexity associated with reinforcement learning.
Thus, it is unclear whether reinforcement learning or supervised learning is the most appropriate optimization framework for algorithmic trading.

\subsubsection{Interpretability and trust}

Interpretability is the ability to determine the cause and effect of a model. Due to their over-parameterized nature, deep neural networks possess a remarkable representation capacity, enabling them to solve a wide range of complex machine learning problems, but at the cost of being difficult to interpret. 
Neural networks are black boxes, acceptable for low-risk commercial applications such as movie recommendations and advertisements. Interpreting these predictions is of little concern as long as the models have sufficient predictive power. 
However, deep learning's opaque nature prevents its adoption in critical applications, as the failure of a commodity trading model could result in substantial financial losses and threaten global energy security. 
Setting aside people generally being risk-averse and unwilling to bet large sums of money on a black box, is the aversion to applying deep learning in critical domains such as commodity trading reasonable? Does understanding the model even matter as long as it delivers satisfactory backtest performance?

This question can be answered by reviewing statistical learning theory. Generally, machine learning models are tested under the assumption that observations are drawn from the same underlying distribution, the data-generating distribution, and that observations are IID. 
In this setting, the test error serves as a proxy for the generalization error, i.e., the expected error on new observations. 
However, the dynamics of the financial markets are constantly changing. 
Fierce competition in financial markets creates a cycle in which market participants attempt to understand the underlying price dynamics. As market participants better understand market dynamics, they adjust their trading strategies to exploit that knowledge, further changing market dynamics. Due to the constantly changing dynamics of the market, models that worked in the past may no longer work in the future as inefficiencies are arbitraged away\footnote{Volatility, for example, used to be a reliable indicator of future returns \cite{isichenko2021quantitative}.}. 
Therefore, it is important to be cautious when interpreting backtest errors as generalization errors, as it is unlikely that observations sampled at different points in time are drawn from the same probability distribution. 
Even if one disregards all the flaws of a backtest\footnote{E.g., not accounting for market impact and lookahead bias.}, the backtest, at best, only reflects performance on historical data. 
In no way is this intended to discourage backtesting. However, naively interpreting backtest performance as an assurance of future performance is dangerous. 
Referring back to section \ref{c:at-backtest}; a backtest is a historical simulation of how the model would have performed should it have been run over a past period. 
Even exceptional results from the most flawlessly executed backtest can never guarantee that the model generalizes to the current market. 
Furthermore, the results should be interpreted cautiously if no ex-ante logical foundation exists to explain them. 
Deep neural networks are highly susceptible to overfitting to random noise when trained on noisy financial time series. 
It is, however, difficult to determine if the agent has detected a legitimate signal if the model is not interpretable. 
Even if the model detects a legitimate signal in the backtests, other market participants may discover the same signal and render the model obsolete. 
Again, determining this is difficult without knowing what inefficiencies the model exploits, and deploying it until it displays a sustained period of poor performance will be costly.

In response to the question of whether or not understanding a model matters if it performs well on a backtest, the answer is an emphatic \textit{yes}.
Blindly taking backtest performance of a black box as an assurance of future performance in a noisy and constantly changing environment can prove costly. 
Thus, the aversion to adopting deep learning in algorithmic trading is reasonable. 
Ensuring that the trading decisions are explainable and the models are interpretable is essential for commercial and regulatory acceptance.
To address this challenge, models should be created with a certain degree of interpretability. This way, stakeholders can get insight into which inefficiencies the model exploits, evaluate its generalizability, and identify its obsolescence \textit{before} incurring significant losses. 
The use of deep learning in algorithmic trading can still be viable with techniques such as explainable AI and model monitoring.

\afterpage{\blankpage}
\newpage
\section{Future work}\label{future-work}

The methods presented in this thesis leave room for improvement in further work. 
More investigation should be done to evaluate the effectiveness of existing methods in different contexts. 
Further investigation will enable a deeper understanding of the model and its generalizability and provide an opportunity to identify potential areas for improvement. 
Considering the lack of real-world market data, one option is to use generative adversarial networks (GANs) to generate synthetic markets \cite{yoon2019time}. GANs can generate unlimited data, which can be used to train and test the model and its generalizability. 
Additionally, the lack of established baselines could be improved upon. While the buy-and-hold baseline is well understood and trusted, it is unrealistic in this context, as futures contracts expire. 
Although it presents its own challenges, developing a baseline more appropriate for futures trading would improve the current model. 
Furthermore, a greater level of interpretability is required to achieve real-world adoption. Therefore, combining algorithmic trading research with explainable AI is imperative to improve existing methods' interpretability.

Incorporating non-traditional data sources, such as social media sentiment or satellite images, may prove beneficial when forecasting market returns. Alternative data can provide a more comprehensive and holistic view of market trends and dynamics, allowing for more accurate predictions. By leveraging alternative data, algorithmic trading agents can gain an edge over their competitors and make better-informed decisions.
Using deep learning techniques such as natural language processing and computer vision to analyze text or image data in an algorithmic trading context is promising.
Neural networks are generally effective in problems with complex structures and high signal-to-noise ratios. Thus, it may be more appropriate to use deep learning to extract features from images or text rather than analyzing price series.

Lastly, the methods presented in this thesis are limited to trading a single instrument. They are, however, compatible with portfolio optimization with minimal modifications. Further research in this area would be interesting, as it better utilizes the potential of the reinforcement learning framework and the scalability of data-driven decision-making.

\afterpage{\blankpage}
\newpage
\section{Conclusion}\label{sec:conclusion}

This thesis investigates the effectiveness of deep reinforcement learning methods in commodities trading. Previous research in algorithmic trading, state-of-the-art reinforcement learning, and deep learning algorithms was examined, and the most promising methods were implemented and tested. 
This chapter summarizes the thesis' most important contributions, results, and conclusions.

This thesis formalizes the commodities trading problem as a continuing discrete-time stochastic dynamical system. 
The system employs a novel time-discretization scheme that is reactive and adaptive to market volatility, providing better statistical properties of the sub-sampled financial time series. 
Two policy gradient algorithms, an actor-based and an actor-critic-based, are proposed to optimize a transaction-cost- and risk-sensitive agent. 
Reinforcement learning agents parameterized using deep neural networks, specifically CNNs and LSTMs, are used to map observations of historical prices to market positions.

The models are backtested on the front month TTF Natural Gas futures contracts from 01-01-2017 to 01-01-2023. 
The backtest results indicate the viability of deep reinforcement learning agents in commodities trading. 
On average, the deep reinforcement learning agents produce an $83 \%$ higher Sharpe ratio out-of-sample than the buy-and-hold baseline. 
The backtests suggest that deep RL models can adapt to the unprecedented volatility caused by the energy crisis during 2021-2022. 
Introducing a risk-sensitivity term functioning as a trade-off hyperparameter between risk and reward produces satisfactory results, where the agents reduce risk as the risk-sensitivity term increases. 
The risk-sensitivity term allows the stakeholder to control the risk of an algorithmic trading agent in volatile markets. 
The direct policy gradient algorithm produces significantly higher Sharpe ($49\%$ on average) than the deterministic actor-critic algorithm, suggesting that an actor-based policy gradient method is more suited to algorithmic trading in an online, continuous time setting. 
The parametric function approximator based on the CNN architecture performs slightly better ($5\%$ higher Sharpe on average) than the LSTM, possibly due to the problem of vanishing gradients for the LSTM.

The algorithmic trading problem is made analytically tractable by simplifying assumptions that remove market frictions.
Performance may be inflated due to these assumptions and should be viewed with a high degree of caution.

%\afterpage{\newpage}
%\afterpage{\blankpage}
%\appendix
%\section{Appendix}

\afterpage{\blankpage}
\newpage
\printglossary[type=\acronymtype]
\addcontentsline{toc}{section}{List of Abbreviations}

\afterpage{\blankpage}
\newpage
\bibliographystyle{alpha} % We choose the "alpha" reference style
\bibliography{sample} % Entries are in the sample.bib file

\end{document}